\documentclass[a4paper,onecolumn,11pt,unpublished]{quantumarticle}
\pdfoutput=1
\usepackage[utf8]{inputenc}
\usepackage[english]{babel}
\usepackage[T1]{fontenc}

\usepackage{amsmath}
\usepackage{amssymb}
\usepackage{amsthm}
\usepackage{mathtools}
\usepackage{bm}
\usepackage{quantikz}
\usepackage{siunitx}
\usepackage{booktabs}
\usepackage{multirow}

\definecolor{LB}{RGB}{134,41,198}

\newcommand{\osc}{\operatorname{osc}(f)}

\newcommand{\norm}[1]{\left\lVert#1\right\rVert}
\newcommand{\eTrotter}{\mathcal{E}_{\mathrm{Trotter}}}
\newcommand{\don}{\epsilon_{\mathrm{online}}}
\newcommand{\eonline}{\mathcal{E}_{\mathrm{online}}}

\newtheorem{theorem}{Theorem}
\newtheorem{lemma}{Lemma}
\newtheorem{proposition}{Proposition}

\usepackage{graphicx}
\usepackage{subcaption}

\usepackage{csquotes}
\usepackage{hyperref}
\hypersetup{
    colorlinks=true,
    allcolors=black
}
\usepackage[nameinlink]{cleveref}

\makeatletter

\@ifundefined{newcounteralias}{%
  \newcounter{circuit}%
  \let\c@circuit\c@equation
  
}{%
  \newcounteralias{circuit}{equation}%
}
\let\bra\relax
\let\ket\relax
\let\braket\relax
\let\ketbra\relax

\DeclarePairedDelimiter{\bra}{\langle}{\rvert}
\DeclarePairedDelimiter{\ket}{\lvert}{\rangle}
\DeclarePairedDelimiterX{\braketLR}[2]{\langle}{\rangle}{#1\,\delimsize\vert\,#2}
\DeclarePairedDelimiterX{\Matrixel}[3]{\langle}{\rangle}{%
  #1\,\delimsize\vert\,#2\,\delimsize\vert\,#3%
}
\newcommand{\braket}[2]{\braketLR*{#1}{#2}}
\newcommand{\ketbra}[2]{\ket*{#1}\!\bra*{#2}}
\newcommand{\matrixel}[3]{\Matrixel*{#1}{#2}{#3}}

\makeatother

\crefname{circuit}{circuit}{circuits}
\Crefname{circuit}{Circuit}{Circuits}

\crefformat{circuit}{#2circuit~(#1)#3}
\Crefformat{circuit}{#2Circuit~(#1)#3}

\newenvironment{qcircuit}[1][]%
{%
  \refstepcounter{circuit}%
  \ifx\relax#1\relax\else\label{#1}\fi
  \begin{equation*}
    \tag{\theequation}
}%
{%
  \end{equation*}
  \ignorespacesafterend
}

\begin{document}
\title{Fully quantum training of parametric circuits with Quantum Hamiltonian Descent}

\author{Marco Burato} \email{marco.burato@unifi.it}
\affiliation{Department of Physics and Astronomy, University of Florence,
via G. Sansone 1, I-50019 Sesto Fiorentino (FI), Italy}

\author{Leonardo Banchi} \email{leonardo.banchi@unifi.it}
\affiliation{Department of Physics and Astronomy, University of Florence,
via G. Sansone 1, I-50019 Sesto Fiorentino (FI), Italy}
\affiliation{ INFN Sezione di Firenze, via G. Sansone 1, I-50019, Sesto Fiorentino (FI), Italy }

\begin{abstract}
Parametric quantum circuits are usually trained by classical, local optimizers, which use the quantum computer only to evaluate the cost function, or its gradient, one point at a time.
Quantum Hamiltonian Descent (QHD) takes a different route: 
in this quantum algorithm for continuous optimization,
the parameters become the position of a quantum particle, with the cost function as its potential, and two time schedules control kinetic and potential energy so that the wavefunction explores the whole landscape and tunnels towards the global minimum.
QHD, however, assumes a classically computable cost function, and its digital implementation is poorly understood: standard error bounds grow exponentially with the number of qubits per parameter, and the schedules are borrowed from classical optimization.
Here we propose to train parametric circuits with QHD entirely on the quantum computer, through an online step in which the qubits encoding the parameters coherently control the circuit on a second register.
We prove that, for cost functions with a smooth periodic extension,
 including those of parametric circuits, the simulation error admits a bound independent of the grid resolution, and that for circuits the bound grows at most linearly with the number of parameters.
We also show that the ideal dynamics depends on the two schedules only through a single curve, which controls the adiabaticity of the search, and that the remaining freedom can be used to minimize the gate count.
Simulations on benchmark functions and parametric circuits show that, at equal cost, schedules designed in this way markedly increase the probability of finding the global minimum.
\end{abstract}

\maketitle


\section{Introduction}%
\label{Introduction}

Training a parametric quantum circuit is a non-convex optimization problem.
In variational quantum algorithms and quantum machine learning models \cite{cerezo2021variational,bharti2022noisy,cerezo2022challenges}, a cost function $f(\bm x)$, typically the expectation value of an observable on a state prepared by a circuit with rotation angles $\bm x$, is minimized by a classical optimizer, with gradients estimated from measurements through the parameter-shift rule \cite{mitarai2018quantum,schuld2019evaluating, banchi2021measuring,banchi2026overshifted}.
The resulting landscapes are rugged, with many local minima \cite{anschuetz2022quantum} and narrow gorges \cite{arrasmith2022equivalence}, and in large circuits they may flatten into barren plateaus \cite{mcclean2018barren}.
Therefore, the success of gradient-based optimizers is often limited. 
Moreover, the quantum computer is only used to evaluate $f$ and its gradient at one point at a time, while the optimization itself is entirely classical.
Gradient descent is the de facto standard in training deep neural networks thanks to the backpropagation algorithm, but getting backpropagation-like scaling in the quantum setting is challenged by the destructive nature of quantum measurements \cite{abbas2023quantum}.
Several proposals have explored alternative hybrid quantum-classical training schemes \cite{Ostaszewski2021structure,tognini2025solving}, or 
the use of quantum resources in the optimization itself, for instance through quantum algorithms for gradient-based methods \cite{rebentrost2019quantum,li2026learning}.

Quantum Hamiltonian Descent (QHD) \cite{leng2023quantum} takes a different route.
Building on the variational formulation of accelerated gradient methods in terms of a Bregman Lagrangian \cite{wibisono2016variational}, QHD quantizes the corresponding classical dynamics: the optimization variables become the position of a quantum particle, whose wavefunction evolves under a Hamiltonian $\hat H(t)=a(t)\,\hat{\bm p}^2/2+b(t)\,f(\hat{\bm x})$, in which the kinetic term is gradually switched off and the cost function is switched on as a potential.
Starting from a uniform superposition over the domain, the state explores the whole landscape at once, can tunnel through barriers between valleys, and, if the evolution is slow enough, ends up concentrated around the global minimum, as in adiabatic quantum computation \cite{albash2018adiabatic}.
QHD has been reported to compare favorably with classical solvers on a variety of non-convex benchmarks \cite{leng2023quantum}, and has been extended to non-smooth problems \cite{leng2025nonsmooth}.

Two questions are central to implementing QHD on a digital quantum computer.
The first one concerns the cost and the accuracy of the simulation, and arises regardless of whether we are optimizing ``classical functions'' or parametric quantum circuits.
The parameters are discretized on a grid of $2^q$ points per dimension, encoded in $q$ qubits, and the evolution is simulated with spectral methods \cite{childs2022quantum} and product formulas.
The error of a product formula is controlled by the commutator between the kinetic and potential terms, which contains the momentum operator: its norm grows with the grid resolution, so that standard bounds, based on operator norms, would predict an error growing exponentially with $q$.
Moreover, QHD depends on two time schedules, $a(t)$ and $b(t)$, which have so far been borrowed directly from classical optimization algorithms \cite{leng2023quantum}, rather than designed on the basis of the cost of the simulation or of the structure of the landscape.
The second question concerns the cost function.
In the original formulation $f$ is assumed to be known and efficiently computable on a classical computer, so that the potential step can be implemented by reversible arithmetic.
When $f$ is defined by a parametric quantum circuit, instead, it is only accessible through measurements, and estimating it on the whole grid would require a number of circuit evaluations exponential in the total number of qubits.

In this work we propose to use QHD to train parametric quantum circuits, running the whole optimization on the quantum computer, and we address both questions.
We simulate QHD in the interaction picture of the kinetic term \cite{low2018hamiltonian}, which is integrated exactly with quantum Fourier transforms, and Trotterize the rotated potential.
We prove that, for cost functions whose periodic extension is sufficiently smooth, the resulting error bound does not grow with $q$.
The reason is that the error depends on the momentum of the actual state, not on the largest momentum of the grid, and the kinetic energy of the state stays bounded independently of the grid, as long as the ratio $r=b/a$ of the schedules is non-decreasing.
Generalizing a change of variables of Ref.~\cite{leng2025nonsmooth}, we then observe that, in terms of the kinetic time $\tau(t)=\int_0^t a(t')\,dt'$, the ideal dynamics depends on the schedules only through the curve $r(\tau)$, and that the remaining freedom, a gauge, can be used to minimize the number of Trotter steps at fixed error bound.
This separates the design of the schedules into two independent problems: the choice of the curve, which determines the outcome of the algorithm and should spend kinetic time in the search window where tunneling between the valleys decides the success probability, and the choice of the gauge, which determines the gate count.

For cost functions defined by parametric circuits, which we call ``quantum functions'', we introduce an online implementation of the potential step.
The register on which QHD is simulated, which encodes the parameters, controls the parametric rotations of the circuit, which acts on a second register that is reset after each step; in this way, the phases $e^{-i\varphi f(\hat{\bm x})}$ of the potential step are applied without ever estimating $f$.
The error of this step amounts to a loss of coherence in the position basis, with a bound that depends neither on $q$ nor on the number of parameters $d$.
Since quantum functions built from Pauli rotations are trigonometric polynomials, the Trotter error bound also becomes explicit, and grows at most linearly with $d$.
The idea of replacing the parameters of a quantum circuit with a quantum register goes back at least to Ref.~\cite{banchi2020convex}, where a program state is optimized so that the circuit simulates a target quantum channel \cite{pirandola2017fundamental}. There, however, the optimization over the program state was carried out classically, whereas here the parameter register itself evolves under the QHD dynamics.

The paper is organized as follows.
In \cref{sec:Quantum Hamiltonian Descent}, after reviewing QHD and its convergence properties, we describe its digital simulation, the error bound, the reparametrization of time and the choice of the schedules.
In \cref{sec:Optimization of Classical Functions} we test the method on standard benchmark functions for global optimization, and in \cref{sec:Optimization of Quantum Functions} we introduce the online procedure for quantum functions, whose numerical results are presented in \cref{sec:results}.
We conclude in \cref{sec:conclusions}.
Proofs and technical details are collected in the appendices.

\section{Quantum Hamiltonian Descent}%
\label{sec:Quantum Hamiltonian Descent}

QHD arises from the ``quantization'' of the Bregman Lagrangian formalism, which is a classical method for finding the minimum $\boldsymbol{x}^*$ of a function $f(\boldsymbol{x})$; in fact, several classical algorithms for the minimization of functions can be interpreted as time-discrete versions of dynamical systems governed by a suitable Bregman Lagrangian \cite{leng2023quantum,wibisono2016variational}.

Let $\chi\subseteq\mathbb{R}^d$ be the domain of the function $f$ whose minimum we want to find, and let $h: \chi\to\mathbb{R}$ be a convex and differentiable function through which we define the Bregman divergence:
\begin{equation}
    D_h(\boldsymbol{x},\boldsymbol{y})\coloneqq h(\boldsymbol{x})-h(\boldsymbol{y})-\nabla h(\boldsymbol{y})\cdot(\boldsymbol{x}-\boldsymbol{y}) \ .
\end{equation}
After choosing the functions $\alpha(t),\ \beta(t),\ \gamma(t)$ so that they satisfy the ideal scaling conditions \cite{wibisono2016variational}
$\dot{\beta}\leq e^{\alpha(t)}$ and $\dot{\gamma}=e^{\alpha(t)}$,
we can define the Bregman Lagrangian:
\begin{equation}
\mathcal{L}(\boldsymbol{x}, \boldsymbol{v}, t)
\coloneqq e^{\alpha(t) + \gamma(t)}\!
\left[
D_h\!\left(\boldsymbol{x} + e^{-\alpha(t)} \boldsymbol{v},\, \boldsymbol{x}\right)
- e^{\beta(t)} f(\boldsymbol{x})
\right]\ .
\end{equation}
One can prove that the dynamics of the system associated with this Lagrangian converges to the solution $\boldsymbol{x}^*$ if the function $f$ is convex; in this case, denoting by $\boldsymbol{x}(t)$ the trajectory identified by the Euler--Lagrange equations, we have:
\begin{equation}\label{eq:classical_convergence_rate}
  f(\boldsymbol{x}(t))-f(\boldsymbol{x}^*)\leq O\left(e^{-\beta(t)}\right) \ .
\end{equation}

Choosing the simplest convex function $h(\boldsymbol{x})=\frac{1}{2}\left|\boldsymbol{x}\right|^2=\frac{1}{2}\sum_{i=1}^d x_i^2$, we obtain the following Bregman Lagrangian:
\begin{equation}
    \mathcal{L}(\boldsymbol{x},\boldsymbol{v},t)
    =
    \underset{\text{kinetic energy}}{\underbrace{e^{\gamma(t)-\alpha(t)}\frac{\left|\boldsymbol{v}\right|^2}{2}}}-\underset{\text{potential energy}}{\underbrace{e^{\alpha(t)+\beta(t)+\gamma(t)}f(\boldsymbol{x})}}\ .
\end{equation}
This choice of $h(\boldsymbol{x})$ leads us back to a typical physical system, simplifies the derivation of the Hamiltonian, and will make its implementation through quantum gates easier.
By performing the Legendre transform of this Bregman Lagrangian, we obtain the corresponding Hamiltonian:
\begin{equation}
    H(\boldsymbol{x},\boldsymbol{p},t)
    =
    e^{\alpha(t)-\gamma(t)}\frac{\left|\boldsymbol{p}\right|^2}{2}+e^{\alpha(t)+\beta(t)+\gamma(t)}f(\boldsymbol{x}) \ .
\end{equation}
The QHD Hamiltonian directly follows by applying canonical quantization or, alternatively, Feynman's path integral method \cite[App.~A.2]{leng2023quantum}:
\begin{equation}\label{eq:QHD_Hamiltonian}
    \hat{H}(t)=a(t)\frac{\hat{\boldsymbol{p}}^2}{2}+b(t)f(\hat{\boldsymbol{x}}) \ ,
\end{equation}
where we use the ``hat'' symbol to denote operators and we have defined the \emph{time schedules}
\begin{align}
  a(t)&\coloneqq e^{\alpha(t)-\gamma(t)} \ , &
  b(t)&\coloneqq e^{\alpha(t)+\beta(t)+\gamma(t)} \ ,
\end{align}
which govern the relative importance of the kinetic and potential terms.
On top of the convergence for convex problems, the following result has been proved for optimizing general functions through QHD \cite{leng2023quantum}.

\begin{theorem}[Global convergence]\label{thm:convergenza_globale_QHD}
    Assume that the following conditions hold:
    \begin{enumerate}
        \item Conditions on the function.
        Let $f$ be continuous, with $f\in C^2$ in a neighborhood of the unique global minimizer $\boldsymbol{x}^*$, and assume that $f(\boldsymbol{x})\to+\infty$ for $\left|\boldsymbol{x}\right|\to\infty$.
        We denote by
        \begin{equation}
            f_q(\boldsymbol{x})=f(\boldsymbol{x}^*)+\frac{1}{2}\left(\boldsymbol{x}-\boldsymbol{x}^*\right)\cdot\frac{\partial^2f}{\partial\boldsymbol{x}\partial\boldsymbol{x}}\cdot\left(\boldsymbol{x}-\boldsymbol{x}^*\right)
        \end{equation}
        the quadratic approximation of $f$ in a neighborhood of the global minimum.
        \item Conditions on the Hamiltonian.
        \begin{itemize}
        \item The QHD Hamiltonian varies ``slowly'' in time (see \cref{App:dimostrazione_teorema_convergenza_globale,eq: approx adiabatica}, and \cref{app:curve,eq:local_adiabatic}).
        \item There are no energy-level \textit{crossings} for $t>0$.
        \item The time schedules are such that
        \[
        \lim_{t\to\infty}\frac{a(t)}{b(t)}=0 \ .
        \]
        \item For sufficiently long times, when $a(t)\ll b(t)$, the first $\mathcal{N}$ eigenstates of the QHD Hamiltonian can be approximated by the first $\mathcal{N}$ eigenstates of the quadratic Hamiltonian
        \[
        \hat{H}_q(t)=a(t)\,\hat{T}_0+b(t)\,f_q(\hat{\boldsymbol{x}})\ ,
        \]
        with $\hat{T}_0\coloneqq\hat{\boldsymbol{p}}^2/2$ the kinetic operator.
        \end{itemize}

        \item Conditions on the state.
        The initial state must belong to the subspace spanned by the first $\mathcal{N}$ eigenstates of the QHD Hamiltonian.
        The evolution of the state is governed by the QHD Hamiltonian.
    \end{enumerate}
    Then,
    \[
    \lim_{t\to\infty}\matrixel{\psi(t)}{f(\hat{\boldsymbol{x}})}{\psi(t)}=f(\boldsymbol{x}^*) \ .
    \]
\end{theorem}

The general idea of the proof is rather simple: one applies the adiabatic approximation to show that the state always remains in the subspace spanned by the first $\mathcal{N}$ instantaneous eigenstates of the Hamiltonian.
Since, for sufficiently long times, the QHD Hamiltonian is approximated by $\hat{H}_q(t)$, which describes a system confined in an increasingly tight potential well localized around $\boldsymbol{x}^*$, the probability of measuring the system at that position tends to $1$.
In particular, for a $d$-dimensional domain, one proves that
\begin{equation}\label{eq:risultato_convergenza_QHD}
    \left|\matrixel{\psi(t)}{f(\hat{\boldsymbol{x}})}{\psi(t)}-f(\boldsymbol{x}^*)\right|\leq \frac{d\,\mathcal{N}^3\omega}{4} \sqrt{\frac{a(t)}{b(t)}}\to 0 \ ,
\end{equation}
where $\omega = \max_i \left|\omega_i\right|$ and 
$\omega_i^2$ denote the eigenvalues of the Hessian matrix of $f$ evaluated at the global minimizer. 
In \cref{App:dimostrazione_teorema_convergenza_globale} we extend the one-dimensional proof of \cite[App.~B.2]{leng2023quantum} to $d$ dimensions.

\subsection{Simulation of QHD on digital quantum computers}
\label{sec:gate simul}

In order to make use of the QHD method for optimizing real problems with digital quantum computers, we need to transform it into the QHD algorithm via space and time discretizations.
For the space discretization, we restrict the domain of $f$ to a hypercube $\boldsymbol{x}_0+[0,L)^d$ of side $L$, which should contain the global minimum, and we sample each coordinate on a uniform grid of $2^q$ points with spacing $L/2^q$, encoded in $q$ qubits.
We thus use a total of $dq$ qubits and a grid of $2^{dq}$ points $\boldsymbol{x}_r$, and we assign each point to a state $\ket{\boldsymbol{x}_r}$ of the computational basis.
Then, through the Fourier spectral method \cite{childs2022quantum}, we approximate the QHD Hamiltonian in the continuous space with the following Hamiltonian on the Hilbert space of the qubits:
\begin{equation}\label{eq:H_discretized}
    \hat{H}(t)=a(t)\,\hat{\boldsymbol{F}}_{\!s}\,\hat{\boldsymbol{L}}\,\hat{\boldsymbol{F}}_{\!s}^\dagger+b(t)\,f(\hat{\boldsymbol{x}}) \ ,
\end{equation}
with
\begin{align}
    &\hat{\boldsymbol{F}}_{\!s}
    \coloneqq
    \bigotimes_{\ell=1}^d \hat{F}_{\!s} \ ,
    &
    \hat{F}_{\!s}
    &\coloneqq
    \sum_{r,s=0}^{2^q-1} \frac{e^{i \frac{2\pi}{2^q} (r-2^{q-1})s}}{\sqrt{2^q}} \ketbra{x_s}{x_r} \ ,\label{eq:def_F_operator}
    \\
    &\hat{\boldsymbol{L}}
    \coloneqq
    \sum_{\ell=1}^d \hat{\mathbb{I}}^{\otimes (\ell-1)}\otimes\hat{L}\otimes\hat{\mathbb{I}}^{\otimes (d-\ell)} \ ,
    &
    \hat{L} 
    &\coloneqq 
    \sum_{p=0}^{2^q-1}\frac{1}{2}\left[\frac{2\pi\left(p-2^{q-1}\right)}{L}\right]^2\ketbra{x_p}{x_p} \ ,\label{eq:def_L_operator}
    \\
    &f(\hat{\boldsymbol{x}})
    \coloneqq
    \sum_{r=0}^{2^{dq}-1} f(\boldsymbol{x}_r)\ketbra{\boldsymbol{x}_r}{\boldsymbol{x}_r} \ . &
\end{align}
The approximation error typically decreases exponentially with the number of points along each dimension, so it is often negligible.

Having space-discretized the QHD Hamiltonian, we now decompose the evolution operator into a sequence of quantum gates.
We consider the simplest product formula, with a single potential step per time step.
To this end, we describe the evolution in the interaction picture with respect to the kinetic term \cite{low2018hamiltonian}, so that the kinetic part of the Hamiltonian is integrated exactly.
Denoting by $\hat{T}_0\coloneqq\frac{\hat{\boldsymbol{p}}^2}{2}=\hat{\boldsymbol{F}}_{\!s}\,\hat{\boldsymbol{L}}\,\hat{\boldsymbol{F}}_{\!s}^\dagger$ the discretized kinetic operator and by $A(t)=\int_0^t a(s)\,ds$ the kinetic time, the exact evolution operator factorizes as
\begin{gather}
  \hat{U}(T,0)\coloneqq\mathcal{T}\left[e^{-i\int_0^T \hat{H}(t)\,dt}\right]
  =e^{-iA(T)\frac{\hat{\boldsymbol{p}}^2}{2}}\,\mathcal{T}\left[e^{-i\int_0^T\hat{H}_I(t)\,dt}\right] \ ,\label{eq:U_interaction_main}
  \\
  \hat{H}_I(t)\coloneqq b(t)\,e^{iA(t)\frac{\hat{\boldsymbol{p}}^2}{2}}f(\hat{\boldsymbol{x}})\,e^{-iA(t)\frac{\hat{\boldsymbol{p}}^2}{2}} \ ,\label{eq:H_interaction_main}
\end{gather}
so that only the interaction-picture Hamiltonian $\hat{H}_I(t)$, i.e.~the potential rotated by the free evolution, has to be time-discretized.
We Trotterize this evolution by splitting $[0,T]$ into $R$ intervals $\mathcal{I}_j$ of length $\varDelta t=T/R$ and replacing $\hat{H}_I(t)$ in each of them with its value at a sampling point $t_j\in\mathcal{I}_j$.
We write this first-order approximation as
\begin{align}
  \hat{U}(T,0)
  &\simeq e^{-iA(T)\frac{\hat{\boldsymbol{p}}^2}{2}}\,e^{-i\varDelta t\,\hat{H}_I(t_R)}\cdots e^{-i\varDelta t\,\hat{H}_I(t_1)}
  \nonumber\\ &=
  \hat{T}(t_R)\,\hat{V}(t_R)\cdots\hat{T}(t_1)\,\hat{V}(t_1)\,\hat{T}(t_0)\eqqcolon\hat{U}_{\mathrm{Trotter}}(T,0) \ ,\label{eq:trotter_interaction_main}
\end{align}
where
\begin{equation}\label{eq:trotter_steps}
  \hat{T}(t_j)\coloneqq\hat{\boldsymbol{F}}_{\!s}\,e^{-i\left[A(t_{j+1})-A(t_j)\right]\hat{\boldsymbol{L}}}\,\hat{\boldsymbol{F}}_{\!s}^\dagger \ ,
  \quad
  \hat{V}(t_j)\coloneqq e^{-i\varDelta t\,b(t_j)f(\hat{\boldsymbol{x}})} \ ,
  \quad
  t_j\coloneqq\varDelta t\left(j-\frac{1}{2}\right) \ ,
\end{equation}
for $j=1,\ldots,R$, with $t_0\coloneqq0$ and $t_{R+1}\coloneqq T$.
The algorithm is thus specified by the kinetic angles $\alpha_j=A(t_{j+1})-A(t_j)$ and the potential angles $\varphi_j=b(t_j)\varDelta t$.
The first kinetic step $\hat{T}(t_0)$ acts trivially on the usual initial state, i.e.~the uniform superposition of all computational-basis states, which is the zero-momentum eigenvector, so the algorithm starts with the potential step $\hat{V}(t_1)$.

In \cref{app:trotter_error} we bound the error of this decomposition on the actual trajectory, up to an irrelevant global phase.
The following bound is first-order in $\varDelta t$ and holds for every time-sampling scheme, even though with midpoint sampling this is a second-order method \cite{blanes2009magnus}.
We find, see \cref{eq:trotter_error_final,eq:trotter_error_smooth,eq:trotter_error_schedules},
\begin{align}
  \eTrotter
  &\coloneqq\min_{\theta\in\mathbb{R}}\left\|\left[e^{i\theta}\hat U_{\mathrm{Trotter}}(T,0)-\hat U(T,0)\right]\ket{\psi(0)}\right\|
  \nonumber\\ &\lesssim
  \varDelta t\left[\frac{\osc}{4}\left(b(T)-b(0)\right)+\int_0^T\left(c_1\sqrt{a\,b^3}+c_2\,a\,b\right)dt\right]
  =\mathcal O\left(R\,\varDelta t^2\right) \ ,
  \label{eq:trotter error}
\end{align}
where $a=a(t)$ and $b=b(t)$ in the integral, and $\osc\coloneqq\max f-\min f$ is the oscillation of the cost function over the domain.
The coefficients $c_1$ and $c_2$ depend only on the cost function and on the grid, not on the schedules.
For a generic $f$, they are proportional to Fourier sums of the discretized cost function, whose dependence on $q$ is set by the regularity of the periodic extension $F$ of $f$ from the domain, and which can grow even exponentially with $q$ if $F$ is discontinuous,  see \cref{eq:c12_generic,tab:regimes} in 
\cref{app:trotter_error}.
If, instead, $F$ is sufficiently smooth, $c_1$ and $c_2$ are bounded independently of $q$ and, for large $q$, they reduce to
\begin{equation}\label{eq:c12_smooth}
  c_1\simeq\frac{1}{2}\sqrt{2\,\osc}\,\max\left|\nabla F\right| \ ,
  \qquad
  c_2\simeq\frac{1}{4}\max\left|\Delta F\right| \ ,
\end{equation}
i.e.~to the maxima of the gradient and of the Laplacian that appear in \cref{eq:commutator_continuum} below,  see \cref{eq:trotter_error_smooth}.
That, for sufficiently smooth cost functions, the error does not grow with $q$ is not obvious: the error of a first-order product formula is controlled by the commutator between the two terms of the Hamiltonian, which, in the continuum, reads
\begin{equation}\label{eq:commutator_continuum}
  \left[\frac{\hat{\boldsymbol p}^2}{2},f(\hat{\boldsymbol x})\right]=-i\,\nabla f(\hat{\boldsymbol x})\cdot\hat{\boldsymbol p}-\frac{1}{2}\,\Delta f(\hat{\boldsymbol x}) \ .
\end{equation}
The first term contains the momentum operator, which is unbounded in the continuum and, on the grid, has eigenvalues up to $\pi 2^q/L$: a standard bound based on the operator norm of this commutator would give an error growing exponentially with the number of qubits $q$.
The analysis of \cref{app:trotter_error} avoids this problem in three steps:
(i) we move to the interaction picture with respect to the kinetic term, which is integrated exactly, so that only the rotated potential $\hat H_I(t)$ of \cref{eq:H_interaction_main} is Trotterized;
(ii) by a telescoping sum the error of each step is written as $\hat H_I(s)-\hat H_I(t_j)$ acting on the \emph{exact} state at time $s$, so we never need the operator norm of the commutator, but only its action on the exact state $\ket{\psi(s)}$.
By \cref{eq:commutator_continuum}, this is controlled by the momentum spread $\norm{\hat{\boldsymbol p}\ket{\psi(s)}}$ of the actual state, rather than by the largest momentum of the grid.
Finally, (iii) we show that, if the initial state has zero momentum and $r=b/a$ is non-decreasing, then $\matrixel{\psi(t)}{\hat T_0}{\psi(t)}\leq\osc\,r(t)$ at all times, for every $q$.
Intuitively, in kinetic time the energy of the system, with the potential shifted to $r\left(f-\min f\right)\geq0$, changes only through the time dependence of $r$, at the rate $\frac{dr}{d\tau}\langle f-\min f\rangle\leq\frac{dr}{d\tau}\osc$ if $r$ is non-decreasing.
The energy supplied up to $\tau$ cannot exceed $r(\tau)\,\osc$.
The momentum scale $2^q$ of the grid is thus replaced by the intrinsic scale $\sqrt{\osc\,r}$.

To obtain this result, the main hypotheses are therefore that $\ket{\psi(0)}$ is the uniform superposition, that $b$ and $r=b/a$ are non-decreasing, and that the cost function $f$ extends to a sufficiently smooth periodic function on the domain, so that the gradient and the Laplacian in \cref{eq:commutator_continuum} are bounded also on the grid.\footnote{The assumption on the initial state is not strictly necessary: the result still holds, with an additional term, for every initial state whose kinetic energy does not grow with the number of qubits $q$ (see \cref{eq:energy_bound}).
Instead, discontinuities in the cost function or in its derivatives can make the error bound grow with $q$, even exponentially, because they can inject high-momentum components into the state.
Nonetheless, if the wavefunction is sufficiently small where these discontinuities occur, in the actual error there is no observable dependence on $q$.}
In \cref{eq:trotter error}, the schedules enter only through $b(T)-b(0)$ and through the integrals of $ab$ and $\sqrt{ab^3}$.
The dependence on $d$, instead, is contained in $\osc$, $c_1$ and $c_2$, i.e.~it is set by the features of the cost function, such as its oscillation and the maxima of its gradient and Laplacian: for a generic $f$, our bound gives no explicit information on the dependence on $d$.

For the quantum functions of \cref{sec:Optimization of Quantum Functions}, the dependence on the gradient and the Laplacian of $f$ can be removed, and our bound becomes linear in $d$.
These functions are trigonometric polynomials of period $2\pi$ and degree at most $\nu$ in each variable \cite{schuld2020effect}, $\nu$ being the maximum number of rotation gates that depend on the same parameter.
Hence, multiplication by $f$ shifts the momentum of the state by at most $\nu$ along each direction, with amplitudes bounded by $\osc/2$ (\cref{prop:trig}): $f$ itself cannot generate high-momentum components, and the commutator \cref{eq:commutator_continuum} is controlled by $\osc$, $\nu$ and the momentum spread of the state only.
The coefficients of \cref{eq:trotter error} are then (see \cref{eq:trotter_error_schedules})
\begin{equation}\label{eq:trotter_error_quantum_main}
  c_1=\frac{\osc}{4}\,\nu(\nu+1)\sqrt{2\,d\,\osc} \ ,
  \quad
  c_2=\frac{\osc}{24}\,\nu(\nu+1)(2\nu+1)\,d \ ,
\end{equation}
for every $q>\log_2\nu+1$, and $\osc$ 
is at most twice the norm of the observable that defines the quantum function (see \cref{sec:Optimization of Quantum Functions}).
Thus the Trotter error for quantum cost functions is
\begin{equation}\label{eq:trotter_error_quantum_scaling}
  \eTrotter\leq\mathcal O\left(d\,R\,\varDelta t^2\right) \ .
\end{equation}
This bound only relies on the trigonometric structure of quantum functions, and requires the domain to be a full period, $[0,2\pi)^d$, as in the encoding \cref{eq:encoding_main}, and integer frequencies, as for Pauli rotations; otherwise, the periodic extension of $f$ is discontinuous and the error bound acquires a dependence on $q$, which can even be exponential (see \cref{app:trotter_quantum}).
When the potential step of a quantum function is implemented with the online procedure that we propose in \cref{sec:online_procedure}, an additional error, the online procedure error, is introduced.

Recent methods based on quantum signal processing and quantum singular value transformation \cite{Low2019hamiltonian,gilyen2019quantum} can be used to significantly reduce the error in Hamiltonian simulation.
However, their extension to time-dependent Hamiltonians
\cite{low2018hamiltonian, poulin2011quantum, berry2020time, chen2021quantum, an2022time, kieferova2019simulating} is less understood.
Therefore, in this paper we focus on Suzuki--Trotter decompositions and leave the exploration of alternative methods to future work.

In \cref{app:kinetic_step} we show that the kinetic step $\hat{T}(t_j)$ can be efficiently simulated with $\mathcal{O}\left(d\,q^2\right)$ elementary gates.
For classical functions, the potential step can be implemented efficiently with polynomial overhead when $f(\boldsymbol x)$ admits an efficient classical representation, e.g.~using software packages like \verb+Bennett.jl+ \cite{Bennett.jl}.
Moreover, in \cref{sec:Optimization of Quantum Functions} we will show that the potential step can also be efficiently simulated, up to a controlled error, when dealing with quantum functions.

\subsection{Time reparametrization and choice of the schedules}
\label{sec:schedules}

The two schedules are not independent parameters of the algorithm.
Writing the Hamiltonian \cref{eq:QHD_Hamiltonian} as $\hat H(t)=a(t)\left[\hat{\boldsymbol p}^2/2+r(t)\,f(\hat{\boldsymbol x})\right]$ with $r\coloneqq b/a$, and using the \emph{kinetic time} $\tau\coloneqq A(t)=\int_0^ta(s)\,ds$, the Schr\"odinger equation becomes exactly \cite[Sec.~4.1]{leng2025nonsmooth}
\begin{equation}\label{eq:tau_dynamics}
  i\,\frac{d}{d\tau}\ket{\psi}=\left[\frac{\hat{\boldsymbol p}^2}{2}+r(\tau)\,f(\hat{\boldsymbol x})\right]\ket{\psi} \ ,
\end{equation}
where, with the usual abuse of notation, $r(\tau)\coloneqq r(t(\tau))=b(t(\tau))/a(t(\tau))$ is the ratio expressed as a function of kinetic time.
The \textit{ideal} dynamics depends on the schedules only through the curve $r(\tau)$, and the gauge $a(t)$ only sets the speed at which the curve is run through in physical time; this freedom does not affect the result of the ideal algorithm, and we can use it to minimize the number of gates.
We stress that the freedom in the gauge is a feature of the digital simulation: if QHD were run as an analog, continuous-time evolution, $a(t)$ would be physical, since the energy scale of the hardware is bounded, and the gauge could not be chosen freely.

As for the gauge, we choose it in order to minimize the number of Trotter steps sufficient to guarantee, according to the bound \cref{eq:trotter error}, an error $\leq\varepsilon^*$.
With $g(r)\coloneqq c_1r^{3/2}+c_2r$, in \cref{app:schedules} we show that the minimum
\begin{equation}\label{eq:R_gauge}
  R_{\mathrm{min}}\simeq\frac{1}{\varepsilon^*}\left(\int_0^{\tau_{\mathrm{f}}}\sqrt{g(r(\tau))}\,d\tau\right)^2
\end{equation}
is achieved if and only if $a\sqrt{g(r)}$ is constant along the curve.
Thus, the optimal gauge is 
\begin{equation}
a(t(\tau))=\frac{\lambda}{\sqrt{g(r(\tau))}} \ ,
\end{equation}
where $\lambda$ can be fixed arbitrarily due to the symmetry of the Trotterized evolution: under $(a,b,t)\to(\lambda a,\lambda b,t/\lambda)$ at fixed $R$, the angles $\alpha_j=\int_{t_j}^{t_{j+1}} a(t)\,dt$ and $\varphi_j=b(t_j)\,\varDelta t$, and hence the error bound \cref{eq:trotter error}, are unchanged (see \cref{eq:angles}).

As for the curve, we consider
\begin{equation}\label{eq:r_curve}
  r(\tau)=r_0\, e^{k\tau} \ .
\end{equation}
We discuss this choice and other details about the curve $r(\tau)$ in \cref{app:curve}.
In \cref{app:angles} we explain how to calculate the kinetic and potential angles needed for the Hamiltonian simulation \cref{eq:trotter_interaction_main}, given $R$ and the set of parameters in $r(\tau)$.

The decoupling between curve and gauge is what makes the choice of the schedules a tractable problem.
The curve is chosen on physical grounds, to keep the evolution adiabatic where the minima are searched for (\cref{app:curve}), and different curves can be compared on the ideal dynamics alone.
The gauge is then chosen to minimize the error of the simulation, without changing the result of the ideal algorithm.
Which gauge is optimal depends on which error dominates: the gauge above minimizes the bound on the Trotter error, which is the relevant one when the potential steps are exact, as for classical functions, while for quantum functions the online procedure error of \cref{sec:online_procedure} dominates and leads to a different gauge.

\section{Optimization of Classical Functions}%
\label{sec:Optimization of Classical Functions}

Before studying the optimization of ``quantum functions'', namely functions estimated through measurements on quantum hardware, we start our analysis with the simpler problem of optimizing ``classical functions'' $f:\chi\subseteq \mathbb{R}^d\to \mathbb{R}$,
by which we mean functions with an explicit analytical form that may arise in both classical and quantum contexts.
The adjective ``classical'' is used here solely to distinguish them from those requiring quantum hardware for their evaluation.
The functions under consideration are selected from a set of benchmark functions, available in the literature \cite{jamil_yang_testfunctions,jamil2013literature,surjanovic2013vlse}, featuring narrow valleys, broad plateaus, many nearly degenerate minima, etc., which make the search for the global minimum extremely challenging for many classical algorithms. 

In order to benchmark the performance of QHD
we define two quantities that allow us to assess the outcome of the simulations:
\begin{itemize}
  \item \textit{success probability}
  \begin{equation}
    P_{\mathrm{suc}}(t) \coloneqq \sum_{\boldsymbol{x} \in \mathcal{B}} p(\boldsymbol{x},t) \ ,
    \label{eq:Psuc_def}
  \end{equation}
  where $\mathcal{B}$ is the ``basin of attraction'' of the global minimum, namely the set of points $\boldsymbol{x}$ in the domain for which following the steepest descent leads to the global-minimum point.
  We denote by $p(\boldsymbol{x},t)$ the probability distribution of the state evaluated at time $t$ and at the point $\boldsymbol{x}$ of the sampling grid\footnote{Although, for brevity, the notation may suggest a probability density, we emphasize that $p(\boldsymbol{x},t)$ is a discrete distribution and $\ket{\boldsymbol{x}}$ is a computational-basis state of the qubit register.}:
  \begin{equation}
    p(\boldsymbol{x},t) \coloneqq \left|\braket{\boldsymbol{x}}{\psi(t)}\right|^2 \ .
  \end{equation}
  
  \noindent In the simulations, after a phase in which the minima are searched for, the wavefunction evolves in an almost classical manner, that is, ``\textit{tunneling}'' between the potential wells ceases and the success probability stabilizes around a definite value.
  If the global-search phase has already ended by time $T$, then $P_{\mathrm{suc}}(T)$ is an estimate of the probability with which the algorithm would return the global-minimum point if the measurement were performed at $t\to\infty$.

  \item \textit{coefficient} $\mathbb{E}$
  \begin{equation}\label{eq:E_def}
    \mathbb{E}(t) \coloneqq \frac{\matrixel{\psi(t)}{f(\hat{\boldsymbol{x}})}{\psi(t)} - f(\boldsymbol{x}^{*})}{\bar f - f(\boldsymbol{x}^{*})}\ ,
  \end{equation}
  where $\bar f \coloneqq 2^{-dq}\sum_{\boldsymbol{x}} f(\boldsymbol{x})$ is the average value of the function over the grid.
  The coefficient $\mathbb{E}$ is a normalized version of the expectation value bounded in \cref{eq:risultato_convergenza_QHD}.
  In the exactly adiabatic case, this expectation value tends to $f(\boldsymbol{x}^*)$, i.e.~$\mathbb{E}\to0$.
  The normalization is chosen so that $\mathbb{E}=1$ for the uniform superposition, i.e.~for the initial state, for which a measurement returns a point uniformly distributed over the whole domain, and hence a very poor estimate of the minimum point: values $\mathbb{E}\simeq1$ indicate that the algorithm has made little progress.
  Conversely, when $\mathbb{E}\ll1$, the wavefunction is concentrated where the function $f(\boldsymbol{x})$ takes its lowest values; this is another measure of the success of the simulation, because collapse around local minima may also be a desirable outcome when such minima are nearly degenerate with the global minimum: in other words, identifying $\boldsymbol{x}'\neq\boldsymbol{x}^*$ is not necessarily a failure, provided that $f(\boldsymbol{x}')\simeq f(\boldsymbol{x}^*)$ and $f(\boldsymbol{x}')\ll\bar{f}$.
\end{itemize}

\begin{table}[h]
  \centering
  \caption{Final success probability $P_{\mathrm{suc}}$ and final coefficient $\mathbb{E}$ for the benchmark functions in $d=1$ and $d=2$, obtained with the same protocol for all functions and dimensions, in the native units of each function: $q=8$ qubits per dimension, curve $r(\tau)=r_0\,e^{k\tau}$ of \cref{eq:r_curve} with rate $k=0.1$, from $r_0=10^{-4}$ to $r_{\mathrm{f}}=10^{2}$, i.e.~$\tau_{\mathrm{f}}\simeq138$, run through in the optimal gauge $a\propto g(r)^{-1/2}$ of \cref{sec:schedules}, with $c_1$ and $c_2$ given by \cref{eq:c12_smooth} with the maxima of $\left|\nabla f\right|$ and $\left|\Delta f\right|$ over the domain, and $R=2^{15}$ Trotter steps.
  Each case is repeated with $2R$ steps as a check.
  Symbols:
  ``/'', the function is defined only for $d=2$;
  $^{\dagger}$, non-negligible Trotterization error;
  $^{\ddagger}$, incomplete convergence: $P_{\mathrm{suc}}$ and $\mathbb{E}$ are still changing at the end of the curve.
  Unmarked values are converged both in $R$ and in $r$, so that $P_{\mathrm{suc}}<1$ is due to the lack of adiabaticity of the evolution at the rate $k$.}
  \label{tab:standard_results}
  \setlength{\tabcolsep}{6pt}
  \begin{tabular}{|l|l|cc|cc|}
    \hline
    & & \multicolumn{2}{c|}{$d=1$} & \multicolumn{2}{c|}{$d=2$} \\
    \hline
    \textbf{Landscape} & \textbf{Function} & $P_{\mathrm{suc}}$ & $\mathbb{E}$ & $P_{\mathrm{suc}}$ & $\mathbb{E}$ \\
    \hline
    \multirow{5}{*}{Valley shaped} & \texttt{csendes}$^{\dagger}$ & 1.00 & 0.029 & 1.00 & 0.029\\
     & \texttt{alpine1} & 1.00 & 0.039 & 1.00 & 0.039\\
     & \texttt{hosaki} & / & / & 1.00 & 0.047\\
     & \texttt{camel3}$^{\dagger}$ & / & / & 0.98 & 0.010\\
     & \texttt{styblinski tang} & 0.89 & 0.049 & 0.79 & 0.049\\
    \hline
    \multirow{9}{*}{Many minima} & \texttt{rastrigin}$^{\dagger}$ & 1.00 & 0.053 & 1.00 & 0.053\\
     & \texttt{ackley} & 1.00 & 0.086 & 1.00 & 0.11\\
     & \texttt{alpine2} & 0.99 & 0.021 & 0.92 & 0.063\\
     & \texttt{levy} & 0.97 & 0.011 & 0.91 & 0.018\\
     & \texttt{levy13}$^{\dagger\ddagger}$ & / & / & 0.77 & 0.092\\
     & \texttt{holder table}$^{\dagger}$ & / & / & 0.75 & 0.24\\
     & \texttt{bohachevsky2}$^{\ddagger}$ & / & / & 0.74 & 0.074\\
     & \texttt{shubert}$^{\dagger}$ & / & / & 0.25 & 0.84\\
     & \texttt{griewank} & 0.36 & 0.029 & 0.084 & 0.060\\
    \hline
    \multirow{2}{*}{Narrow gorge} & \texttt{michalewicz} & 1.00 & 0.20 & 1.00 & 0.23\\
     & \texttt{dropwave}$^{\ddagger}$ & / & / & 0.013 & 0.41\\
    \hline
    \multirow{2}{*}{Barren plateau} & \texttt{easom} & / & / & 1.00 & 0.10\\
     & \texttt{xin she yang3} & 0.30 & 0.46 & 0.055 & 0.65\\
    \hline
  \end{tabular}
\end{table}

\begin{figure}[h]
  \centering
  \includegraphics[width=\linewidth]{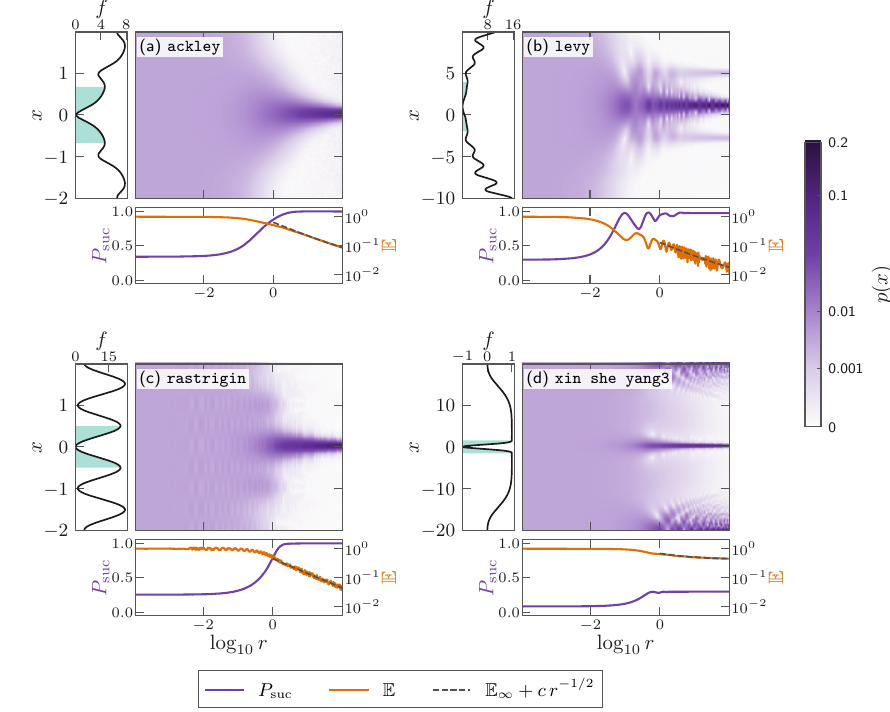}
  \caption{Evolution of QHD along the curve $r(\tau)$ for four benchmark functions in $d=1$, with the protocol of \cref{tab:standard_results} ($R=2^{15}$).
  In each panel, left: cost function $f(x)$, with the basin of the global minimum shaded; top: probability distribution $p(x)$ of the state as a function of $\log_{10}r$ (power-law color scale, common to all panels); bottom: success probability $P_{\mathrm{suc}}$ (left axis) and coefficient $\mathbb{E}$ (right axis, logarithmic scale), with the least-squares fit $\mathbb{E}_\infty+c\,r^{-1/2}$ over $10\le r\le r_{\mathrm{f}}$ (dashed line).}
  \label{fig:casi_1d}
\end{figure}

In \cref{tab:standard_results} we report a summary of the final success probability and final coefficient $\mathbb{E}$ over all the functions considered.
In \cref{fig:casi_1d} we show the evolution of the state along the curve $r(\tau)$ for four functions in $d=1$: three successful cases (\texttt{ackley}, \texttt{levy}, \texttt{rastrigin}) and a failure (\texttt{xin she yang3}).
In all cases the state is initially close to the uniform distribution, and nothing happens as long as $r\,\osc$ is small compared with the kinetic excitation energy.
In the search window, around $r\sim10^{-1}$--$1$, the probability flows through tunneling between the valleys of $f$ and $P_{\mathrm{suc}}$ is decided: for \texttt{ackley} and \texttt{rastrigin} the whole state ends up in the valley of the global minimum, whereas for \texttt{levy} $P_{\mathrm{suc}}$ oscillates as the population is exchanged among several valleys, before freezing at $0.97$.
Beyond the search window, $P_{\mathrm{suc}}$ is frozen and QHD only refines the position within each valley: $\mathbb{E}$ decreases as $\mathbb{E}_\infty+c\,r^{-1/2}$, as expected from \cref{eq:risultato_convergenza_QHD}, with $\mathbb{E}_\infty\simeq0$ when the whole state is in the global basin.
For \texttt{xin she yang3}, instead, the global minimum is a narrow well on a broad plateau, and only $30\%$ of the probability enters it; the rest remains spread over the plateau and over the valleys at the boundaries of the domain, so that $\mathbb{E}$ saturates at $\mathbb{E}_\infty\simeq0.43$.
Since the results of \cref{tab:standard_results} without symbols are converged in the number of Trotter steps and in $r$, a success probability $P_{\mathrm{suc}}<1$ is due to the lack of adiabaticity of the evolution, i.e.~to diabatic transitions between instantaneous eigenstates in the search window.
From this perspective, improving QHD performance amounts to controlling diabatic transitions between eigenstates: this can be done by slowing down the curve $r(\tau)$ in the search window, at the cost of a larger number of Trotter steps.

\begin{figure}[h]
  \centering
  \includegraphics[width=\linewidth]{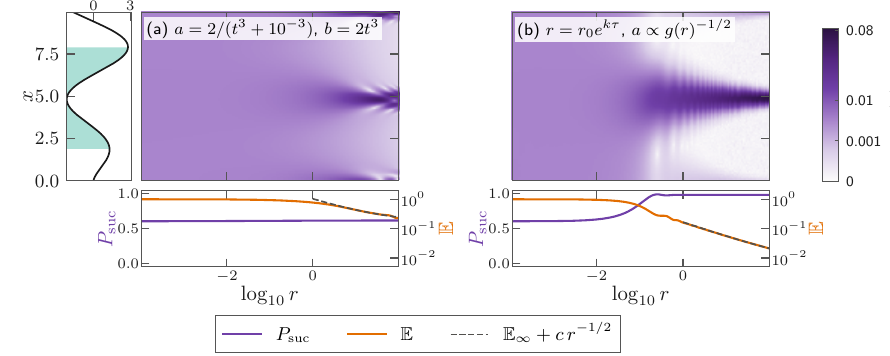}
  \caption{Evolution of QHD for \texttt{alpine2} in $d=1$ with two choices of the schedules, at the same number of qubits $q=8$ and Trotter steps $R=2^{15}$, and over the same range of $r$: (a) schedules $a(t)=2/\left(t^3+10^{-3}\right)$, $b(t)=2t^3$ of \cite{leng2023quantum}, with Trotter steps uniform in $t$, up to $T\simeq2.15$, where $r=b/a$ reaches $r_{\mathrm{f}}=10^2$; (b) protocol of \cref{tab:standard_results}, i.e.~curve $r(\tau)=r_0\,e^{k\tau}$ in the optimal gauge of \cref{sec:schedules}.
  Left: cost function $f(x)$, with the basin of the global minimum shaded; top: probability distribution $p(x)$ of the state as a function of $\log_{10}r$ (power-law color scale, common to both panels); bottom: success probability $P_{\mathrm{suc}}$ (left axis) and coefficient $\mathbb{E}$ (right axis, logarithmic scale), with the least-squares fit $\mathbb{E}_\infty+c\,r^{-1/2}$ over $10\le r\le r_{\mathrm{f}}$ (dashed line).}
  \label{fig:confronto_schedule}
\end{figure}

The choice of the curve $r(\tau)$ discussed in \cref{sec:schedules,app:curve} is a direct way to control diabatic transitions at a given cost.
In \cref{fig:confronto_schedule} we compare our protocol with the schedules $a(t)=2/\left(t^3+10^{-3}\right)$, $b(t)=2t^3$, which follow from the choice $\alpha(t)=\ln(2/t)$, $\beta(t)=\gamma(t)=2\ln t$ of \cite[App.~B.3]{leng2023quantum}, i.e.~from the Bregman Lagrangian of the continuous-time limit of Nesterov's accelerated gradient method \cite{wibisono2016variational}, with $a$ regularized at $t=0$.
Using \texttt{alpine2} in $d=1$ as an example, the two simulations have the same cost, $R=2^{15}$ Trotter steps on $q=8$ qubits, and span the same range of $r$: we run the schedules of \cite{leng2023quantum} up to $T\simeq2.15$, where $r(T)=10^2=r_{\mathrm{f}}$.
Both are converged in the number of Trotter steps: with $2R$ steps, $P_{\mathrm{suc}}$ and $\mathbb{E}$ change by less than $3\%$.
Therefore, the difference between the two panels is entirely due to the ideal dynamics \cref{eq:tau_dynamics}, i.e.~to the curve $r(\tau)$, because by the decoupling of \cref{sec:schedules} the gauge does not affect the result.
With the schedules of \cite{leng2023quantum}, out of a total kinetic time $\tau_{\mathrm{f}}=A(T)\simeq242$, about $220$ are spent at $r<10^{-4}$, where nothing happens, and only $\varDelta\tau\simeq3.6$ in the interval $10^{-2}\le r\le1$, where $P_{\mathrm{suc}}$ is decided; along the curve $r(\tau)=r_0\,e^{k\tau}$, the same interval takes $\varDelta\tau=\ln\left(10^2\right)/k\simeq46$.
As a consequence, in panel (a) the search window is crossed almost suddenly: the state is still close to the uniform distribution when tunneling between the valleys is suppressed, and $P_{\mathrm{suc}}=0.62$ remains close to the weight $0.61$ of the global basin in the uniform distribution, with $\mathbb{E}=0.23$.
In panel (b), instead, the probability flows into the valley of the global minimum, so that $P_{\mathrm{suc}}=0.99$, $\mathbb{E}=0.021$.

With schedules borrowed from classical optimization, the local minima retain a large fraction of the probability, and the remedy is to spend more kinetic time in the search window.
For classical functions, this remedy has a moderate cost since the potential steps are exact and, with midpoint sampling, the Trotter error is of second order in $\varDelta t$ (\cref{app:trotter_error}): the same number of steps, $R=2^{15}$, is sufficient for both panels of \cref{fig:confronto_schedule} and for the unmarked results of \cref{tab:standard_results}, although the curve of panel (b) is much slower in the search window.
The choice of the curve can thus be driven by the ideal dynamics alone.
As we will see in \cref{sec:online_procedure}, this is no longer the case for quantum functions, whose potential steps are implemented approximately.

\section{Optimization of Quantum Functions}%
\label{sec:Optimization of Quantum Functions}

We use the term ``quantum function'' to denote a function defined as
\begin{equation}
    f(\boldsymbol{x})\coloneqq\matrixel{\psi(\boldsymbol{x})}{\hat{O}}{\psi(\boldsymbol{x})} \ ,
\end{equation}
where the state $\ket{\psi(\boldsymbol{x})} = \hat W(\boldsymbol{x})\ket{0}^{\otimes m}$ is prepared by a parametric unitary $\hat{W}(\boldsymbol{x})$ acting on $m$ qubits initialized in $\ket{0}$, and $\hat{O}$ is a Hermitian observable.
Many problems involving parametric quantum circuits can be expressed in terms of the optimization of such functions \cite{bharti2022noisy}.

To optimize a quantum function with QHD, we use two registers with distinct roles.
The \emph{simulation register} $S$ is the one on which QHD is simulated, as in \cref{sec:gate simul}, and whose measurement at the end of the algorithm gives an estimate of the minimizer $\boldsymbol{x}^*$.
It is made up of $d$ sub-registers $S_\ell$ of $q$ qubits each, where $d$ is the number of components of $\boldsymbol{x}=\left(x_1,\ldots,x_d\right)$, and the computational basis of $S_\ell$ encodes the possible values of $x_\ell$ with $q$ bits of precision; the computational basis of the whole register, with $dq$ qubits, is thus $\left\{\ket{\boldsymbol{x}_r}\right\}_{r=0}^{2^{dq}-1}$.
The \emph{measure register} $M$ consists of the $m$ qubits on which the circuit $\hat W(\boldsymbol x)$ acts, and on which $\hat{O}$ would be measured to estimate $f$ in a standard variational algorithm.
The quantum function is defined on this register, and, denoting by the subscript $M$ its states and operators, we write
\begin{equation}
    f(\boldsymbol{x})=\matrixel{0_{\!M}}{\hat{W}_{\!M}^\dagger(\boldsymbol{x})\,\hat{O}_{\!M}\,\hat{W}_{\!M}(\boldsymbol{x})}{0_{\!M}} \ .
\end{equation}

We assume that $f(\boldsymbol x)$ cannot be computed efficiently with classical hardware or by analytical techniques, due to the complexity of the preparation $\hat{W}(\boldsymbol{x})$, the number of parameters and the nature of the observable $\hat{O}$.
Moreover, estimating $f$ through measurements at every point of the grid would require a number of circuit evaluations exponential in $dq$.
Therefore, the potential step of \cref{sec:gate simul} cannot be implemented with reversible arithmetic, as for classical functions.

\subsection{Online procedure}
\label{sec:online_procedure}

The idea behind the online procedure is to skip the phase of getting a classical description of the quantum function, i.e.~its values on the grid or its analytical expression, and to directly apply the phases $e^{-i\varphi_j f(\hat{\boldsymbol x})}$ of the potential step during the quantum simulation.
This is done by coupling the simulation register to the measure register, which acts as an ancilla, such that the reduced evolution on $S$ is approximately equivalent to the ideal potential step.
First, we note that
\begin{equation}
    e^{-i\hat{H}\varDelta t}\,\rho\, e^{i\hat{H}\varDelta t} = \rho -i\varDelta t \left(\hat{H}\rho-\rho\hat{H}\right) + \mathcal O\left(\varDelta t^{2}\right) \ .
\end{equation}
If the state of $S+M$ is described by $\rho=\rho_{\!S}\otimes \ketbra{0_{\!M}}{0_{\!M}}$ and its evolution is guided by the Hamiltonian $\hat{H}$, then the first-order approximation of the state of $S$ after a time step $\varDelta t$ is:
\begin{equation}\label{eq:effective_evolution}
  \rho_{\!S}' \simeq
     e^{- i \hat{H}_{\!S}^{\mathrm{eff}}\varDelta t}\ 
     \rho_{\!S}\ 
     e^{i \hat{H}_{\!S}^{\mathrm{eff}}\varDelta t}\ ,
\end{equation}
where we have defined the effective Hamiltonian on $S$ as
\begin{equation}
  \hat{H}_{\!S}^{\mathrm{eff}}
  \coloneqq  \matrixel{0_{\!M}}{\hat{H}_{\!S\!M}}{0_{\!M}} \ .
\end{equation}
In order for the evolution in \cref{eq:effective_evolution} to be the same as that in the ideal potential step, we require that $\hat{H}_{\!S}^{\mathrm{eff}}=b(t)\,f(\hat{\boldsymbol{x}})$.
This is achieved by the following Hamiltonian on $S+M$:
\begin{equation}
  \hat{H}_{\!S\!M}(t)
  = b(t)\,\hat{W}_{\!S\!M}^\dagger\,
    (\hat{\mathbb{I}}_{\!S}\otimes\hat{O}_{\!M})\,
    \hat{W}_{\!S\!M}\ ,
\end{equation}
with
\begin{equation}
  \hat{W}_{\!S\!M}
  \coloneqq  \sum_{r}
      \ketbra{\boldsymbol{x}_r}{\boldsymbol{x}_r}_{\!S}
      \otimes \hat{W}_{\!M}(\boldsymbol{x}_r)\ .
      \label{eq:W extended}
\end{equation}
All in all, the online procedure implements (a first-order approximation in $\varDelta t$ of) the potential step as
\begin{equation}
    e^{-i\varphi_j f(\hat{\boldsymbol{x}})}\simeq \matrixel{0_{\!M}}{\hat{V}_{\!S\!M}(t_j)}{0_{\!M}} \ ,
\end{equation}
with $\varphi_j=b(t_j)\varDelta t$ the potential angle of the step and
\begin{equation}\label{eq:online_potential_operator}
    \hat{V}_{\!S\!M}(t)\coloneqq e^{- i \varDelta t\,\hat{H}_{\!S\!M}(t)}
  = \hat{W}_{\!S\!M}^\dagger\,
    e^{- i \varDelta t\, b(t)\,(\hat{\mathbb{I}}_{\!S}\otimes\hat{O}_{\!M})}\,
    \hat{W}_{\!S\!M}\ .
\end{equation}
In order for this approximation to be valid, the state of the whole system before each potential step must be in the form
\begin{equation}
    \rho=\rho_{\!S}\otimes \ketbra{0_{\!M}}{0_{\!M}} \ ,
\end{equation}
and since the evolution will not generally leave the measure register in $\ket{0_{\!M}}$, this register must be reset to $\ket{0_{\!M}}$ between each potential step and the next.

The QHD algorithm implemented via the online procedure can be represented by \cref{circ:QHD_online}, where $\hat{P}_{\!S}$ is the preparation of the simulation register --- typically a Hadamard gate on each qubit, to prepare the state in the uniform superposition of all computational-basis states --- $\hat{T}(t_j)$ is the kinetic step, whose implementation is described in \cref{app:kinetic_step}, and $\hat{V}_{\!S\!M}$ is the online potential step.
\begin{qcircuit}[circ:QHD_online]
\begin{quantikz}[row sep=0.3cm, column sep=0.3cm]
\lstick{$\ket{0_{\!S}}$} & \qwbundle{dq}
  & \gate{\hat P_{\!S}}
  & \gate[wires=2]{\hat V_{\!S\!M}(t_1)}
  & \qw
  & \gate{\hat T(t_1)}
  & \ \cdots\ 
  & \gate[wires=2]{\hat V_{\!S\!M}(t_R)}
  & \qw
  & \gate{\hat T(t_R)}
  & \meter{}\\
\lstick{$\ket{0_{\!M}}$} & \qwbundle{m} 
  & \qw
  & \ghost{\hat V_{\!S\!M}(t_1)}
  & \gate{\text{Reset}}
  & \qw
  & \ \cdots\ 
  & \ghost{\hat V_{\!S\!M}(t_R)}
  & \gate{\text{Reset}}
  & \qw
  & 
\end{quantikz}
\end{qcircuit}
\noindent 
The potential block, defined in \cref{eq:online_potential_operator}, can be decomposed as in \cref{circ:online_step}.
There, the operator $\hat{W}_{\!S\!M}$ of \cref{eq:W extended} and its inverse are represented as gates on $M$ controlled by $S$, since they apply $\hat{W}_{\!M}(\boldsymbol{x}_r)$ conditionally on the computational-basis state $\ket{\boldsymbol{x}_r}$ of the simulation register, while the rotation $e^{-i\varDelta t\,b(t)\,\hat{O}_{\!M}}$ acts on the measure register only.
\begin{qcircuit}[circ:online_step]
    \begin{quantikz}[row sep=0.5cm, column sep=0.3cm, baseline=(current bounding box.center)]
      \lstick{$S$}
        & \qwbundle{dq}
        & \gate[wires=2]{\hat{V}_{\!S\!M}(t)}
        & \qw \\
      \lstick{$M$}
        & \qwbundle{m}
        & \ghost{}
        & \qw
    \end{quantikz}
    =
    \begin{quantikz}[row sep=0.5cm, column sep=0.3cm, baseline=(current bounding box.center)]
      \lstick{$S$}
        & \qwbundle{dq}
        & \ctrl{1}
        & \qw
        & \ctrl{1}
        & \qw \\
      \lstick{$M$}
        & \qwbundle{m}
        & \gate{\hat{W}_{\!M}}
        & \gate{e^{- i \varDelta t\, b(t)\, \hat{O}_{\!M}}}
        & \gate{\hat{W}^{\dagger}_{\!M}}
        & \qw
    \end{quantikz} \ .
\end{qcircuit}
\noindent If $\hat W(\boldsymbol x)$ consists of a polynomial number of gates and $\hat{O}$ is, e.g., a Pauli string, so that $e^{-i\varDelta t\,b(t)\,\hat{O}_{\!M}}$ is a single multi-qubit Pauli rotation, the online potential step is implemented with a number of gates polynomial in $m$, $d$ and $q$, as we now show.

Assume that $\hat W(\boldsymbol x)$ is constructed by composing a polynomial number of fixed ($\boldsymbol x$-independent) entangling gates and
single-qubit rotations $\hat{R}_{\boldsymbol{n}}(x)$ about an axis $\boldsymbol{n}$, where $x\in[0,2\pi)$ is a component of $\boldsymbol{x}$.
Then $\hat W_{\!S\!M}$ from \cref{eq:W extended} can be decomposed as a combination of fixed operations acting on $M$ only and controlled rotations.
As in \cref{sec:gate simul}, $x$ is discretized on a grid, here over the full period $[0,2\pi)$, and encoded in the simulation register via computational basis states
\begin{equation}\label{eq:encoding_main}
  \ket{x_r}=\ket{\frac{2\pi}{2^q}r}=\ket{c_{q-1}\,\ldots\,c_0} \ ,\quad \text{with }
    r=\sum_{j=0}^{q-1} 2^j c_j \ .
\end{equation}
Then we have
\begin{equation}
  \hat{R}_{\boldsymbol{n}}(x_r)=\hat{R}_{\boldsymbol{n}}\!\left(\sum_{j=0}^{q-1} c_j \frac{2\pi}{2^{q-j}}\right) = \prod_{j=0}^{q-1}\hat{R}_{\boldsymbol{n}}\!\left(c_j \frac{2\pi}{2^{q-j}}\right) \ .
\end{equation}
Therefore, the $j$-th qubit of the simulation register controls the rotation by the angle $\frac{2\pi}{2^{q-j}}$.
Having explicitly characterized its action on a generic computational-basis state, by linearity \cref{circ:implemetazione_rotazione_controllata} shows how to implement the controlled rotations that appear in the controlled operator $\hat{W}_{\!S\!M}$ of \cref{circ:online_step}:
\begin{qcircuit}[circ:implemetazione_rotazione_controllata]
  \begin{quantikz}[row sep=0.5cm, column sep=0.3cm]
          \lstick{$S$}  & \qwbundle{q} & \ctrl{1}& \\
          \lstick{$M$} & & \gate{\hat R_{\boldsymbol{n}}}&
  \end{quantikz}
  \quad \coloneqq \quad
  \begin{quantikz}[row sep=0.4cm, column sep=0.25cm]
   &[0.7cm]
    \lstick{$c_{q-1}$}\wireoverride{n} &
    \ctrl{4} & \qw & \qw & \qw & \qw \\
  &[0.7cm]
    \lstick{$c_{q-2}$}\wireoverride{n} &
    \qw & \ctrl{3} & \qw & \qw & \qw \\
      & [0.7cm]\lstick{$\vdots$}
      \wireoverride{n} 
      & \wireoverride{n} 
      & \wireoverride{n}
      & \wireoverride{n}
      & \wireoverride{n}
      & \wireoverride{n}
      \\
  &[0.7cm]
    \lstick{$c_{0}$}\wireoverride{n} &
    \qw & \qw & \qw & \ctrl{1} & \qw \\
  & \lstick{$M$}\wireoverride{n} &
    \gate{\hat R_{\boldsymbol{n}}(\pi)} &
    \gate{\hat R_{\boldsymbol{n}}\left(\frac{\pi}{2}\right)} &
    \ \cdots \  &
    \gate{\hat R_{\boldsymbol{n}}\left(\frac{\pi}{2^{q-1}}\right)} &
    \qw
  \end{quantikz} \ .
\end{qcircuit}
This circuit is readily generalizable to problems in arbitrary dimension $d$, where each rotation that depends on $x_\ell$ is controlled by the sub-register $S_\ell$: every parametric rotation of $\hat W(\boldsymbol x)$ is thus replaced by $q$ controlled rotations, and $\hat{W}_{\!S\!M}$ requires $\mathcal{O}(q)$ times as many elementary gates as $\hat W(\boldsymbol x)$.

The approximation error in \cref{eq:effective_evolution} has a simple structure.
The Kraus operators of the online step with potential angle $\varphi$, derived in \cref{eq:kraus_online} of \cref{app:online_error}, are diagonal in the computational basis,
\begin{equation}
  \hat K_m=\sum_rK_m(\boldsymbol x_r)\ketbra{\boldsymbol x_r}{\boldsymbol x_r} \ ,
  \qquad
  K_m(\boldsymbol x_r)=\matrixel{m}{\hat W_{\!M}^\dagger(\boldsymbol x_r)\,e^{-i\varphi\hat{O}_{\!M}}\,\hat W_{\!M}(\boldsymbol x_r)}{0_{\!M}} \ ,
\end{equation}
where $\ket{m}$ runs over the computational basis of the measure register.
Hence, the online step, i.e.~the channel $\Phi^{\mathrm{on}}(\rho)=\sum_m\hat K_m\,\rho\,\hat K_m^\dagger$, acts on the matrix elements of the state $\rho$ of the simulation register as
\begin{equation}\label{eq:online_hadamard}
  \matrixel{\boldsymbol x_r}{\Phi^{\mathrm{on}}(\rho)}{\boldsymbol x_s}=\matrixel{\boldsymbol x_r}{\rho}{\boldsymbol x_s}\sum_mK_m(\boldsymbol x_r)\,K_m^*(\boldsymbol x_s) \ .
\end{equation}
Since $\sum_m\left|K_m(\boldsymbol x_r)\right|^2=1$ by unitarity, the factor in \cref{eq:online_hadamard} equals $1$ for $r=s$ and, by the Cauchy--Schwarz inequality, has modulus at most $1$ for $r\neq s$: each online step leaves the populations of the grid points exactly unchanged and damps the coherences between them.
The online procedure thus causes a loss of coherence, which the subsequent kinetic steps turn into an error on the populations, and hence on $P_{\mathrm{suc}}$ and $\mathbb{E}$.

The error in \cref{eq:effective_evolution} is $\mathcal O\left(\varDelta t^2\right)$ per step, i.e.~$\mathcal O\left(R\,\varDelta t^2\right)$ in total.
More precisely, defining the online procedure error $\don$ as the trace distance between the final states of the online and of the unitary Trotterized evolutions, in \cref{app:online_error} we derive the following upper bound for the online procedure,
\begin{equation}\label{eq:online_bound_main}
    \don \lesssim \sum_{j=1}^R \varphi_j^2=\sum_{j=1}^R\left[b(t_j)\,\varDelta t\right]^2 \ ,
\end{equation}
which does not show any dependence on $d$ and $q$.
The bound relies on an expansion to second order in the potential angles, so it requires $\varphi_j\ll1$, and replaces the variance of $\hat{O}$ on the states $\ket{\psi(\boldsymbol x_r)}$ by its upper bound $\norm{\hat{O}}^2$, which equals $1$ for the Pauli strings considered here.
For quantum functions, the error analysis of \cref{sec:gate simul} has therefore one additional ingredient, the online procedure error $\don$ of \cref{eq:online_bound_main}.
Combining, via the triangle inequality, the Trotter error \cref{eq:trotter error} and the online procedure error \cref{eq:online_bound_main}, at first order in $\varDelta t$, the QHD error via the online procedure, i.e.~the trace distance between the final state of the online evolution and the exact (untrotterized) final state, is
\begin{equation}\label{eq:error QHD}
  \eonline \lesssim \varDelta t\left[\frac{\osc}{4}\left(b(T)-b(0)\right)+\int_0^T\left(c_1\sqrt{a\,b^3}+c_2\,a\,b+b^2\right)dt\right] \ ,
\end{equation}
with $c_1$ and $c_2$ as in \cref{eq:trotter_error_quantum_main}.
\Cref{eq:error QHD} contains no dependence on $q$ and an at most linear dependence on $d$.
The two contributions have a different structure: the Trotter error depends on the cost function through the coefficients \cref{eq:trotter_error_quantum_main}, i.e.~only through its trigonometric structure, and on both schedules, through the integrals of $a\,b$ and $\sqrt{a\,b^3}$; instead, the online procedure error depends on the cost function only through $\norm{\hat{O}}$, and on the schedules only through the potential angles $\varphi_j=b(t_j)\,\varDelta t$, i.e.~it does not depend on the kinetic schedule $a$.

The gauge analysis of \cref{sec:schedules} applies unchanged with $g(r)\mapsto g(r)+r^2$, where the last term comes from the online procedure error.
This is always the leading contribution in our simulations, because with midpoint time sampling the Trotter error is second order in $\varDelta t$ (see \cref{app:trotter_error}), while the online procedure error is first order, $\sum_j\varphi_j^2\simeq\varDelta t\int_0^Tb^2\,dt$ (\cref{eq:online_error_schedules}).
Hence, by the argument of \cref{app:schedules} with $g(r)=r^2$, the gauge that minimizes \cref{eq:online_bound_main} at fixed curve is $a\propto1/r$, i.e.~$b$ is constant.
In this gauge the cost length $s(\tau)$ of the curve \cref{eq:r_curve}, defined in \cref{eq:cost_length} and proportional to the physical time $t$ (see \cref{eq:dt_ds}), is $s(\tau)=\int_0^\tau r(\tau')\,d\tau'=\left[r(\tau)-r_0\right]/k$.
Thus the ratio $r$ grows linearly in physical time, and the Trotter steps of \cref{app:angles} are uniform in $r$, with angles
\begin{equation}\label{eq:angles_online}
  \alpha_j=\frac{1}{k}\ln\left(\frac{r_0+k\,s_{j+1}}{r_0+k\,s_j}\right) \ ,
  \qquad
  \varphi_j=\frac{r_{\mathrm{f}}-r_0}{k\,R} \ .
\end{equation}
All the potential angles are equal, and the bound \cref{eq:online_bound_main} becomes
\begin{equation}\label{eq:online_bound_curve}
  \don\lesssim\frac{\left(r_{\mathrm{f}}-r_0\right)^2}{k^2R} \ .
\end{equation}
The (dominant) online procedure error decreases as $1/R$, but it grows as $1/k^2$ and as $r_{\mathrm{f}}^2$: improving the success probability with a slower curve requires a quadratically larger number of steps.
Therefore, if QHD with the online procedure were used to refine the solution up to the large ratio $r_{\mathrm{f}}$ fixed by the target accuracy $\delta_f$ through \cref{eq:risultato_convergenza_QHD}, the cost would be dominated by the online simulation of the potential, since $r_{\mathrm{f}}\propto\delta_f^{-2}$ and hence $R\propto\delta_f^{-4}$: this supports the hybrid strategy of \cref{app:curve}, in which QHD stops at a moderate $r_{\mathrm{f}}$ (in \cref{sec:results} we stop the curve at $r_{\mathrm{f}}=10$) and the solution is refined classically.

This is why the error analysis, together with the decoupling between curve and gauge, is essential for quantum functions.
For classical functions, the potential steps are exact, and the schedules can be chosen by looking at the ideal dynamics alone (\cref{sec:Optimization of Classical Functions}).
With the online procedure, instead, the ideal dynamics and the cost of the simulation compete: a slower curve (which increases $P_{\mathrm{suc}}$) and a larger $r_{\mathrm{f}}$ (which decreases $\mathbb{E}$) both make the simulation more expensive.
The error analysis turns this trade-off into a design rule: the curve, i.e.~$k$ and $r_{\mathrm{f}}$, is chosen on the basis of the ideal dynamics, which does not depend on the gauge and can be studied with exact potential steps, e.g.~in classical simulations of small instances; the gauge, constant $b$, minimizes the dominant error without affecting the ideal dynamics.
Finally, \cref{eq:online_bound_curve} gives the number of steps $R\simeq\left(r_{\mathrm{f}}-r_0\right)^2/\left(k^2\varepsilon^*\right)$ sufficient for a target online procedure error $\varepsilon^*$, before the algorithm is run on quantum hardware.

\subsection{Results}
\label{sec:results}

We test QHD with the online procedure of \cref{sec:online_procedure} on five quantum functions.
Their circuits, observables and analytic expressions are given in \cref{app:quantum_instances}; the analytic expressions are only used to analyze the results, since the algorithm accesses $f$ through the circuits.
The online procedure is simulated exactly, by applying the channel \cref{eq:online_hadamard} to the density matrix of the simulation register.

\begin{figure}[htbp]
  \centering
  \includegraphics[width=\linewidth]{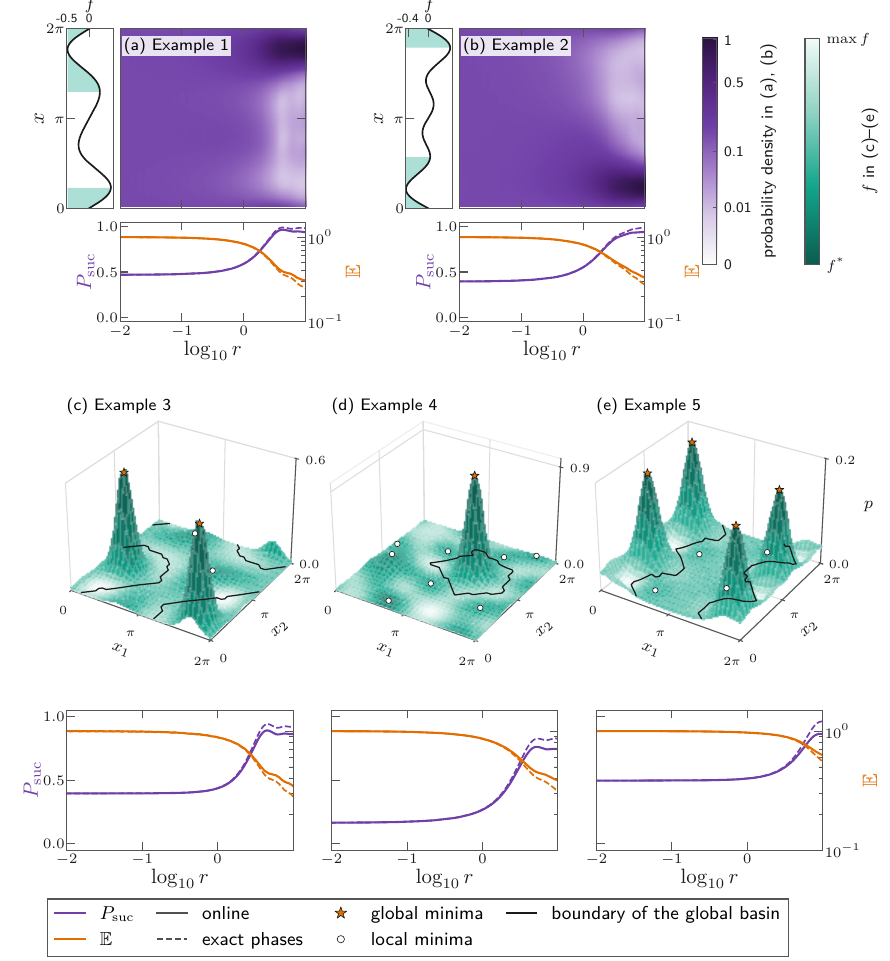}
  \caption{QHD with the online procedure on the quantum functions of \cref{tab:quantum_results}, with the same protocol.
  (a), (b) Examples 1 and 2 ($d=1$), as in \cref{fig:casi_1d}: cost function $f(x)$, with the basin of the global minimum shaded (left), and probability distribution of the state as a function of $\log_{10}r$ (power-law color scale).
  (c)--(e) Examples 3--5 ($d=2$): probability distribution of the final state, at $r=r_{\mathrm{f}}$, on the simulation grid (height), colored according to the value of $f$; stars and circles mark the global and local minima, and the black line the boundary of the basin of the global minimum.
  Bottom of each panel: success probability $P_{\mathrm{suc}}$ (left axis) and coefficient $\mathbb{E}$ (right axis, logarithmic scale) along the curve, with the online procedure (solid lines) and with exact phases (dashed lines).}
  \label{fig:casi_online}
\end{figure}

\begin{table}[t]
  \centering
  \caption{Final success probability $P_{\mathrm{suc}}$ and final coefficient $\mathbb{E}$ for the quantum functions of \cref{app:quantum_instances}, obtained with the same protocol for all instances: curve $r(\tau)=r_0\,e^{k\tau}$ of \cref{eq:r_curve} with $k=0.1$, from $r_0=10^{-4}$ to $r_{\mathrm{f}}=10$, run through in the gauge with constant $b$ of \cref{sec:online_procedure}, and $R=2^{15}$ Trotter steps.
  Columns: $d$, number of parameters; $q$, qubits per parameter; $m$, qubits of the measure register; minima, number of local minima of $f$ in $[0,2\pi)^d$, with the global ones in parentheses.
  The Trotterized evolution with exact potential steps (exact phases) is compared with the online procedure.
  The results with exact phases are checked with $2R$ steps: $P_{\mathrm{suc}}$ and $\mathbb{E}$ change by less than $1\%$, so that the Trotter error is negligible.
  Symbols: $^{\ddagger}$, $P_{\mathrm{suc}}$ is still changing at $r_{\mathrm{f}}$ and reaches $1.00$ at $r=10^2$.}
  \label{tab:quantum_results}
  \setlength{\tabcolsep}{5pt}
  \begin{tabular}{|l|cccc|cc|cc|}
    \hline
    & & & & & \multicolumn{2}{c|}{exact phases} & \multicolumn{2}{c|}{online} \\
    \textbf{Instance} & $d$ & $q$ & $m$ & minima & $P_{\mathrm{suc}}$ & $\mathbb{E}$ & $P_{\mathrm{suc}}$ & $\mathbb{E}$ \\
    \hline
    Example 1 & 1 & 8 & 3 & 2 (1) & 0.98 & 0.25 & 0.94 & 0.31 \\
    Example 2 & 1 & 8 & 3 & 3 (1) & 0.99 & 0.28 & 0.94 & 0.33 \\
    Example 3 & 2 & 5 & 2 & 4 (2) & 0.92 & 0.28 & 0.86 & 0.34 \\
    Example 4 & 2 & 5 & 3 & 9 (1) & 0.83 & 0.32 & 0.74 & 0.39 \\
    Example 5$^{\ddagger}$ & 2 & 5 & 3 & 8 (4) & 0.96 & 0.56 & 0.86 & 0.63 \\
    \hline
  \end{tabular}
\end{table}

In \cref{fig:casi_online} we report the evolution of the state along the curve and in \cref{tab:quantum_results} the final success probability and coefficient $\mathbb{E}$.
With $R=2^{15}$, while the Trotter error is negligible, the online procedure error is not: the state evolved with the online procedure achieves lower success probabilities than the one evolved with exact phases, and $\mathbb{E}$ is correspondingly larger; according to \cref{eq:online_bound_curve}, the bound on this difference decreases as $1/R$.
As for classical functions, $P_{\mathrm{suc}}<1$ is due to the lack of adiabaticity of the evolution at the rate $k$: the lowest value is found for Example 4, which has nine minima and the smallest basin of the global minimum; with a slower curve, $k=0.03$, the success probability with exact phases reaches at least $0.99$ for Examples 1--4 (not shown).
For Example 5, instead, the search is not over at $r_{\mathrm{f}}$: $P_{\mathrm{suc}}$ is still increasing at the end of the curve and reaches $1.00$ at $r=10^2$ (not shown), while the probability is shared among the four degenerate global minima (\cref{fig:casi_online}e).
On the other hand, slower curves and longer evolutions heavily increase the online procedure error: by \cref{eq:online_bound_curve}, the curve with $k=0.03$ would require $\left(0.1/0.03\right)^2\simeq11$ times more steps for the same bound, and continuing up to $r=10^2$ about $100$ times more.
This is why the values of $\mathbb{E}$ with exact phases, between $0.25$ and $0.56$, are generally larger than for classical functions of \cref{tab:standard_results}: the curve has to stop at a moderate ratio (\cref{sec:online_procedure}), where the ground state is still wide within each valley.

\subsection{Cost functions defined by datasets}
\label{sec:datasets}

In supervised quantum machine learning, the cost function is often an empirical average over a training set $\mathcal Z$,
\begin{equation}\label{eq:dataset_cost}
  f(\boldsymbol x)=\frac{1}{|\mathcal Z|}\sum_{z\in\mathcal Z}\ell(\boldsymbol x,z) \ ,
\end{equation}
where each data point $z$, for instance an input together with its label, enters the circuit $\hat W(\boldsymbol x,z)$ through fixed gates that are known classically, and $\ell(\boldsymbol x,z)$ is the corresponding loss.
We have not studied this case, and here we only discuss how it fits into our framework.

The online procedure of \cref{sec:online_procedure} applies to losses that are expectation values, $\ell(\boldsymbol x,z)=\matrixel{0}{\hat W^\dagger(\boldsymbol x,z)\,\hat O_z\,\hat W(\boldsymbol x,z)}{0}$, possibly with an observable $\hat O_z$ that depends on the label.
Losses that are powers of expectation values, such as the squared loss, can be brought to this form on several copies of the measure register, all controlled by the same simulation register, since $\langle\hat O\rangle^2=\langle\hat O\otimes\hat O\rangle$ on two copies of the state.

The product-formula structure of our simulation makes the sum over the dataset natural.
All the terms $\ell(\hat{\boldsymbol x},z)$ are diagonal in the computational basis of the simulation register and commute with each other, so that the potential step factorizes exactly,
\begin{equation}
  e^{-i\varphi_j f(\hat{\boldsymbol x})}=\prod_{z\in\mathcal Z}e^{-i\varphi_j\,\ell(\hat{\boldsymbol x},z)/|\mathcal Z|} \ ,
\end{equation}
and each factor can be implemented by an online step with the circuit $\hat W(\boldsymbol x,z)$, without additional Trotter error and without coherent access to the dataset.
Processing the whole dataset in a single potential step multiplies by $|\mathcal Z|$ the number of online steps per potential step.
However, since the online procedure error is quadratic in the angles, the bound \cref{eq:online_bound_main} decreases by the same factor, so that $R$ can be reduced by $|\mathcal Z|$ at equal error, as long as the Trotter error does not become dominant.
Alternatively, one can apply at each Trotter step the online step of a single data point $z_j$ with the full angle $\varphi_j$, as in stochastic gradient descent and in randomized Hamiltonian simulation \cite{campbell2019random}.
If $z_j$ is drawn independently, at each step and in each run, from the distribution $p(z)=1/|\mathcal Z|$, the statistics of the final measurement are determined by the channel averaged over $z_j$, so that \cref{prop:online_step} from \cref{app:online_error} applies with $\norm{\ket{\delta_r}}^2$ replaced by
\begin{equation}
  \mathbb{E}_z\norm{\ket{\delta_{z,r}}}^2+\mathrm{Var}_z\,\ell(\boldsymbol x_r,z)\leq\max_z\norm{\hat O_z}^2 \ ,
\end{equation}
where $\ket{\delta_{z,r}}$ is the vector $\ket{\delta_r}$ of \cref{app:online_error} for the data point $z$.
Hence, for Pauli strings, the bound \cref{eq:online_bound_main} holds unchanged, and the number of steps does not depend on $|\mathcal Z|$.
This bound holds for the average over runs, each with a new sequence: in a single run, the $R$ sampled data points reproduce $p(z)$ only up to fluctuations of order $R^{-1/2}$.
The term $\mathrm{Var}_z\,\ell$ describes a dephasing in the computational basis, of the same kind as that of the online procedure.
An intermediate strategy uses at each Trotter step a minibatch of $n_{b}$ data points, each with angle $\varphi_j/n_{b}$.
The effects of the dephasing due to sampling on the dynamics, such as heating or an implicit regularization as in stochastic gradient descent, are left to future work, together with the study of optimal schedules. 

%

\section{Conclusions}
\label{sec:conclusions}

We have studied the implementation of Quantum Hamiltonian Descent on digital quantum computers, and its application to the training of parametric quantum circuits, also called ``quantum functions'' here.
On the simulation side, we have shown that the Trotterized evolution in the interaction picture of the kinetic term admits an error bound that, for cost functions with a sufficiently smooth periodic extension, does not grow with the number $q$ of qubits per parameter, even though the norm of the momentum operator, and hence of the commutator that controls product formulas, grows exponentially with $q$.
The key ingredient is a bound on the kinetic energy of the exact state, which holds for every cost function and every grid as long as the ratio $r=b/a$ of the schedules is non-decreasing.
For generic cost functions, the dependence of the error bound on $q$ is set by the regularity of the periodic extension of $f$, while for quantum functions, which are trigonometric polynomials, the bound holds for every $q$ and grows at most linearly with the number of parameters $d$.
On the algorithmic side, the reparametrization of time in terms of kinetic time shows that QHD is determined by a single curve $r(\tau)$, and that the remaining gauge freedom can be fixed to minimize the number of Trotter steps.
With a curve that spends kinetic time in the search window, rather than in the regions where nothing happens, the success probability improves substantially at equal cost, as we have shown for a standard benchmark function.
For quantum functions, the online potential step makes QHD run entirely on the quantum computer: the cost function is never estimated classically, and the bound on the error of the step depends neither on $q$ nor on $d$.
Since this error is of first order in the time step, whereas the Trotter error is of second order with midpoint sampling, the online error dominates the gate count; this favors a hybrid strategy, in which QHD is used for the global search and the solution is then refined by a local optimizer.

Several directions remain open.
Our bounds are first order in the time step and rely on a worst-case estimate of the kinetic energy, so that they are pessimistic; a second-order analysis, based on the same energy bound, should give tighter estimates.
The online step can be replaced by a coherent implementation of $e^{-i\varphi f(\hat{\bm x})}$: when $\hat{O}$ is a Pauli string, the operator $\hat W_{\!S\!M}^\dagger\left(\hat{\mathbb I}_S\otimes\hat{O}_{\!M}\right)\hat W_{\!S\!M}$, introduced with the online procedure, is a block encoding of $f(\hat{\bm x})$, and quantum signal processing \cite{Low2019hamiltonian,gilyen2019quantum} could implement the potential step without loss of coherence.
Combined with the exact integration of the kinetic term in the interaction picture, which avoids any dependence of the cost on the norm of the momentum operator, this would retain the grid-independent error analysis presented here.
Finally, the hardness of QHD resides in the search window, where the gaps of the Hamiltonian depend on the landscape: understanding how these gaps behave for the landscapes of parametric circuits, including those affected by barren plateaus, and comparing QHD with classical optimizers at equal number of circuit evaluations, are natural next steps.

\paragraph{Note added.}
After completing this work, we became aware of Ref.~\cite{acedo2026coherent}, which appeared on arXiv on 3 September 2026 and independently proposes to train parametric quantum circuits through a QHD-like Hamiltonian evolution of a quantum register encoding the parameters, with the loss function accessed through a coherent evaluation of the circuit over the parameter grid.
The main ideas of the present work, in particular the coherent evaluation of quantum cost functions within QHD, were developed in the Master's thesis of one of us \cite{burato2026thesis}, defended on 23 April 2026.
The two works are complementary: Ref.~\cite{acedo2026coherent} constructs the loss Hamiltonian through block encodings and simulates the evolution by qubitization with an extra clock dimension, with applications to batched learning and parameter estimation, while a complete complexity analysis is left for future work.
Here, on the other hand, we focus on error bounds for the Trotterized simulation that are independent of the grid resolution and at most linear in the number of parameters, on the design of the schedules, and on an online potential step that does not require block-encoding techniques, while the application 
to machine learning problems is left to future work.

\cleardoublepage

\appendix
\crefalias{section}{appendix}
\crefalias{subsection}{appendix}

\section{Proof of the Global Convergence Theorem}
\label{App:dimostrazione_teorema_convergenza_globale}

    In this appendix we prove \cref{thm:convergenza_globale_QHD}, extending to $d$ dimensions and to an anisotropic Hessian the one-dimensional proof of \cite[App.~B.2]{leng2023quantum}.
    First, without loss of generality, we set $f(\boldsymbol{x}^*)=0$.
    Since $f$ is continuous and $f(\boldsymbol{x})\to+\infty$ for $\left|\boldsymbol{x}\right|\to\infty$, $\hat{H}(t)$ has a compact resolvent, and hence a purely discrete spectrum, at any time.
    Let $E_n(t)$ denote its eigenvalues and $\ket{n_t}$ its eigenvectors.
    We choose to represent the state of the system in the basis of the Hamiltonian eigenvectors:
    \[
    \ket{\psi(t)}=\sum_nc_n(t) e^{i\theta_n(t)} \ket{n_t}\ ,\text{ with }\theta_n(t)=-\int_0^tE_n(t')dt' \ .
    \]
    We substitute this expression for the state into the Schrödinger equation and take the inner product with $\bra{m_t}$:
    \begin{align*}
        0&=\bra{m_t}i\frac{\partial\ket{\psi(t)}}{\partial t} - \bra{m_t}\hat{H}(t)\ket{\psi(t)}\\
        &=\bra{m_t}i\frac{\partial}{\partial t}\left[\sum_nc_n(t) e^{i\theta_n(t)} \ket{n_t}\right] - \bra{m_t}\hat{H}(t)\left[\sum_nc_n(t) e^{i\theta_n(t)} \ket{n_t}\right]\\
        &=i\bra{m_t}\sum_n\left[\dot{c}_ne^{i\theta_n}\ket{n_t}+ic_n\dot{\theta}_ne^{i\theta_n}\ket{n_t}+c_ne^{i\theta_n}\frac{\partial\ket{n_t}}{\partial t}\right] - \bra{m_t}\sum_nc_n e^{i\theta_n}E_n\ket{n_t}\\
        &=\!i\,\dot{c}_me^{i\theta_m}\!+\!\bra{m_t}\sum_nc_nE_ne^{i\theta_n}\ket{n_t}\!+\!i\sum_nc_ne^{i\theta_n}\bra{m_t}\frac{\partial\ket{n_t}}{\partial t}\! -\! \bra{m_t}\sum_nc_n e^{i\theta_n}E_n\ket{n_t}\\
        &=i\,e^{i\theta_m}\left[\dot{c}_m+\sum_nc_ne^{i(\theta_n-\theta_m)}\bra{m_t}\frac{\partial\ket{n_t}}{\partial t}\right] \ ,
    \end{align*}
    so that the term in square brackets vanishes.
    We now differentiate the eigenvalue equation for the Hamiltonian and take the inner product with $\bra{m_t}$:
    \begin{align*}
        \bra{m_t}\frac{d}{dt}\left[\hat{H}(t)\ket{n_t}\right]&=\bra{m_t}\frac{d}{dt}\left[E_n(t)\ket{n_t}\right]\\
        \matrixel{m_t}{\dot{\hat{H}}}{n_t}+\bra{m_t}\hat{H}(t)\frac{\partial\ket{n_t}}{\partial t} &=\bra{m_t}\dot{E_n}\ket{n_t}+E_n(t)\bra{m_t}\frac{\partial \ket{n_t}}{  \partial t}\\
        \matrixel{m_t}{\dot{\hat{H}}}{n_t}+E_m(t)\bra{m_t}\frac{\partial \ket{n_t}}{  \partial t} &= E_n(t)\bra{m_t}\frac{\partial \ket{n_t}}{  \partial t}\\
        \bra{m_t}\frac{\partial \ket{n_t}}{  \partial t}&=\frac{\matrixel{m_t}{\dot{\hat{H}}}{n_t}}{E_n(t)-E_m(t)} \quad \forall m\neq n\ .
    \end{align*}
    Substituting this result into the previous equation yields a system of coupled differential equations:
    \begin{equation}\label{eq: approx adiabatica}
        \dot{c}_m=-c_m(t)\bra{m_t}\frac{\partial \ket{m_t}}{  \partial t}-\sum_{n\neq m}c_n(t)e^{i(\theta_n(t)-\theta_m(t))}\frac{\matrixel{m_t}{\dot{\hat{H}}}{n_t}}{E_n(t)-E_m(t)} \ .
    \end{equation}
    By the assumptions of \cref{thm:convergenza_globale_QHD} on the Hamiltonian, i.e.~slow variation and no level crossings, we can invoke the adiabatic approximation and neglect the second term, which is responsible for the mixing between orthogonal eigenstates induced by the time dependence of the Hamiltonian.
    We observe that
    \[
    \Re\left[\bra{m_t}\frac{\partial{\ket{m_t}}}{\partial t}\right]=\frac{1}{2}\left[\bra{m_t}\frac{\partial{\ket{m_t}}}{\partial t}+\frac{\partial\bra{m_t}}{\partial t}\ket{m_t}\right]=\frac{1}{2}\frac{d}{dt}\braket{m_t}{m_t}=0 \ .
    \]
    Therefore,
    \[
    c_m(t)\simeq c_m(0)e^{-i\int_0^t\mathcal{A}_m(s)ds}\ ,\text{ where } \mathcal{A}_m(t)=\Im\left(\bra{m_t}\frac{\partial \ket{m_t}}{  \partial t}\right)\in\mathbb{R} \ .
    \]
    In this way, we have shown that the amplitudes of the expansion coefficients of $\ket{\psi(t)}$ in the instantaneous eigenbasis of the Hamiltonian do not change.
    Hence, the state remains constrained to the subspace generated by the first $\mathcal{N}$ eigenstates.

    We now consider the expectation value of $f(\hat{\boldsymbol{x}})$ on the state at sufficiently long times, for which the eigenstates of the QHD Hamiltonian, denoted by $\ket{n_t}$, can be approximated by those of $\hat{H}_q(t)$, denoted by $\ket{e_n(t)}$.
    The approximation we use consists in replacing $f(\hat{\boldsymbol{x}})$ by $f_q(\hat{\boldsymbol x}) =\frac{\hat{V}_q(t)}{b(t)}$, 
where $\hat{V}_q(t) = b(t) f_q(\hat{\boldsymbol x}) $ is the 
potential in $\hat{H}_q(t)$; this approximation is valid because these operators are evaluated on harmonic-oscillator eigenstates, which are localized in a neighborhood of $\boldsymbol{x}^*$ where $f\simeq f_q$.

    \noindent We denote by $\varepsilon_m(t)$ the eigenvalues of the $d$-dimensional anisotropic harmonic oscillator of mass $1/a(t)$:
    \begin{equation}
        \varepsilon_m(t)=\sum_{i=1}^d \sqrt{a(t)\,b(t)\,\omega^2_i}\ \left(m_i+\frac{1}{2}\right) \ ,
    \end{equation}
    where, as in \cref{sec:Quantum Hamiltonian Descent}, $\omega^2_i$ are the eigenvalues of the Hessian matrix of $f$ evaluated at the global minimizer, and $\{m_1,\ldots,m_d\}$ is the set of quantum numbers identifying the $m$-th energy level.
    Moreover, choosing $m=0$ as the index of the ground state, we have $m\geq\max\{m_i\}$, and therefore, with $\omega=\max_i|\omega_i|$,
        \[
        \varepsilon_m(t)\leq d \,\omega\sqrt{a(t)\,b(t)}\left(m+\frac{1}{2}\right) \ .
        \]
    We begin by upper-bounding the expectation value of interest, using the virial theorem $\matrixel{e_m(t)}{\hat{V}_q}{e_m(t)}=\varepsilon_m/2$ and the triangle inequality:
    \begin{align*}
        \left|\matrixel{\psi(t)}{f(\hat{\boldsymbol{x}})}{\psi(t)}\right|
        &\simeq
        \left\lvert
        \begin{aligned}[c]
        &\sum_m |c_m|^2\matrixel{e_m(t)}{f(\hat{\boldsymbol{x}})}{e_m(t)}+ \\
        &\quad+\sum_{m\neq n}
        c_m^*c_n\,e^{i(\theta_n-\theta_m)}\,
        \matrixel{e_m(t)}{f(\hat{\boldsymbol{x}})}{e_n(t)}
        \end{aligned}
        \right\rvert\\
        &\simeq
        \left\lvert
        \begin{aligned}[c]
        &\sum_m|c_m|^2\matrixel{e_m(t)}{\frac{\hat{V}_q(t)}{b(t)}}{e_m(t)}+ \\
        &\quad+\sum_{m\neq n}
        c_m^*c_n\,e^{i(\theta_n-\theta_m)}\,
        \matrixel{e_m(t)}{\frac{\hat{V}_q}{b(t)}}{e_n(t)}
        \end{aligned}
        \right\rvert\\
        &=
        \left\lvert
        \begin{aligned}[c]
        &\sum_m|c_m|^2\,\frac{\varepsilon_m}{2\,b(t)}+ \\
        &\quad+\sum_{m\neq n}
        c_m^*c_n\,e^{i(\theta_n-\theta_m)}\,
        \matrixel{e_m(t)}{\frac{\hat{V}_q}{b(t)}}{e_n(t)}
        \end{aligned}
        \right\rvert\\
        &\leq\sum_m\left||c_m|^2\frac{\varepsilon_m}{2\,b(t)}\right|+\sum_{m\neq n}\left|c_m^*c_ne^{i(\theta_n-\theta_m)}\matrixel{e_m(t)}{\frac{\hat{V}_q}{b(t)}}{e_n(t)}\right| \ .
  \end{align*}
We now use the consequences of the adiabaticity assumption: we can truncate the sums by considering only the eigenspace of the first $\mathcal{N}$ eigenstates.
Within this eigenspace, we can upper-bound the expression by setting $\left|c_n(t)\right|=1\ \forall n<\mathcal{N}$, obtaining
\begin{align*}
  \left|\matrixel{\psi(t)}{f(\hat{\boldsymbol{x}})}{\psi(t)}\right|\leq \sum_{m<\mathcal{N}}\frac{\varepsilon_m}{2\,b(t)}+\sum_{\substack{m,n<\mathcal{N}\\ m\neq n}}\left|\matrixel{e_m(t)}{\frac{\hat{V}_q}{b(t)}}{e_n(t)}\right| \ .
\end{align*}
We observe that, since $\hat{V}_q$ is a non-negative Hermitian operator, the Cauchy--Schwarz inequality and the virial theorem give
\begin{align*}
  \left|\matrixel{e_m(t)}{\hat{V}_q}{e_n(t)}\right|
  &=
  \left|\matrixel{e_m(t)}{\sqrt{\hat{V}_q^\dagger}\sqrt{\hat{V}_q}}{e_n(t)}\right|\\
  &\leq\sqrt{\matrixel{e_m(t)}{\sqrt{\hat{V}_q^\dagger}\sqrt{\hat{V}_q}}{e_m(t)}\matrixel{e_n(t)}{\sqrt{\hat{V}_q^\dagger}\sqrt{\hat{V}_q}}{e_n(t)}}\\
  &=\sqrt{\matrixel{e_m(t)}{\hat{V}_q}{e_m(t)}\matrixel{e_n(t)}{\hat{V}_q}{e_n(t)}}\\
  &=\frac{\sqrt{\varepsilon_m\,\varepsilon_n}}{2} \ .
\end{align*}
Substituting this result and applying the Cauchy--Schwarz inequality to the sum over $n$ gives
\begin{align*}
  \left|\matrixel{\psi(t)}{f(\hat{\boldsymbol{x}})}{\psi(t)}\right|
  &\leq
  \frac{1}{2\,b(t)}\left[\sum_{m<\mathcal{N}}\varepsilon_m+\sum_{\substack{m,n<\mathcal{N}\\ m\neq n}}\sqrt{\varepsilon_m\,\varepsilon_n}\right]= \frac{1}{2\,b(t)}\sum_{m,n<\mathcal{N}}\sqrt{\varepsilon_m\,\varepsilon_n}\\
  &\leq \frac{d\,\omega}{2}\sqrt{\frac{a(t)}{b(t)}}\sum_{m,n<\mathcal{N}}\sqrt{\left(m+\frac{1}{2}\right)\left(n+\frac{1}{2}\right)}\\
  &=\frac{d\,\omega}{2}\sqrt{\frac{a(t)}{b(t)}}\left(\sum_{n=0}^{\mathcal{N}-1} \sqrt{n+\frac{1}{2}}\right)^2\\
  &\leq \frac{d\,\omega}{2}\sqrt{\frac{a(t)}{b(t)}}\left[\sum_{n=0}^{\mathcal{N}-1}\left(n+\frac{1}{2}\right)\right]\left[\sum_{n=0}^{\mathcal{N}-1}1\right]\\
  &= \frac{d\,\omega}{2}\sqrt{\frac{a(t)}{b(t)}}\left[\frac{\mathcal{N}(\mathcal{N}-1)}{2}+\frac{\mathcal{N}}{2}\right]\mathcal{N}\\
  &= \frac{d\,\mathcal{N}^3\omega}{4}\sqrt{\frac{a(t)}{b(t)}}\underset{t\to\infty}{\longrightarrow}0 \ .
\end{align*}
This proves the claim of the theorem.

\section{Trotterization error bound}
\label{app:trotter_error}

In this appendix we prove the bounds on the Trotterization error stated in \cref{sec:gate simul}.
We first move to the interaction picture with respect to the kinetic term and, by a telescoping sum, write the error as a sum of single-step errors acting on the \emph{exact} state: this reduces the problem to bounding the operator norm of $f(\hat{\boldsymbol{x}})$ and the norm of the commutator $\left[\hat T_0,f(\hat{\boldsymbol{x}})\right]$ applied to the exact trajectory (\cref{app:trotter_decomposition}).
We then bound this commutator in three cases: generic cost functions (\cref{app:trotter_generic}), cost functions with a smooth periodic extension (\cref{app:trotter_smooth}) and trigonometric polynomials, which include the quantum functions of \cref{sec:Optimization of Quantum Functions} (\cref{app:trotter_quantum}).
In all cases, the key ingredient is a bound on the kinetic energy of the exact state (\cref{lem:kinetic_moments}), which replaces the largest momentum of the grid with the momentum spread of the actual state.
Finally, we combine these results into the bound \cref{eq:trotter error} and give the explicit coefficients $c_1$ and $c_2$ (\cref{app:trotter_final}).

We make the following minimal hypotheses on the time schedules and on the initial state:
\begin{itemize}
  \item[(H1)] $a(t)$ and $b(t)$ are continuously differentiable on $[0,T]$, with $a(t)>0$ and $b(t)\geq0$;
  \item[(H2)] $b(t)$ is non-decreasing, i.e.~$\dot b(t)\geq0$;
  \item[(H3)] $b(t)/a(t)$ is non-decreasing;
  \item[(H4)] the initial state $\ket{\psi(0)}$ is the uniform superposition of all computational-basis states, i.e.~the zero-momentum eigenvector, so that $\hat T_0\ket{\psi(0)}=0$.
\end{itemize}
These hypotheses are generally satisfied by sensible choices of the QHD schedules, such as those used in this work, for which $a(t)$ decreases and $b(t)$ does not decrease with time, and by the usual initial state of QHD.
Hypotheses (H1) and (H2) enter the error decomposition, while (H1), (H3) and (H4) are used in \cref{lem:kinetic_moments} to bound the kinetic energy of the exact state.
For the refined bounds of \cref{prop:smooth,prop:trig}, we will also make assumptions on the regularity of the cost function, \cref{eq:smooth_hypothesis}, and on its Fourier spectrum, respectively.

\subsection{Error decomposition}
\label{app:trotter_decomposition}

Let $\ket{\psi(t)}$ be the exact trajectory generated by the space-discretized Hamiltonian \cref{eq:H_discretized}, which we write as $\hat{H}(t)=a(t)\hat T_0+b(t)f(\hat{\boldsymbol{x}})$, with $\hat T_0=\hat{\boldsymbol{p}}^2/2$ the discretized kinetic operator of \cref{sec:gate simul}:
\begin{equation}\label{eq:exact_trajectory}
  \ket{\psi(t)}=\hat{U}(t,0)\ket{\psi(0)} \ ,
  \qquad
  \hat{U}(t,0)\coloneqq\mathcal{T}\left[e^{-i\int_0^t\hat{H}(s)\,ds}\right] \ .
\end{equation}
We describe the evolution in the interaction picture with respect to the kinetic term.
Defining
\begin{equation}\label{eq:def_interaction_picture}
  A(t)\coloneqq\int_0^t a(s)\,ds \ ,
  \qquad
  \hat{U}_0(t)\coloneqq e^{-iA(t)\frac{\hat{\boldsymbol{p}}^2}{2}} \ ,
  \qquad
  \ket{\psi_I(t)}\coloneqq\hat{U}_0^\dagger(t)\ket{\psi(t)} \ ,
\end{equation}
the state $\ket{\psi_I(t)}$ evolves under the interaction-picture Hamiltonian
\begin{equation}\label{eq:H_interaction}
  \hat{H}_I(t)\coloneqq b(t)\,\hat{U}_0^\dagger(t)\,f(\hat{\boldsymbol{x}})\,\hat{U}_0(t) \ ,
\end{equation}
so that the exact evolution operator factorizes as
\begin{equation}\label{eq:U_interaction}
  \hat{U}(t,0)=\hat{U}_0(t)\,\hat{U}_I(t,0) \ ,
  \qquad
  \hat{U}_I(t,t')\coloneqq\mathcal{T}\left[e^{-i\int_{t'}^{t}\hat{H}_I(s)\,ds}\right] \ .
\end{equation}

We split the time interval $[0,T]$ into $R=T/\varDelta t$ intervals $\mathcal{I}_j\coloneqq\left[(j-1)\varDelta t,j\varDelta t\right]$ and in each of them we approximate $\hat{H}_I(s)$ with its value at a sampling time $t_j\in\mathcal{I}_j$ (e.g.~the midpoint $t_j=\left(j-\frac{1}{2}\right)\varDelta t$).
We denote by $\ket{\tilde{\psi}(T)}$ the final state of this Trotterized evolution:
\begin{equation}\label{eq:trotterized_state}
  \ket{\tilde{\psi}(T)}=\hat{U}_0(T)\,\hat{W}_R\cdots\hat{W}_1\ket{\psi(0)} \ ,
  \qquad
  \hat{W}_j\coloneqq e^{-i\varDelta t\,\hat{H}_I(t_j)}=\hat{U}_0^\dagger(t_j)\,\hat{V}(t_j)\,\hat{U}_0(t_j) \ ,
\end{equation}
where $\hat{V}(t)=e^{-i\varDelta t\,b(t)f(\hat{\boldsymbol{x}})}$ is the potential step of \cref{eq:trotter_steps}.

We define the Trotterization error as the distance between the final states of these two trajectories, up to a global phase, which is not observable:
\begin{equation}\label{eq:def_trotter_error}
  \eTrotter\coloneqq\min_{\theta\in\mathbb{R}}\left\|e^{i\theta}\ket{\tilde{\psi}(T)}-\ket{\psi(T)}\right\|
  \leq\left\|\left(\hat{W}_R\cdots\hat{W}_1-\hat{U}_I(T,0)\right)\ket{\psi(0)}\right\| \ ,
\end{equation}
where $\left\|\cdot\right\|$ is the norm induced by the scalar product, and the inequality follows from the choice $\theta=0$ and from the unitarity of $\hat{U}_0(T)$.
To evaluate the right-hand side, we also decompose $\hat{U}_I(T,0)$ into the product of $R$ partial evolutions:
\begin{equation}\label{eq:U_steps}
  \hat{U}_I(T,0)=\hat{U}_R\cdots\hat{U}_1 \ ,
  \qquad
  \hat{U}_j\coloneqq\hat{U}_I\left(j\varDelta t,(j-1)\varDelta t\right)=\mathcal{T}\left[e^{-i\int_{\mathcal{I}_j}\hat{H}_I(s)\,ds}\right] \ .
\end{equation}
We then split the final error into a sum over the steps by means of a telescoping sum:
\begin{align}
  \eTrotter
  &\leq\left\|\left(\hat{W}_R\cdots\hat{W}_1-\hat{U}_R\cdots\hat{U}_1\right)\ket{\psi(0)}\right\|
  \nonumber\\
  &=\left\|\sum_{j=1}^R\hat{W}_R\cdots\hat{W}_{j+1}\left(\hat{U}_j-\hat{W}_j\right)\hat{U}_{j-1}\cdots\hat{U}_1\ket{\psi(0)}\right\|
  \nonumber\\
  &\leq\sum_{j=1}^R\left\|\left(\hat{U}_j-\hat{W}_j\right)\ket{\psi_I\left((j-1)\varDelta t\right)}\right\|
  \eqqcolon\sum_{j=1}^R\varepsilon_j \ ,\label{eq:telescoping}
\end{align}
where we used the triangle inequality and the unitarity of the operators $\hat{W}_j$, and we identified $\hat{U}_{j-1}\cdots\hat{U}_1\ket{\psi(0)}$ with the exact state $\ket{\psi_I\left((j-1)\varDelta t\right)}$.
With this ordering of the telescoping sum, each single-step error $\hat{U}_j-\hat{W}_j$ acts on the \emph{exact} state at the beginning of the step.

To compute $\varepsilon_j$, we use the following lemma.
\begin{lemma}\label{lem:duhamel}
  Let $\hat{U}_1(s)$ and $\hat{U}_2(s)$ be evolution operators defined by
  \begin{equation}
    i\frac{\partial}{\partial s}\hat{U}_1(s)=\hat{H}_1(s)\,\hat{U}_1(s) \ ,
    \qquad
    i\frac{\partial}{\partial s}\hat{U}_2(s)=\hat{H}_2(s)\,\hat{U}_2(s) \ ,
    \qquad
    \hat{U}_1(0)=\hat{U}_2(0)=\hat{\mathbb{I}} \ ,
  \end{equation}
  with $\hat{H}_1(s)$ and $\hat{H}_2(s)$ Hermitian.
  Then, for any state $\ket{\psi}$ and any $t\geq0$,
  \begin{equation}\label{eq:duhamel}
    \left(\hat{U}_1(t)-\hat{U}_2(t)\right)\ket{\psi}=i\int_0^t\hat{U}_2(t)\,\hat{U}_2^\dagger(s)\left[\hat{H}_2(s)-\hat{H}_1(s)\right]\hat{U}_1(s)\ket{\psi}\,ds \ .
  \end{equation}
\end{lemma}

\begin{proof}
Let $\ket{\phi(s)}\coloneqq\hat{U}_2(t)\,\hat{U}_2^\dagger(s)\,\hat{U}_1(s)\ket{\psi}$.
Differentiating with respect to $s$, we get
\begin{align*}
  \frac{d}{ds}\ket{\phi(s)}
  &=\hat{U}_2(t)\left[\frac{d\hat{U}_2^\dagger(s)}{ds}\,\hat{U}_1(s)+\hat{U}_2^\dagger(s)\,\frac{d\hat{U}_1(s)}{ds}\right]\ket{\psi}\\
  &=i\,\hat{U}_2(t)\left[\hat{U}_2^\dagger(s)\,\hat{H}_2(s)\,\hat{U}_1(s)-\hat{U}_2^\dagger(s)\,\hat{H}_1(s)\,\hat{U}_1(s)\right]\ket{\psi}\\
  &=i\,\hat{U}_2(t)\,\hat{U}_2^\dagger(s)\left[\hat{H}_2(s)-\hat{H}_1(s)\right]\hat{U}_1(s)\ket{\psi} \ .
\end{align*}
Integrating over $[0,t]$, we get
\begin{align*}
  \int_0^t\frac{d}{ds}\ket{\phi(s)}\,ds
  &=i\int_0^t\hat{U}_2(t)\,\hat{U}_2^\dagger(s)\left[\hat{H}_2(s)-\hat{H}_1(s)\right]\hat{U}_1(s)\ket{\psi}\,ds \ ,\\
  \ket{\phi(t)}-\ket{\phi(0)}
  &=i\,\int_0^t\hat{U}_2(t)\hat{U}_2^\dagger(s)\left[\hat{H}_2(s)-\hat{H}_1(s)\right]\hat{U}_1(s)\ket{\psi}\,ds \ ,\\
  \left(\hat{U}_1(t)-\hat{U}_2(t)\right)\ket{\psi}
  &=i\int_0^t\hat{U}_2(t)\,\hat{U}_2^\dagger(s)\left[\hat{H}_2(s)-\hat{H}_1(s)\right]\hat{U}_1(s)\ket{\psi}\,ds \ ,
\end{align*}
where in the last line we used $\hat{U}_1(0)=\hat{U}_2(0)=\hat{\mathbb{I}}$ and $\hat{U}_2(t)\hat{U}_2^\dagger(t)=\hat{\mathbb{I}}$.
\end{proof}

\noindent From this lemma it immediately follows, by the triangle inequality and the unitarity of $\hat{U}_2(t)\,\hat{U}_2^\dagger(s)$, that
\begin{equation}\label{eq:duhamel_bound}
  \left\|\left(\hat{U}_1(t)-\hat{U}_2(t)\right)\ket{\psi}\right\|\leq\int_0^t\left\|\left[\hat{H}_2(s)-\hat{H}_1(s)\right]\hat{U}_1(s)\ket{\psi}\right\|ds \ .
\end{equation}
We apply \cref{eq:duhamel_bound} to each term $\varepsilon_j$ of \cref{eq:telescoping}, with the following identifications:
\begin{itemize}
  \item $t\to\varDelta t$ is the duration of the step, and $s\in[0,\varDelta t]$ is the time elapsed since its beginning;
  \item $\hat{H}_1(s)\to\hat{H}_I\left((j-1)\varDelta t+s\right)$, so that $\hat{U}_1(s)=\hat{U}_I\left((j-1)\varDelta t+s,(j-1)\varDelta t\right)$ and $\hat{U}_1(\varDelta t)=\hat{U}_j$;
  \item $\hat{H}_2(s)\to\hat{H}_I(t_j)$, independent of $s$, so that $\hat{U}_2(s)=e^{-is\hat{H}_I(t_j)}$ and $\hat{U}_2(\varDelta t)=\hat{W}_j$;
  \item $\ket{\psi}\to\ket{\psi_I\left((j-1)\varDelta t\right)}$, so that $\hat{U}_1(s)\ket{\psi}=\ket{\psi_I\left((j-1)\varDelta t+s\right)}$ is the exact trajectory.
\end{itemize}
Shifting the integration variable by $(j-1)\varDelta t$ and summing over the steps, we thus obtain
\begin{equation}\label{eq:trotter_error_bound}
  \eTrotter\leq\sum_{j=1}^R\int_{\mathcal{I}_j}\left\|\left[\hat{H}_I(s)-\hat{H}_I(t_j)\right]\ket{\psi_I(s)}\right\|ds \ .
\end{equation}
We calculate the operator acting on $\ket{\psi_I(s)}$ by means of the fundamental theorem of calculus:
\begin{align*}
  \hat H_I(s)-\hat H_I(t_j)
  &=
  \int_{t_j}^s \frac{d\hat H_I}{du}\,du
  \\ &=
  \int_{t_j}^s \left[\dot b(u) \hat U_0^\dagger(u) f(\hat{\boldsymbol{x}}) \hat U_0(u)+ b(u) i \frac{dA}{du}\hat U_0^\dagger(u)\left[\hat T_0, f(\hat{\boldsymbol{x}})\right]\hat U_0(u)\right]\,du
  \\ &=
  \int_{t_j}^s \hat U_0^\dagger(u)\left(\dot b(u) f(\hat{\boldsymbol{x}}) + i a(u)b(u) \left[\hat T_0, f(\hat{\boldsymbol{x}})\right]\right)\hat U_0(u)\,du \ .
\end{align*}
We apply this result to \cref{eq:trotter_error_bound}:
\begin{align}
  \eTrotter
  &\leq
  \sum_{j=1}^R\int_{\mathcal{I}_j}ds\left\| \int_{t_j}^s \hat U_0^\dagger(u)\left(\dot b(u) f(\hat{\boldsymbol{x}}) + i a(u)b(u) \left[\hat T_0, f(\hat{\boldsymbol{x}})\right]\right)\hat U_0(u)\,du \ket{\psi_I(s)}\right\| 
  \nonumber \\ &\leq
  \sum_{j=1}^R\int_{\mathcal{I}_j}ds\left| \int_{t_j}^s du \left[\dot b(u) \left\|f(\hat{\boldsymbol{x}})\right\| + a(u)b(u) \left\|\left[\hat T_0, f(\hat{\boldsymbol{x}})\right]\hat U_0(u)\hat U_0^\dagger(s)\ket{\psi(s)}\right\|\right] \right|
  \nonumber \\ &\leq
  \sum_{j=1}^R\int_{\mathcal{I}_j}ds\,\left|s-t_j\right| \left[\sup_{\mathcal{I}_j}\dot b \left\|f(\hat{\boldsymbol{x}})\right\| + \sup_{\mathcal{I}_j}ab \, \sup_{\sigma\in\mathbb R}\left\|\left[\hat T_0, f(\hat{\boldsymbol{x}})\right]e^{i\sigma\hat T_0}\ket{\psi(s)}\right\|\right]
  \ ,\label{eq:trotter_error_two_norms}
\end{align}
where we used hypotheses (H1) and (H2), and $\norm{\hat A}\coloneqq\sup_{\norm{\ket{\chi}}=1}\norm{\hat A\ket{\chi}}$ denotes the operator norm.
To complete the bound, we have to estimate two norms.

The first norm is the operator norm of $f(\hat{\boldsymbol{x}})$, which is diagonal in the position basis, so that $\norm{f(\hat{\boldsymbol{x}})}=\max_{\boldsymbol{r}}\left|f(\boldsymbol{x}_{\boldsymbol{r}})\right|$.
Shifting the cost function by a constant, $f\to f-c$, multiplies both $\ket{\psi(T)}$ and $\ket{\tilde\psi(T)}$ by global phases, which are absorbed by the minimization over $\theta$ in \cref{eq:def_trotter_error}, and does not change the commutator $\left[\hat T_0,f(\hat{\boldsymbol{x}})\right]$.
Therefore, $\eTrotter$ does not change either, and we can choose $c=\left(\max f+\min f\right)/2$, so that
\begin{equation}\label{eq:normaV}
  \norm{f(\hat{\boldsymbol x})}=\frac{\max f-\min f}{2}=\frac{1}{2}\osc \ ,
\end{equation}
where $\max f$ and $\min f$ are taken over the grid points and $\osc\coloneqq\max f-\min f$ is the oscillation of the cost function.
In general, this term depends only on the cost function, and not on the parameters $d$ and $q$.

It remains to bound the norm $\norm{\left[\hat T_0, f(\hat{\boldsymbol{x}})\right]e^{i\sigma\hat T_0}\ket{\psi(s)}}$, which depends on the class of the cost function.
We bound it for generic cost functions in \cref{app:trotter_generic}, for cost functions with a smooth periodic extension in \cref{app:trotter_smooth}, and for quantum functions in \cref{app:trotter_quantum}; the resulting error bounds are collected in \cref{app:trotter_final}.

\subsection{Bounds for generic cost functions}
\label{app:trotter_generic}

We first bound the commutator acting on a generic vector (\cref{lem:fourier_bound}) and the kinetic energy along the exact trajectory (\cref{lem:kinetic_moments}), and we then combine the two results.

We expand the cost function in a discrete Fourier series.
Let $N\coloneqq2^q$ be the number of grid points along each dimension, and let us label the grid points by the multi-index $\boldsymbol{r}=\left(r_1,\ldots,r_d\right)\in\left\{0,\ldots,N-1\right\}^d$; then
\begin{equation}\label{eq:f_DFT}
  f(\boldsymbol{x}_{\boldsymbol{r}})=\sum_{\boldsymbol{k}\in K_N^d}\tilde f_{\boldsymbol{k}}\,e^{i\frac{2\pi}{N}\boldsymbol{k}\cdot\boldsymbol{r}} \ ,
  \qquad
  \tilde f_{\boldsymbol{k}}=\frac{1}{N^d}\sum_{\boldsymbol{r}}f(\boldsymbol{x}_{\boldsymbol{r}})\,e^{-i\frac{2\pi}{N}\boldsymbol{k}\cdot\boldsymbol{r}} \ ,
\end{equation}
where $K_N^d$ is the $d$-fold Cartesian product of
\begin{equation}\label{eq:def_K_N}
  K_N\coloneqq\left\{-\frac{N}{2},-\frac{N}{2}+1,\ldots,\frac{N}{2}-1\right\} \ .
\end{equation}
We denote by $\left\{\ket{\boldsymbol{m}}\,\middle|\,\boldsymbol{m}\in K_N^d\right\}$ the eigenvectors of $\hat{\boldsymbol{p}}$, namely
\begin{equation}\label{eq:momentum_eigenvectors}
  \hat p_{\ell}\ket{\boldsymbol{m}}=\frac{2\pi m_{\ell}}{L}\ket{\boldsymbol{m}} \ ,
  \qquad
  \hat T_0\ket{\boldsymbol{m}}=\frac{2\pi^2}{L^2}\left|\boldsymbol{m}\right|^2\ket{\boldsymbol{m}} \ ,
  \qquad
  \braket{\boldsymbol{x}_{\boldsymbol{r}}}{\boldsymbol{m}}=\frac{e^{i\frac{2\pi}{N}\boldsymbol{m}\cdot\boldsymbol{r}}}{N^{d/2}} \ ,
\end{equation}
and we define the cyclic shift in momentum
\begin{equation}\label{eq:def_shift_momentum}
  \hat S_{\boldsymbol{k}}\coloneqq e^{i\frac{2\pi}{N}\boldsymbol{k}\cdot\hat{\boldsymbol{r}}} \ ,
  \qquad
  \hat{\boldsymbol{r}}\coloneqq\sum_{\boldsymbol{r}}\boldsymbol{r}\,\ketbra{\boldsymbol{x}_{\boldsymbol{r}}}{\boldsymbol{x}_{\boldsymbol{r}}} \ ,
\end{equation}
such that $\hat S_{\boldsymbol{k}}\ket{\boldsymbol{m}}=\ket{\boldsymbol{m}\oplus\boldsymbol{k}}$, where $\oplus$ denotes the component-by-component sum modulo $N$, with values in $K_N$.
With these definitions, \cref{eq:f_DFT} reads $f(\hat{\boldsymbol{x}})=\sum_{\boldsymbol{k}}\tilde f_{\boldsymbol{k}}\hat S_{\boldsymbol{k}}$.

\begin{lemma}\label{lem:fourier_bound}
  For every function $g$ on the grid and every vector $\ket{\chi}$
  \begin{equation}\label{eq:fourier_bound}
    \norm{\left[\hat T_0, g(\hat{\boldsymbol{x}})\right]\ket{\chi}}
    \leq
    \sum_{\boldsymbol{k}\in K_N^d}\left|\tilde g_{\boldsymbol{k}}\right|\left[\frac{2\pi}{L}\left|\boldsymbol{k}\right|\norm{\hat{\boldsymbol{p}}\ket{\chi}}+\frac{2\pi^2}{L^2}\left|\boldsymbol{k}\right|^2\norm{\ket{\chi}}\right] \ .
  \end{equation}
\end{lemma}

\begin{proof}
Just like the cost function $f$, we can write the Fourier expansion of the function $g$ as
\begin{equation}
  g(\hat{\boldsymbol{x}})=\sum_{\boldsymbol{k}}\tilde g_{\boldsymbol{k}}\hat S_{\boldsymbol{k}} \ .
\end{equation}
Writing $\ket{\chi}=\sum_{\boldsymbol{m}}\chi_{\boldsymbol{m}}\ket{\boldsymbol{m}}$, we have
\begin{align}
  \norm{\left[\hat T_0, g(\hat{\boldsymbol{x}})\right]\ket{\chi}}
  &=
  \norm{\sum_{\boldsymbol{k}\in K_N^d}\tilde g_{\boldsymbol{k}}\left[\hat T_0,\hat S_{\boldsymbol{k}}\right]\sum_{\boldsymbol{m}\in K_N^d}\chi_{\boldsymbol{m}}\ket{\boldsymbol{m}}}
  \nonumber\\ &=
  \norm{\sum_{\boldsymbol{k},\boldsymbol{m}\in K_N^d}\tilde g_{\boldsymbol{k}}\,\chi_{\boldsymbol{m}}\left(\hat T_0\ket{\boldsymbol{m}\oplus\boldsymbol{k}}-\hat S_{\boldsymbol{k}}\hat T_0\ket{\boldsymbol{m}}\right)}
  \nonumber\\ &=
  \norm{\sum_{\boldsymbol{k},\boldsymbol{m}\in K_N^d}\tilde g_{\boldsymbol{k}}\,\chi_{\boldsymbol{m}}\,\frac{2\pi^2}{L^2}\left(\left|\boldsymbol{m}\oplus\boldsymbol{k}\right|^2-\left|\boldsymbol{m}\right|^2\right)\ket{\boldsymbol{m}\oplus\boldsymbol{k}}}
  \nonumber\\ &\leq
  \frac{2\pi^2}{L^2}\sum_{\boldsymbol{k}}\left|\tilde g_{\boldsymbol{k}}\right|\,\norm{\sum_{\boldsymbol{m}}\chi_{\boldsymbol{m}}\left(\left|\boldsymbol{m}\oplus\boldsymbol{k}\right|^2-\left|\boldsymbol{m}\right|^2\right)\ket{\boldsymbol{m}\oplus\boldsymbol{k}}}
  \nonumber\\ &=
  \frac{2\pi^2}{L^2}\sum_{\boldsymbol{k}}\left|\tilde g_{\boldsymbol{k}}\right|\sqrt{\sum_{\boldsymbol{m}}\left|\chi_{\boldsymbol{m}}\right|^2\left(\left|\boldsymbol{m}\oplus\boldsymbol{k}\right|^2-\left|\boldsymbol{m}\right|^2\right)^2} \ ,\label{eq:commutator_fourier}
\end{align}
where we used the triangle inequality on the sum over $\boldsymbol{k}$ and, in the last step, the fact that for fixed $\boldsymbol{k}$ the map $\boldsymbol{m}\mapsto\boldsymbol{m}\oplus\boldsymbol{k}$ is a bijection of $K_N^d$, so that the vectors $\ket{\boldsymbol{m}\oplus\boldsymbol{k}}$ are orthonormal.
One easily verifies that
\begin{equation}\label{eq:shift_momentum_bound}
  \left|\left|\boldsymbol{m}\oplus\boldsymbol{k}\right|^2-\left|\boldsymbol{m}\right|^2\right|\leq\sum_{\ell=1}^d\left(2\left|k_{\ell}\right|\left|m_{\ell}\right|+\left|k_{\ell}\right|^2\right) \ .
\end{equation}
We now observe that, given a vector $\boldsymbol{v}$ with components $v_{\boldsymbol{m}}$, $\boldsymbol{m}\in K_N^d$, the quantity $\sqrt{\sum_{\boldsymbol{m}}\left|\chi_{\boldsymbol{m}}\right|^2v_{\boldsymbol{m}}^2}$ is a (weighted $\ell^2$) norm, and therefore it satisfies the triangle inequality.
Inserting \cref{eq:shift_momentum_bound} into \cref{eq:commutator_fourier}, we thus obtain
\begin{align}
  \norm{\left[\hat T_0, g(\hat{\boldsymbol{x}})\right]\ket{\chi}}
  &\leq
  \frac{2\pi^2}{L^2}\sum_{\boldsymbol{k}}\left|\tilde g_{\boldsymbol{k}}\right|\sqrt{\sum_{\boldsymbol{m}}\left|\chi_{\boldsymbol{m}}\right|^2\left[\sum_{\ell=1}^d\left(\left|k_{\ell}\right|^2+2\left|m_{\ell}k_{\ell}\right|\right)\right]^2}
  \nonumber\\ &\leq
  \frac{2\pi^2}{L^2}\sum_{\boldsymbol{k}}\left|\tilde g_{\boldsymbol{k}}\right|\sum_{\ell=1}^d\left[\sqrt{\sum_{\boldsymbol{m}}\left|\chi_{\boldsymbol{m}}\right|^2\left(\left|k_{\ell}\right|^2\right)^2}+\sqrt{\sum_{\boldsymbol{m}}\left|\chi_{\boldsymbol{m}}\right|^2\left(2\left|m_{\ell}k_{\ell}\right|\right)^2}\right]
  \nonumber\\ &=
  \frac{2\pi^2}{L^2}\sum_{\boldsymbol{k}}\left|\tilde g_{\boldsymbol{k}}\right|\sum_{\ell=1}^d\left[\left|k_{\ell}\right|^2\norm{\ket{\chi}}+2\left|k_{\ell}\right|\sqrt{\sum_{\boldsymbol{m}}\left|\chi_{\boldsymbol{m}}\right|^2\left|m_{\ell}\right|^2}\right]
  \nonumber\\ &=
  \frac{2\pi^2}{L^2}\sum_{\boldsymbol{k}}\left|\tilde g_{\boldsymbol{k}}\right|\sum_{\ell=1}^d\left(\left|k_{\ell}\right|^2\norm{\ket{\chi}}+2\left|k_{\ell}\right|\norm{\hat p_{\ell}\ket{\chi}}\frac{L}{2\pi}\right)
  \nonumber\\ &=
  \sum_{\boldsymbol{k}}\left|\tilde g_{\boldsymbol{k}}\right|\sum_{\ell=1}^d\left(\frac{2\pi}{L}\left|k_{\ell}\right|\norm{\hat p_{\ell}\ket{\chi}}+\frac{2\pi^2}{L^2}\left|k_{\ell}\right|^2\norm{\ket{\chi}}\right) \ ,
\end{align}
where we used $\sqrt{\sum_{\boldsymbol{m}}\left|\chi_{\boldsymbol{m}}\right|^2}=\norm{\ket{\chi}}$ and, from \cref{eq:momentum_eigenvectors},
\begin{equation}
  \norm{\hat p_{\ell}\ket{\chi}}^2=\left(\frac{2\pi}{L}\right)^2\sum_{\boldsymbol{m}}\left|\chi_{\boldsymbol{m}}\right|^2\left|m_{\ell}\right|^2 \ .
\end{equation}
We conclude by applying the Cauchy--Schwarz inequality to the $d$-component vectors with components $\left|k_{\ell}\right|$ and $\norm{\hat p_{\ell}\ket{\chi}}$, which gives $\sum_{\ell}\left|k_{\ell}\right|\norm{\hat p_{\ell}\ket{\chi}}\leq\left|\boldsymbol{k}\right|\norm{\hat{\boldsymbol{p}}\ket{\chi}}$.
\end{proof}

To control the norms of $\hat{\boldsymbol{p}}$ appearing in \cref{eq:fourier_bound}, we bound the kinetic energy along the exact trajectory.
To this end, we define
\begin{equation}
  \hat G\coloneqq f(\hat{\boldsymbol{x}})-\min f \ ,
  \qquad
  \hat{\mathcal{E}}(t)\coloneqq a(t)\,\hat T_0+b(t)\,\hat G=\hat H(t)-b(t)\min f \ ,
\end{equation}
so that $0\leq\hat G\leq\osc$ as operators.

\begin{lemma}\label{lem:kinetic_moments}
  Under hypotheses (H1) and (H3), along the exact trajectory \cref{eq:exact_trajectory}, $\ket{\psi(t)}=\hat U(t,0)\ket{\psi(0)}$,
  \begin{equation}\label{eq:energy_bound}
    \norm{\hat{\mathcal{E}}(t)\ket{\psi(t)}}\leq a(t)\norm{\hat T_0\ket{\psi(0)}}+b(t)\,\osc \ .
  \end{equation}
  In particular, under hypothesis (H4),
  \begin{equation}\label{eq:kinetic_energy_bound}
    \matrixel{\psi(t)}{\hat T_0}{\psi(t)}\leq\frac{b(t)}{a(t)}\,\osc \ ,
    \qquad
    \norm{\hat T_0\ket{\psi(t)}}\leq2\,\frac{b(t)}{a(t)}\,\osc \ .
  \end{equation}
\end{lemma}

\begin{proof}
Since $\hat{\mathcal{E}}(t)$ differs from $\hat H(t)$ by a multiple of the identity, it commutes with $\hat H(t)$, and the vector $\hat{\mathcal{E}}(t)\ket{\psi(t)}$ evolves as
\begin{equation}
  \frac{d}{dt}\left(\hat{\mathcal{E}}\ket{\psi}\right)
  =-i\hat H\left(\hat{\mathcal{E}}\ket{\psi}\right)+\dot{\hat{\mathcal{E}}}\ket{\psi} \ ,
  \qquad
  \dot{\hat{\mathcal{E}}}
  =\dot a\,\hat T_0+\dot b\,\hat G
  =\frac{\dot a}{a}\,\hat{\mathcal{E}}+a\,\frac{d}{dt}\left(\frac{b}{a}\right)\hat G \ .
\end{equation}
The rescaled vector $\ket{w(t)}\coloneqq\hat{\mathcal{E}}(t)\ket{\psi(t)}/a(t)$ thus satisfies
\begin{equation}
  \frac{d}{dt}\ket{w}=-i\hat H\ket{w}+\frac{d}{dt}\left(\frac{b}{a}\right)\hat G\ket{\psi} \ ,
\end{equation}
whose solution is
\begin{equation}
  \ket{w(t)}=\hat U(t,0)\ket{w(0)}+\int_0^t\hat U(t,s)\,\frac{d}{ds}\left(\frac{b}{a}\right)\hat G\ket{\psi(s)}\,ds \ .
\end{equation}
Using the unitarity of $\hat U$, $\frac{d}{ds}\left(\frac{b}{a}\right)\geq0$ by hypothesis (H3) and $\norm{\hat G\ket{\psi(s)}}\leq\osc$, we get
\begin{equation}
  \norm{\ket{w(t)}}
  \leq\norm{\ket{w(0)}}+\left[\frac{b(t)}{a(t)}-\frac{b(0)}{a(0)}\right]\osc
  \leq\norm{\hat T_0\ket{\psi(0)}}+\frac{b(t)}{a(t)}\,\osc \ ,
\end{equation}
where in the last step we used $\ket{w(0)}=\hat T_0\ket{\psi(0)}+\frac{b(0)}{a(0)}\hat G\ket{\psi(0)}$.
Multiplying by $a(t)$ we obtain \cref{eq:energy_bound}.
If $\hat T_0\ket{\psi(0)}=0$, as guaranteed by hypothesis (H4), \cref{eq:kinetic_energy_bound} follows from $\hat G\geq0$ and $\norm{\hat G\ket{\psi}}\leq\osc$:
\begin{gather*}
  \matrixel{\psi}{\hat T_0}{\psi}
  =\frac{\matrixel{\psi}{\hat{\mathcal{E}}}{\psi}-b\matrixel{\psi}{\hat G}{\psi}}{a}
  \leq\frac{\norm{\hat{\mathcal{E}}\ket{\psi}}}{a}
  \leq\frac{b}{a}\,\osc \ ,
  \\
  \norm{\hat T_0\ket{\psi}}
  =\frac{\norm{\hat{\mathcal{E}}\ket{\psi}-b\,\hat G\ket{\psi}}}{a}
  \leq\frac{\norm{\hat{\mathcal{E}}\ket{\psi}}+b\,\osc}{a}
  \leq2\,\frac{b}{a}\,\osc \ . \qedhere
\end{gather*}
\end{proof}

We now consider the vector $\ket{\phi}\coloneqq e^{i\sigma\hat T_0}\ket{\psi(s)}$ appearing in \cref{eq:trotter_error_two_norms}.
Since $e^{i\sigma\hat T_0}$ is diagonal in the momentum basis, $\ket{\phi}$ has the same momentum distribution as $\ket{\psi(s)}$, and the first bound in \cref{eq:kinetic_energy_bound} gives, independently of $\sigma$,
\begin{equation}\label{eq:p_bound}
  \norm{\hat{\boldsymbol{p}}\ket{\phi}}=\sqrt{2\matrixel{\psi(s)}{\hat T_0}{\psi(s)}}\leq\sqrt{2\,\osc\,\frac{b(s)}{a(s)}} \ .
\end{equation}
Substituting \cref{eq:p_bound} into \cref{eq:fourier_bound} with $g=f$ and $\ket{\chi}=\ket{\phi}$, we get an error bound whose dependence on $d$ and $q$ is contained in
\begin{equation}\label{eq:def_Gamma}
  \Gamma_1^{(q)}\coloneqq\sum_{\boldsymbol{k}\in K_N^d}\left|\tilde f_{\boldsymbol{k}}\right|\left|\boldsymbol{k}\right| \ ,
  \qquad
  \Gamma_2^{(q)}\coloneqq\sum_{\boldsymbol{k}\in K_N^d}\left|\tilde f_{\boldsymbol{k}}\right|\left|\boldsymbol{k}\right|^2 \ .
\end{equation}
This gives the bound for generic cost functions, which holds for every cost function and every $q$:
\begin{equation}\label{eq:commutator_bound_final}
  \sup_{\sigma\in\mathbb{R}}\norm{\left[\hat T_0, f(\hat{\boldsymbol{x}})\right]e^{i\sigma\hat T_0}\ket{\psi(s)}}
  \leq
  \frac{2\pi}{L}\sqrt{2\,\osc\,\frac{b(s)}{a(s)}}\,\sum_{\boldsymbol{k}}\left|\tilde f_{\boldsymbol{k}}\right|\left|\boldsymbol{k}\right|+\frac{2\pi^2}{L^2}\sum_{\boldsymbol{k}}\left|\tilde f_{\boldsymbol{k}}\right|\left|\boldsymbol{k}\right|^2 \ .
\end{equation}

\subsection{Bounds for smooth cost functions}
\label{app:trotter_smooth}

Since the sums \cref{eq:def_Gamma} can be difficult to evaluate, we consider a continuum-space version of the error bound.
This is useful since, in the continuum, the commutator is given by \cref{eq:commutator_continuum},
\[
\left[\frac{\hat{\boldsymbol p}^2}{2},f(\hat{\boldsymbol x})\right]=-i\,\nabla f(\hat{\boldsymbol x})\cdot\hat{\boldsymbol p}-\frac{1}{2}\,\Delta f(\hat{\boldsymbol x})
\]
which, for well-behaved functions, can be bounded in terms of $\max\left|\nabla f\right|$ and $\max\left|\Delta f\right|$, without summing over the Fourier modes.
On the grid, \cref{eq:commutator_continuum} holds exactly as long as the momentum of the state plus the frequency of the cost function does not leave $K_N^d$, i.e.~as long as $\boldsymbol{m}\oplus\boldsymbol{k}=\boldsymbol{m}+\boldsymbol{k}$.
We exploit this by splitting the low- and high-frequency components of $f$ with a cutoff $K$.
Let the grid points be $\boldsymbol{x}_{\boldsymbol{r}}=\boldsymbol{x}_0+\boldsymbol{r}L/N$, let $\norm{\boldsymbol{v}}_\infty\coloneqq\max_{\ell}\left|v_{\ell}\right|$ denote the maximum norm of a vector $\boldsymbol{v}$ and, for an integer $0\leq K<N/2$, let
\begin{equation}\label{eq:def_fK}
  f_K(\boldsymbol{x})\coloneqq\sum_{\norm{\boldsymbol{k}}_\infty\leq K}\tilde f_{\boldsymbol{k}}\,e^{i\frac{2\pi}{L}\boldsymbol{k}\cdot\left(\boldsymbol{x}-\boldsymbol{x}_0\right)}
\end{equation}
be the trigonometric interpolant of $f$ restricted to the frequencies $\norm{\boldsymbol{k}}_\infty\leq K$.
Correspondingly, we denote by $\hat P_K$ the projector onto the low-momentum subspace, i.e.~onto the momentum eigenvectors $\ket{\boldsymbol{m}}$ with $\norm{\boldsymbol{m}}_\infty\leq N/2-K-1$.
We also define the complementary projector $\hat Q_K\coloneqq\hat{\mathbb{I}}-\hat P_K$.
We split $\Gamma_i^{(q)}=\Gamma_i^{(q),\leq K}+\Gamma_i^{(q),>K}$, where $\Gamma_i^{(q),\leq K}$ and $\Gamma_i^{(q),>K}$ contain the terms of \cref{eq:def_Gamma} with $\norm{\boldsymbol{k}}_\infty\leq K$ and $\norm{\boldsymbol{k}}_\infty>K$, respectively, and we denote by $\max\left|h\right|$ the maximum of $\left|h(\boldsymbol{x})\right|$ over the domain.

\begin{proposition}\label{prop:cutoff}
  For every cost function $f$, every $q$, every integer $0\leq K<N/2$ and every normalized $\ket{\phi}$,
  \begin{align}\label{eq:prop_cutoff}
    \norm{\left[\hat T_0,f(\hat{\boldsymbol{x}})\right]\ket{\phi}}
    &\leq
    \max\left|\nabla f_K\right|\norm{\hat{\boldsymbol{p}}\ket{\phi}}+\frac{1}{2}\max\left|\Delta f_K\right|
    +\frac{2\pi}{L}\Gamma_1^{(q),>K}\norm{\hat{\boldsymbol{p}}\ket{\phi}}+\frac{2\pi^2}{L^2}\Gamma_2^{(q),>K}
    \nonumber\\ &\qquad
    +\frac{2\pi}{L}\Gamma_1^{(q),\leq K}\norm{\hat{\boldsymbol{p}}\hat Q_K\ket{\phi}}+\frac{2\pi^2}{L^2}\Gamma_2^{(q),\leq K}\norm{\hat Q_K\ket{\phi}} \ .
  \end{align}
\end{proposition}

\begin{proof}
We split $f=f_K+\left(f-f_K\right)$ on the grid and $\ket{\phi}=\hat P_K\ket{\phi}+\hat Q_K\ket{\phi}$, so that
\begin{align}\label{eq:split}
  \left[\hat T_0,f(\hat{\boldsymbol{x}})\right]\ket{\phi}
  &=\left[\hat T_0,f_K(\hat{\boldsymbol{x}})\right]\hat P_K\ket{\phi}
  +\left[\hat T_0,f_K(\hat{\boldsymbol{x}})\right]\hat Q_K\ket{\phi}
  \nonumber\\ &\quad
  +\left[\hat T_0,f(\hat{\boldsymbol{x}})-f_K(\hat{\boldsymbol{x}})\right]\ket{\phi} \ .
\end{align}
If $\norm{\boldsymbol{k}}_\infty\leq K$ and $\norm{\boldsymbol{m}}_\infty\leq N/2-K-1$, then $\hat S_{\boldsymbol{k}}\ket{\boldsymbol{m}}=\ket{\boldsymbol{m}+\boldsymbol{k}}$ and
\begin{equation}
  \left[\hat T_0,\hat S_{\boldsymbol{k}}\right]\ket{\boldsymbol{m}}
  =\frac{2\pi^2}{L^2}\left(\left|\boldsymbol{m}+\boldsymbol{k}\right|^2-\left|\boldsymbol{m}\right|^2\right)\ket{\boldsymbol{m}+\boldsymbol{k}}
  =\frac{2\pi^2}{L^2}\left(2\,\boldsymbol{m}\cdot\boldsymbol{k}+\left|\boldsymbol{k}\right|^2\right)\ket{\boldsymbol{m}+\boldsymbol{k}} \ .
\end{equation}
The same matrix elements are obtained from the right-hand side of \cref{eq:commutator_continuum} with $f_K$ in place of $f$, using $\partial_{\ell}f_K(\hat{\boldsymbol{x}})=\sum_{\boldsymbol{k}}\frac{2\pi i k_{\ell}}{L}\tilde f_{\boldsymbol{k}}\hat S_{\boldsymbol{k}}$, $\Delta f_K(\hat{\boldsymbol{x}})=-\sum_{\boldsymbol{k}}\frac{4\pi^2\left|\boldsymbol{k}\right|^2}{L^2}\tilde f_{\boldsymbol{k}}\hat S_{\boldsymbol{k}}$ and \cref{eq:momentum_eigenvectors}.
Hence, on the support of $\hat P_K$, \cref{eq:commutator_continuum} holds for $f_K$:
\begin{equation}
  \left[\hat T_0,f_K(\hat{\boldsymbol{x}})\right]\hat P_K\ket{\phi}
  =-i\sum_{\ell=1}^d\partial_{\ell}f_K(\hat{\boldsymbol{x}})\,\hat p_{\ell}\hat P_K\ket{\phi}-\frac{1}{2}\Delta f_K(\hat{\boldsymbol{x}})\hat P_K\ket{\phi} \ .
\end{equation}
Since $\partial_{\ell}f_K(\hat{\boldsymbol{x}})$ and $\Delta f_K(\hat{\boldsymbol{x}})$ are diagonal in the position basis, applying the Cauchy--Schwarz inequality to the sum over $\ell$ at each grid point we get
\begin{align}\label{eq:leibniz_grid}
  \norm{\left[\hat T_0,f_K(\hat{\boldsymbol{x}})\right]\hat P_K\ket{\phi}}
  &=\norm{-i\sum_{\ell=1}^d\partial_{\ell}f_K(\hat{\boldsymbol{x}})\,\hat p_{\ell}\hat P_K\ket{\phi}-\frac{1}{2}\Delta f_K(\hat{\boldsymbol{x}})\hat P_K\ket{\phi}}
  \nonumber\\ &\leq
  \sqrt{\sum_{\boldsymbol{r}}\left|\sum_{\ell=1}^d\partial_{\ell}f_K(\boldsymbol{x}_{\boldsymbol{r}})\bra{\boldsymbol{x}_{\boldsymbol{r}}}\hat p_{\ell}\hat P_K\ket{\phi}\right|^2}
  \nonumber\\ &\qquad
  +\frac{1}{2}\sqrt{\sum_{\boldsymbol{r}}\left|\Delta f_K(\boldsymbol{x}_{\boldsymbol{r}})\right|^2\left|\bra{\boldsymbol{x}_{\boldsymbol{r}}}\hat P_K\ket{\phi}\right|^2}
  \nonumber\\ &\leq
  \max\left|\nabla f_K\right|\norm{\hat{\boldsymbol{p}}\hat P_K\ket{\phi}}+\frac{1}{2}\max\left|\Delta f_K\right|\norm{\hat P_K\ket{\phi}}
  \nonumber\\ &\leq
  \max\left|\nabla f_K\right|\norm{\hat{\boldsymbol{p}}\ket{\phi}}+\frac{1}{2}\max\left|\Delta f_K\right| \ .
\end{align}
The other two terms in \cref{eq:split} are bounded by \cref{lem:fourier_bound}, with $g=f-f_K$ and $\ket{\chi}=\ket{\phi}$ for the third one, and with $g=f_K$ and $\ket{\chi}=\hat Q_K\ket{\phi}$ for the second one.
We get \cref{eq:prop_cutoff} by applying the triangle inequality to the norm of \cref{eq:split}.
\end{proof}

For $K=0$, $f_0=\tilde f_{\boldsymbol{0}}$ is a constant, so that $\nabla f_0=\Delta f_0=0$ and $\Gamma_i^{(q),\leq0}=0$, while $\Gamma_i^{(q),>0}=\Gamma_i^{(q)}$.
\Cref{eq:prop_cutoff} then reduces to \cref{lem:fourier_bound} with $g=f$, and hence to the generic bound \cref{eq:commutator_bound_final}.

To specialize \cref{prop:cutoff} to cost functions with a smooth periodic extension, we compare $f$ with the periodic extension $F$ of $f$ from the box $\boldsymbol{x}_0+[0,L)^d$, whose Fourier series is
\begin{equation}\label{eq:fourier_continuum}
  F(\boldsymbol{x})=\sum_{\boldsymbol{\kappa}\in\mathbb{Z}^d}\tilde F(\boldsymbol{\kappa})\,e^{i\frac{2\pi}{L}\boldsymbol{\kappa}\cdot\left(\boldsymbol{x}-\boldsymbol{x}_0\right)} \ .
\end{equation}
In analogy with \cref{eq:def_Gamma}, we define
\begin{equation}\label{eq:def_Gamma_inf}
  \Gamma_i^{(\infty)}\coloneqq\sum_{\boldsymbol{\kappa}\in\mathbb{Z}^d}\left|\tilde F(\boldsymbol{\kappa})\right|\left|\boldsymbol{\kappa}\right|^i \ ,
  \qquad
  i=1,2 \ ,
\end{equation}
and we denote by $\Gamma_i^{(\infty),>K}$ the same sums restricted to $\norm{\boldsymbol{\kappa}}_\infty>K$.

We will now bound the high-momentum components in \cref{eq:prop_cutoff}.
We assume that
\begin{equation}\label{eq:smooth_hypothesis}
  \Gamma_2^{(\infty)}<\infty \ ,
\end{equation}
which is satisfied by every $C^s$ periodic function $F$, with $s>\frac{d}{2}+2$.
Under this assumption, the series \cref{eq:fourier_continuum} and its first and second derivatives converge absolutely and uniformly, and $\Gamma_1^{(\infty)}\leq\Gamma_2^{(\infty)}<\infty$, since $\left|\boldsymbol{\kappa}\right|\leq\left|\boldsymbol{\kappa}\right|^2$ for $\boldsymbol{\kappa}\in\mathbb{Z}^d$.

\emph{Aliasing.}
On the grid we have $f(\boldsymbol{x}_{\boldsymbol{r}})=F(\boldsymbol{x}_{\boldsymbol{r}})$, and $e^{i\frac{2\pi}{L}\boldsymbol{\kappa}\cdot\left(\boldsymbol{x}_{\boldsymbol{r}}-\boldsymbol{x}_0\right)}=e^{i\frac{2\pi}{N}\boldsymbol{\kappa}\cdot\boldsymbol{r}}$ depends only on the class of $\boldsymbol{\kappa}$ modulo $N$.
Inserting \cref{eq:fourier_continuum} into \cref{eq:f_DFT}, we therefore get the aliasing formula
\begin{equation}\label{eq:aliasing}
  \tilde f_{\boldsymbol{k}}=\sum_{\boldsymbol{n}\in\mathbb{Z}^d}\tilde F(\boldsymbol{k}+N\boldsymbol{n}) \ ,
  \qquad
  \boldsymbol{k}\in K_N^d \ .
\end{equation}

\begin{lemma}\label{lem:aliasing_sums}
  Under the assumption \cref{eq:smooth_hypothesis}, for $i=1,2$ and every integer $0\leq K<N/2$,
  \begin{equation}\label{eq:aliasing_Gamma}
    \Gamma_i^{(q)}\leq\Gamma_i^{(\infty)} \ ,
    \qquad
    \Gamma_i^{(q),>K}\leq\Gamma_i^{(\infty),>K} \ .
  \end{equation}
\end{lemma}

\begin{proof}
Since $k_{\ell}\in K_N$ is the representative of minimum modulus of its class modulo $N$, we have $\left|k_{\ell}\right|\leq\left|k_{\ell}+Nn_{\ell}\right|$ for every $n_{\ell}\in\mathbb{Z}$, and hence $\left|\boldsymbol{k}\right|\leq\left|\boldsymbol{k}+N\boldsymbol{n}\right|$.
By \cref{eq:aliasing}, for every subset $S\subseteq K_N^d$ we thus have
\begin{align}
  \sum_{\boldsymbol{k}\in S}\left|\tilde f_{\boldsymbol{k}}\right|\left|\boldsymbol{k}\right|^i
  &=
  \sum_{\boldsymbol{k}\in S}\left|\sum_{\boldsymbol{n}\in\mathbb{Z}^d}\tilde F(\boldsymbol{k}+N\boldsymbol{n})\right|\left|\boldsymbol{k}\right|^i
  \nonumber\\ &\leq
  \sum_{\boldsymbol{k}\in S}\sum_{\boldsymbol{n}\in\mathbb{Z}^d}\left|\tilde F(\boldsymbol{k}+N\boldsymbol{n})\right|\left|\boldsymbol{k}\right|^i
  \nonumber\\ &\leq
  \sum_{\boldsymbol{k}\in S}\sum_{\boldsymbol{n}\in\mathbb{Z}^d}\left|\tilde F(\boldsymbol{k}+N\boldsymbol{n})\right|\left|\boldsymbol{k}+N\boldsymbol{n}\right|^i \ .\label{eq:aliasing_sums}
\end{align}
The map $(\boldsymbol{k},\boldsymbol{n})\mapsto\boldsymbol{\kappa}=\boldsymbol{k}+N\boldsymbol{n}$ is a bijection of $K_N^d\times\mathbb{Z}^d$ onto $\mathbb{Z}^d$, and $\norm{\boldsymbol{k}+N\boldsymbol{n}}_\infty\geq\norm{\boldsymbol{k}}_\infty$.
Therefore, for $S=K_N^d$ the left-hand side of \cref{eq:aliasing_sums} is $\Gamma_i^{(q)}$ and the right-hand side is $\Gamma_i^{(\infty)}$.
For $S=\left\{\boldsymbol{k}\in K_N^d\,\middle|\,\norm{\boldsymbol{k}}_\infty>K\right\}$, instead, the left-hand side is $\Gamma_i^{(q),>K}$, and the right-hand side is a sum over a subset of $\left\{\boldsymbol{\kappa}\in\mathbb{Z}^d\,\middle|\,\norm{\boldsymbol{\kappa}}_\infty>K\right\}$, which is bounded by $\Gamma_i^{(\infty),>K}$.
This proves \cref{eq:aliasing_Gamma}.
\end{proof}

\noindent\Cref{lem:aliasing_sums} means that every discrete $\Gamma$-like sum is upper bounded by its continuum counterpart, which is a finite number independent of $q$.

\begin{lemma}\label{lem:derivatives_fK}
  Under the assumption \cref{eq:smooth_hypothesis}, for every integer $0\leq K<N/2$,
  \begin{equation}\label{eq:aliasing_derivatives}
    \left|\nabla f_K\right|\leq\left|\nabla F\right|+\frac{4\pi}{L}\,\Gamma_1^{(\infty),>K} \ ,
    \qquad
    \left|\Delta f_K\right|\leq\left|\Delta F\right|+\frac{8\pi^2}{L^2}\,\Gamma_2^{(\infty),>K} \ .
  \end{equation}
  The same bounds hold for the maxima over the domain, $\max\left|\nabla f_K\right|$ and $\max\left|\Delta f_K\right|$, which appear in \cref{eq:prop_cutoff}.
\end{lemma}

\begin{proof}
Let us consider the difference $f_K-F=\left(f_K-F_K\right)-\left(F-F_K\right)$, where, as for $f_K$ in \cref{eq:def_fK},
\begin{equation}\label{eq:def_FK}
  F_K(\boldsymbol{x})\coloneqq\sum_{\norm{\boldsymbol{\kappa}}_\infty\leq K}\tilde F(\boldsymbol{\kappa})\,e^{i\frac{2\pi}{L}\boldsymbol{\kappa}\cdot\left(\boldsymbol{x}-\boldsymbol{x}_0\right)}
\end{equation}
denotes the low-frequency version of $F$.
By \cref{eq:aliasing}, we have
\begin{align}
  f_K(\boldsymbol{x})-F_K(\boldsymbol{x})
  &=
  \sum_{\norm{\boldsymbol{k}}_\infty\leq K}\left(\tilde f_{\boldsymbol{k}}-\tilde F(\boldsymbol{k})\right)e^{i\frac{2\pi}{L}\boldsymbol{k}\cdot\left(\boldsymbol{x}-\boldsymbol{x}_0\right)}
  \nonumber\\ &=
  \sum_{\norm{\boldsymbol{k}}_\infty\leq K}\left(\sum_{\boldsymbol{n}\in\mathbb{Z}^d}\tilde F(\boldsymbol{k}+N\boldsymbol{n})-\tilde F(\boldsymbol{k})\right)e^{i\frac{2\pi}{L}\boldsymbol{k}\cdot\left(\boldsymbol{x}-\boldsymbol{x}_0\right)} \ ,\label{eq:fK_minus_FK}
\end{align}
while, by \cref{eq:fourier_continuum},
\begin{equation}\label{eq:F_minus_FK}
  F(\boldsymbol{x})-F_K(\boldsymbol{x})=\sum_{\norm{\boldsymbol{\kappa}}_\infty>K}\tilde F(\boldsymbol{\kappa})\,e^{i\frac{2\pi}{L}\boldsymbol{\kappa}\cdot\left(\boldsymbol{x}-\boldsymbol{x}_0\right)} \ .
\end{equation}
We now calculate
\begin{align}
  \left|\nabla\left(f_K-F_K\right)\right|
  &=
  \left|i\,\frac{2\pi}{L}\sum_{\norm{\boldsymbol{k}}_\infty\leq K}\boldsymbol{k}\left(\sum_{\boldsymbol{n}\in\mathbb{Z}^d}\tilde F(\boldsymbol{k}+N\boldsymbol{n})-\tilde F(\boldsymbol{k})\right)e^{i\frac{2\pi}{L}\boldsymbol{k}\cdot\left(\boldsymbol{x}-\boldsymbol{x}_0\right)}\right|
  \nonumber\\ &\leq
  \frac{2\pi}{L}\sum_{\norm{\boldsymbol{k}}_\infty\leq K}\ \sum_{\boldsymbol{n}\in\mathbb{Z}^d\setminus\left\{\boldsymbol{0}\right\}}\left|\boldsymbol{k}\right|\left|\tilde F(\boldsymbol{k}+N\boldsymbol{n})\right|
  \nonumber\\ &\leq
  \frac{2\pi}{L}\sum_{\norm{\boldsymbol{\kappa}}_\infty>K}\left|\boldsymbol{\kappa}\right|\left|\tilde F(\boldsymbol{\kappa})\right|
  \nonumber\\ &=
  \frac{2\pi}{L}\,\Gamma_1^{(\infty),>K} \ .\label{eq:grad_fK_minus_FK}
\end{align}
Similarly, differentiating \cref{eq:F_minus_FK} term by term, which is allowed by the uniform convergence, we get $\left|\nabla\left(F-F_K\right)\right|\leq\frac{2\pi}{L}\Gamma_1^{(\infty),>K}$.
Analogously, since
\begin{equation*}
  \Delta e^{i\frac{2\pi}{L}\boldsymbol{\kappa}\cdot\left(\boldsymbol{x}-\boldsymbol{x}_0\right)}=-\left(\frac{2\pi}{L}\right)^2\left|\boldsymbol{\kappa}\right|^2e^{i\frac{2\pi}{L}\boldsymbol{\kappa}\cdot\left(\boldsymbol{x}-\boldsymbol{x}_0\right)} \ ,
\end{equation*}
we get
\begin{equation}\label{eq:lapl_fK_minus_FK}
  \left|\Delta\left(f_K-F_K\right)\right|,\ \left|\Delta\left(F-F_K\right)\right|
  \leq
  \left(\frac{2\pi}{L}\right)^2\Gamma_2^{(\infty),>K} \ .
\end{equation}
Thus, by the triangle inequality, we have
\begin{align*}
  \left|\nabla f_K\right|
  &\leq
  \left|\nabla F\right|+\left|\nabla\left(f_K-F_K\right)\right|+\left|\nabla\left(F-F_K\right)\right|
  \leq
  \left|\nabla F\right|+2\,\frac{2\pi}{L}\,\Gamma_1^{(\infty),>K} \ ,
  \\
  \left|\Delta f_K\right|
  &\leq
  \left|\Delta F\right|+\left|\Delta\left(f_K-F_K\right)\right|+\left|\Delta\left(F-F_K\right)\right|
  \leq
  \left|\Delta F\right|+2\left(\frac{2\pi}{L}\right)^2\Gamma_2^{(\infty),>K} \ ,
\end{align*}
which is \cref{eq:aliasing_derivatives}.
Taking the maximum over $\boldsymbol{x}$ of both sides gives the bounds on the maxima.
\end{proof}

Lastly, we have to consider $\norm{\hat{\boldsymbol{p}}\hat Q_K\ket{\phi}}$ and $\norm{\hat Q_K\ket{\phi}}$ in \cref{eq:prop_cutoff}, with $\ket{\phi}=e^{i\sigma\hat T_0}\ket{\psi(s)}$.

\begin{lemma}\label{lem:momentum_tail}
  Under hypotheses (H1), (H3) and (H4), let $\ket{\psi(s)}$ be the exact trajectory of \cref{eq:exact_trajectory}, and let $\ket{\phi}=e^{i\sigma\hat T_0}\ket{\psi(s)}$ with $\sigma\in\mathbb{R}$.
  Then, for every integer $0\leq K<N/2$,
  \begin{equation}\label{eq:momentum_tail}
    \norm{\hat{\boldsymbol{p}}\hat Q_K\ket{\phi}}\leq\frac{4}{p_K}\,\frac{b(s)}{a(s)}\,\osc \ ,
    \qquad
    \norm{\hat Q_K\ket{\phi}}\leq\frac{4}{p_K^2}\,\frac{b(s)}{a(s)}\,\osc \ ,
  \end{equation}
  where
  \begin{equation}\label{eq:def_pK}
    p_K\coloneqq\frac{2\pi}{L}\left(\frac{N}{2}-K\right) \ .
  \end{equation}
\end{lemma}

\begin{proof}
We expand both states in the momentum basis \cref{eq:momentum_eigenvectors},
\begin{equation}
  \ket{\psi(s)}=\sum_{\boldsymbol{m}\in K_N^d}\psi_{\boldsymbol{m}}\ket{\boldsymbol{m}} \ ,
  \qquad
  \ket{\phi}=\sum_{\boldsymbol{m}\in K_N^d}\phi_{\boldsymbol{m}}\ket{\boldsymbol{m}} \ ,
  \qquad
  \phi_{\boldsymbol{m}}=e^{i\sigma\frac{2\pi^2}{L^2}\left|\boldsymbol{m}\right|^2}\psi_{\boldsymbol{m}} \ ,
\end{equation}
so that $\left|\phi_{\boldsymbol{m}}\right|=\left|\psi_{\boldsymbol{m}}\right|$, and we denote by $\boldsymbol{p}_{\boldsymbol{m}}\coloneqq\frac{2\pi}{L}\boldsymbol{m}$ the momentum of $\ket{\boldsymbol{m}}$.
For every vector $\ket{\chi}=\sum_{\boldsymbol{m}}\chi_{\boldsymbol{m}}\ket{\boldsymbol{m}}$ we then have
\begin{equation}\label{eq:momentum_moments}
  \norm{\hat{\boldsymbol{p}}\ket{\chi}}^2=\sum_{\ell=1}^d\norm{\hat p_{\ell}\ket{\chi}}^2=\sum_{\boldsymbol{m}}\left|\chi_{\boldsymbol{m}}\right|^2\left|\boldsymbol{p}_{\boldsymbol{m}}\right|^2 \ ,
  \qquad
  \norm{\hat T_0\ket{\chi}}^2=\frac{1}{4}\sum_{\boldsymbol{m}}\left|\chi_{\boldsymbol{m}}\right|^2\left|\boldsymbol{p}_{\boldsymbol{m}}\right|^4 \ .
\end{equation}
The projector $\hat Q_K$ keeps only the components with $\boldsymbol{m}\in M_K$, where
\begin{equation*}
  M_K\coloneqq\left\{\boldsymbol{m}\in K_N^d\,\middle|\,\norm{\boldsymbol{m}}_\infty\geq\frac{N}{2}-K\right\} \ .
\end{equation*}
For every $\boldsymbol{m}\in M_K$ there is an $\ell$ with $\left|m_{\ell}\right|\geq\frac{N}{2}-K$, so that
\begin{equation}
  \left|\boldsymbol{p}_{\boldsymbol{m}}\right|\geq\frac{2\pi}{L}\left|m_{\ell}\right|\geq p_K \ ,
  \qquad
  \boldsymbol{m}\in M_K \ .
\end{equation}
Hence, on $M_K$ we have $\left|\boldsymbol{p}_{\boldsymbol{m}}\right|^2\leq\left|\boldsymbol{p}_{\boldsymbol{m}}\right|^4/p_K^2$ and $1\leq\left|\boldsymbol{p}_{\boldsymbol{m}}\right|^4/p_K^4$, and, by \cref{eq:momentum_moments},
\begin{align*}
  \norm{\hat{\boldsymbol{p}}\hat Q_K\ket{\phi}}^2
  &=
  \sum_{\boldsymbol{m}\in M_K}\left|\phi_{\boldsymbol{m}}\right|^2\left|\boldsymbol{p}_{\boldsymbol{m}}\right|^2
  \leq
  \frac{1}{p_K^2}\sum_{\boldsymbol{m}\in K_N^d}\left|\psi_{\boldsymbol{m}}\right|^2\left|\boldsymbol{p}_{\boldsymbol{m}}\right|^4
  =
  \frac{4}{p_K^2}\norm{\hat T_0\ket{\psi(s)}}^2 \ ,
  \\
  \norm{\hat Q_K\ket{\phi}}^2
  &=
  \sum_{\boldsymbol{m}\in M_K}\left|\phi_{\boldsymbol{m}}\right|^2
  \leq
  \frac{1}{p_K^4}\sum_{\boldsymbol{m}\in K_N^d}\left|\psi_{\boldsymbol{m}}\right|^2\left|\boldsymbol{p}_{\boldsymbol{m}}\right|^4
  =
  \frac{4}{p_K^4}\norm{\hat T_0\ket{\psi(s)}}^2 \ .
\end{align*}
Taking the square roots and using the second bound in \cref{eq:kinetic_energy_bound}, $\norm{\hat T_0\ket{\psi(s)}}\leq2\,\frac{b(s)}{a(s)}\,\osc$, we get \cref{eq:momentum_tail}.
\end{proof}

We can now combine \cref{prop:cutoff} with \cref{lem:aliasing_sums,lem:derivatives_fK,lem:momentum_tail}.

\begin{proposition}\label{prop:smooth}
  Assume hypotheses (H1), (H3), (H4) and \cref{eq:smooth_hypothesis}.
  Then, for every $s$ and every integer $0\leq K<N/2$,
  \begin{equation}\label{eq:commutator_bound_smooth}
    \sup_{\sigma\in\mathbb{R}}\norm{\left[\hat T_0, f(\hat{\boldsymbol{x}})\right]e^{i\sigma\hat T_0}\ket{\psi(s)}}
    \leq
    \max\left|\nabla F\right|\sqrt{2\,\osc\,\frac{b(s)}{a(s)}}+\frac{1}{2}\max\left|\Delta F\right|+\delta_q(s) \ ,
  \end{equation}
  with
  \begin{align}\label{eq:delta_q}
    \delta_q(s)
    &\coloneqq
    3\left[\frac{2\pi}{L}\Gamma_1^{(\infty),>K}\sqrt{2\,\osc\,\frac{b(s)}{a(s)}}+\frac{2\pi^2}{L^2}\Gamma_2^{(\infty),>K}\right]
    \nonumber\\ &\quad
    +4\,\osc\,\frac{b(s)}{a(s)}\left[\frac{2\pi}{L}\,\frac{\Gamma_1^{(\infty)}}{p_K}+\frac{2\pi^2}{L^2}\,\frac{\Gamma_2^{(\infty)}}{p_K^2}\right] \ ,
  \end{align}
  and $p_K$ defined in \cref{eq:def_pK}.
\end{proposition}

\begin{proof}
We apply \cref{prop:cutoff} to $\ket{\phi}=e^{i\sigma\hat T_0}\ket{\psi(s)}$, which is normalized.
We insert \cref{eq:aliasing_Gamma,eq:aliasing_derivatives,eq:momentum_tail,eq:p_bound} into \cref{eq:prop_cutoff}, also using $\Gamma_i^{(q),\leq K}\leq\Gamma_i^{(q)}\leq\Gamma_i^{(\infty)}$, and we group its terms in pairs:
\begin{align*}
  \max\left|\nabla f_K\right|\norm{\hat{\boldsymbol{p}}\ket{\phi}}+\frac{2\pi}{L}\Gamma_1^{(q),>K}\norm{\hat{\boldsymbol{p}}\ket{\phi}}
  \leq\left(\max\left|\nabla F\right|+3\,\frac{2\pi}{L}\Gamma_1^{(\infty),>K}\right)\sqrt{2\,\osc\,\frac{b(s)}{a(s)}} \ ,
  \\
  \frac{1}{2}\max\left|\Delta f_K\right|+\frac{2\pi^2}{L^2}\Gamma_2^{(q),>K}
  \leq\frac{1}{2}\max\left|\Delta F\right|+3\,\frac{2\pi^2}{L^2}\Gamma_2^{(\infty),>K} \ ,
  \\
  \frac{2\pi}{L}\Gamma_1^{(q),\leq K}\norm{\hat{\boldsymbol{p}}\hat Q_K\ket{\phi}}+\frac{2\pi^2}{L^2}\Gamma_2^{(q),\leq K}\norm{\hat Q_K\ket{\phi}}\leq4\,\osc\,\frac{b(s)}{a(s)}\left[\frac{2\pi}{L}\,\frac{\Gamma_1^{(\infty)}}{p_K}+\frac{2\pi^2}{L^2}\,\frac{\Gamma_2^{(\infty)}}{p_K^2}\right] \ .
\end{align*}
Summing the three lines we get \cref{eq:commutator_bound_smooth}, since the right-hand side does not depend on $\sigma$.
\end{proof}

The first term of $\delta_q$ is the high-frequency tail of the cost function, and the second one comes from the high-momentum tail of the state.
Choosing $K=N/4$, so that $p_K=\frac{\pi N}{2L}$, the second term becomes $\frac{16}{N}\,\osc\,\frac{b}{a}\left[\Gamma_1^{(\infty)}+\frac{2}{N}\Gamma_2^{(\infty)}\right]=\mathcal{O}\left(2^{-q}\right)$.
The first term contains $\Gamma_i^{(\infty),>N/4}$, i.e.~the tails beyond $N/4$ of the series $\Gamma_i^{(\infty)}$, which converge for every $F$ satisfying \cref{eq:smooth_hypothesis}: therefore it vanishes for $q\to\infty$ for all these cost functions.
\Cref{eq:commutator_bound_smooth} has the same form as the continuum bound obtained from \cref{eq:commutator_continuum}: compared to \cref{eq:commutator_bound_final}, $\frac{2\pi}{L}\Gamma_1^{(q)}$ and $\frac{2\pi^2}{L^2}\Gamma_2^{(q)}$ are replaced by $\max\left|\nabla F\right|$ and $\frac{1}{2}\max\left|\Delta F\right|$, up to corrections that vanish for large $q$.

\subsection{Bounds for quantum functions}
\label{app:trotter_quantum}

For the quantum functions of \cref{sec:Optimization of Quantum Functions}, the dependence on $d$ of the error bound is at most linear.
If the parameter $x_{\ell}$ (with $\ell\in\{1,\ldots,d\}$) appears in $\nu_{\ell}$ rotations of the state preparation $\hat W(\boldsymbol{x})$, generated by operators with eigenvalues $\pm1/2$, as for the Pauli rotations $e^{-ix\hat\sigma/2}$, then $f$ is a trigonometric polynomial of period $2\pi$ and degree at most $\nu_{\ell}$ in each variable $x_{\ell}$ \cite{schuld2020effect}.
More generally, $f$ is a trigonometric polynomial of period $2\pi$ whenever the generators have integer eigenvalue differences, and the bounds below hold with $\nu_\ell$ replaced by its degree in $x_\ell$.
We also assume that the domain is a full period, $[0,2\pi)^d$, i.e.~$L=2\pi$, as in the encoding \cref{eq:encoding_main}.
Both assumptions are essential to avoid a dependence on $q$: if the domain covers only a fraction of the period, or if the frequencies are not integers, the periodic extension $F$ is in general discontinuous at the boundary of the domain and, according to \cref{tab:regimes}, $\Gamma_i^{(q)}$ can grow exponentially in $q$.
Under these assumptions, instead, the periodic extension $F=f$ is smooth, and, on the grid with $L=2\pi$ and $\nu_{\ell}<N/2$, which holds for $q>\log_2\nu_{\ell}+1$, there is no aliasing: $\tilde f_{\boldsymbol{k}}=0$ unless $\left|k_{\ell}\right|\leq\nu_{\ell}$ for every $\ell$.

For these functions we bound the commutator by expanding $f$ along one direction at a time, instead of in the full Fourier series of \cref{lem:fourier_bound}.

\begin{proposition}\label{prop:trig}
  Let $\tilde F(\boldsymbol{\kappa})=0$ unless $\left|\kappa_{\ell}\right|\leq\nu_{\ell}$ for every $\ell$, with $\nu_{\ell}<N/2$.
  Then, for every vector $\ket{\chi}$,
  \begin{equation}\label{eq:commutator_trig}
    \norm{\left[\hat T_0,f(\hat{\boldsymbol{x}})\right]\ket{\chi}}
    \leq
    \frac{\osc}{2}\sum_{\ell=1}^d\nu_{\ell}\left(\nu_{\ell}+1\right)\left[\frac{2\pi}{L}\norm{\hat p_{\ell}\ket{\chi}}+\frac{2\pi^2}{L^2}\,\frac{2\nu_{\ell}+1}{3}\norm{\ket{\chi}}\right] \ .
  \end{equation}
\end{proposition}

\begin{proof}
Since $\tilde f_{\boldsymbol{k}}=\tilde F(\boldsymbol{k})$, we have $\tilde f_{\boldsymbol{k}}=0$ unless $\left|k_{\ell}\right|\leq\nu_{\ell}$ for every $\ell$.
For each $\ell$, we write
\begin{align*}
  f(\hat{\boldsymbol{x}})
  &=
  \sum_{\boldsymbol{k}\in K_N^d} \tilde{f}_{\boldsymbol{k}} e^{i\dfrac{2\pi}{N}\boldsymbol{k}\cdot\hat{\boldsymbol{r}}}
  \\ &=
  \sum_{|j|\leq\nu_\ell} \sum_{\boldsymbol{k}: k_\ell=j} \tilde{f}_{\boldsymbol{k}} e^{i\frac{2\pi}{N}\left(\boldsymbol{k}-j\boldsymbol{e}_\ell\right)\cdot\hat{\boldsymbol{r}}}\hat S_{j\boldsymbol{e}_\ell}
  \\ &=
  \sum_{|j|\leq\nu_\ell} g_{\ell,j}(\hat{\boldsymbol{x}})\hat S_{j\boldsymbol{e}_\ell} \ ,
\end{align*}
where $\boldsymbol{e}_{\ell}$ is the $\ell$-th unit vector.
Since $\hat T_0=\sum_{\ell}\hat T_{0,\ell}$, with $\hat T_{0,\ell}\coloneqq\hat p_{\ell}^2/2$, and $g_{\ell,j}(\hat{\boldsymbol{x}})$ commutes with $\hat T_{0,\ell}$, we have
\begin{equation}\label{eq:commutator_directions}
  \left[\hat T_0,f(\hat{\boldsymbol{x}})\right]\ket\chi
  =\sum_{\ell=1}^d\left[\hat T_{0,\ell},f(\hat{\boldsymbol{x}})\right]\ket\chi
  =\sum_{\ell=1}^d\sum_{0<\left|j\right|\leq\nu_{\ell}}g_{\ell,j}(\hat{\boldsymbol{x}})\left[\hat T_{0,\ell},\hat S_{j\boldsymbol{e}_{\ell}}\right]\ket\chi \ ,
\end{equation}
where the terms with $j=0$ vanish, since $\hat S_{\boldsymbol{0}}=\hat{\mathbb{I}}$.
Proceeding as in the proof of \cref{lem:fourier_bound}, with $\boldsymbol{k}=j\boldsymbol{e}_{\ell}$ and $\hat T_{0,\ell}$ in place of $\hat T_0$, we get
\begin{equation}\label{eq:commutator_shift_direction}
  \norm{\left[\hat T_{0,\ell},\hat S_{j\boldsymbol{e}_{\ell}}\right]\ket{\chi}}\leq\frac{2\pi}{L}\left|j\right|\norm{\hat p_{\ell}\ket{\chi}}+\frac{2\pi^2}{L^2}j^2\norm{\ket{\chi}} \ .
\end{equation}
We can also bound $g_{\ell,j}$ by writing it explicitly as a discrete Fourier transform along the $\ell$-th direction.
For $0<\left|j\right|\leq\nu_{\ell}$ and $c\coloneqq\left(\max f+\min f\right)/2$, we have
\begin{align}
  \left|g_{\ell,j}(\boldsymbol{x}_{\boldsymbol{r}})\right|
  &=\left|\sum_{\boldsymbol{k}\,:\,k_{\ell}=j}\tilde f_{\boldsymbol{k}}\,e^{i\frac{2\pi}{N}\left(\boldsymbol{k}-j\boldsymbol{e}_{\ell}\right)\cdot\boldsymbol{r}}\right|
  \nonumber\\ &=
  \left|\sum_{\boldsymbol{k}\,:\,k_{\ell}=j}\ \sum_{\boldsymbol{s}}\frac{f(\boldsymbol{x}_{\boldsymbol{s}})}{N^d}\,e^{-i\frac{2\pi}{N}\boldsymbol{k}\cdot\boldsymbol{s}}\,e^{i\frac{2\pi}{N}\left(\boldsymbol{k}-j\boldsymbol{e}_{\ell}\right)\cdot\boldsymbol{r}}\right|
  \nonumber\\ &=
  \left|\sum_{\boldsymbol{s}}\frac{f(\boldsymbol{x}_{\boldsymbol{s}})}{N^d}\,e^{-i\frac{2\pi}{N}jr_{\ell}}\sum_{\boldsymbol{k}\,:\,k_{\ell}=j}e^{i\frac{2\pi}{N}\boldsymbol{k}\cdot\left(\boldsymbol{r}-\boldsymbol{s}\right)}\right|
  \nonumber\\ &=
  \left|\sum_{\boldsymbol{s}}\frac{f(\boldsymbol{x}_{\boldsymbol{s}})}{N^d}\,e^{-i\frac{2\pi}{N}jr_{\ell}}\,e^{i\frac{2\pi}{N}j\left(r_{\ell}-s_{\ell}\right)}\,N^{d-1}\prod_{m\neq\ell}\delta_{r_m,s_m}\right|
  \nonumber\\ &=
  \left|\sum_{s_{\ell}=0}^{N-1}\frac{1}{N}\,f(\boldsymbol{x}_{\boldsymbol{r}+(s_{\ell}-r_{\ell})\boldsymbol{e}_{\ell}})\,e^{-i\frac{2\pi}{N}js_{\ell}}\right|
  \nonumber\\ &=
  \left|\sum_{s_{\ell}=0}^{N-1}\frac{1}{N}\left[f(\boldsymbol{x}_{\boldsymbol{r}+(s_{\ell}-r_{\ell})\boldsymbol{e}_{\ell}})-c\right]e^{-i\frac{2\pi}{N}js_{\ell}}\right|
  \nonumber\\ &\leq
  \sum_{s_{\ell}=0}^{N-1}\frac{1}{N}\left|f(\boldsymbol{x}_{\boldsymbol{r}+(s_{\ell}-r_{\ell})\boldsymbol{e}_{\ell}})-c\right|
  \nonumber\\ &\leq
  \max_{\boldsymbol{s}}\left|f(\boldsymbol{x}_{\boldsymbol{s}})-c\right|
  =\frac{\osc}{2} \ .\label{eq:g_bound}
\end{align}
The first equality is the definition of $g_{\ell,j}$ given by the expansion above, evaluated at the grid point $\boldsymbol{x}_{\boldsymbol{r}}$, and in the second one we inserted the discrete Fourier transform \cref{eq:f_DFT}, with summation index $\boldsymbol{s}\in\left\{0,\ldots,N-1\right\}^d$.
In the sixth equality we subtracted the constant $c$, which does not change the result since $\sum_{s_{\ell}=0}^{N-1}e^{-i\frac{2\pi}{N}js_{\ell}}=0$ for $0<\left|j\right|<N$.

The operator $g_{\ell,j}(\hat{\boldsymbol{x}})$ is diagonal in the position basis; therefore, for every vector $\ket{\xi}$, by \cref{eq:g_bound},
\begin{align}
  \norm{g_{\ell,j}(\hat{\boldsymbol{x}})\ket{\xi}}^2
  &=\sum_{\boldsymbol{r}}\left|g_{\ell,j}(\boldsymbol{x}_{\boldsymbol{r}})\right|^2\left|\braket{\boldsymbol{x}_{\boldsymbol{r}}}{\xi}\right|^2
  \nonumber\\ &\leq
  \max_{\boldsymbol{r}}\left|g_{\ell,j}(\boldsymbol{x}_{\boldsymbol{r}})\right|^2\sum_{\boldsymbol{r}}\left|\braket{\boldsymbol{x}_{\boldsymbol{r}}}{\xi}\right|^2
  \leq\left(\frac{\osc}{2}\right)^2\norm{\ket{\xi}}^2 \ .\label{eq:g_operator_bound}
\end{align}
\Cref{eq:commutator_trig} then follows from \cref{eq:commutator_directions}, the triangle inequality, \cref{eq:commutator_shift_direction,eq:g_operator_bound}, and the sums $\sum_{0<\left|j\right|\leq\nu}\left|j\right|=\nu(\nu+1)$, $\sum_{0<\left|j\right|\leq\nu}j^2=\nu(\nu+1)(2\nu+1)/3$.
\end{proof}

\noindent We now set $L=2\pi$ and apply \cref{prop:trig} to the normalized vector $\ket{\chi}=e^{i\sigma\hat T_0}\ket{\psi(s)}$.
We bound $\sum_{\ell}\norm{\hat p_{\ell}\ket{\chi}}\leq\sqrt{d}\norm{\hat{\boldsymbol{p}}\ket{\chi}}$ by the Cauchy--Schwarz inequality and $\norm{\hat{\boldsymbol{p}}\ket{\chi}}$ by \cref{eq:p_bound}.
Then, with $\nu\coloneqq\max_{\ell}\nu_{\ell}$, from \cref{eq:commutator_trig} we have
\begin{equation}\label{eq:commutator_quantum}
  \sup_{\sigma\in\mathbb{R}}\norm{\left[\hat T_0,f(\hat{\boldsymbol{x}})\right]e^{i\sigma\hat T_0}\ket{\psi(s)}}
  \leq
  \frac{\osc}{2}\,\nu(\nu+1)\left[\sqrt{2\,d\,\osc\,\frac{b(s)}{a(s)}}+\frac{2\nu+1}{6}\,d\right] \ .
\end{equation}
This result holds for every $q>\log_2\nu+1$ and contains no correction depending on $q$.

\subsection{Final error bounds}
\label{app:trotter_final}

Under hypotheses (H1)--(H4), we now insert \cref{eq:normaV} and the bounds on the commutator into \cref{eq:trotter_error_two_norms}.

For generic cost functions we insert \cref{eq:commutator_bound_final}, and we bound $b(s)/a(s)\leq\sup_{\mathcal{I}_j}b/a$ for $s\in\mathcal{I}_j$, so that the term in square brackets no longer depends on $s$.
The remaining integral is $\int_{\mathcal{I}_j}\left|s-t_j\right|ds=\Theta\,\frac{\varDelta t^2}{2}$, where $\Theta\in\left[\frac{1}{2},1\right]$ is a numerical coefficient that depends on the choice of $t_j\in\mathcal{I}_j$ ($\Theta=\frac{1}{2}$ for the midpoint, and $\Theta=1$ if $t_j$ is an endpoint of $\mathcal I_j$).
We thus obtain
\begin{equation}\label{eq:trotter_error_riemann}
  \eTrotter
  \leq
  \frac{\Theta}{2}\varDelta t\sum_{j=1}^R\varDelta t\left[\frac{\osc}{2}\sup_{\mathcal{I}_j}\dot b+\sup_{\mathcal{I}_j}ab\left(\sqrt{2\osc\sup_{\mathcal{I}_j}\frac{b}{a}}\;\frac{2\pi}{L}\Gamma_1^{(q)}+\frac{2\pi^2}{L^2}\Gamma_2^{(q)}\right)\right] \ .
\end{equation}
By hypothesis (H1), $\dot b$, $ab$ and $b/a$ are continuous, and hence uniformly continuous, on $[0,T]$: each supremum differs from the value of the corresponding function at any point of $\mathcal{I}_j$ by a quantity that vanishes uniformly for $\varDelta t\to0$.
The sum over $j$ is therefore a Riemann sum, which converges to the corresponding integral.
At first order in $\varDelta t$ we get
\begin{equation}\label{eq:trotter_error_final}
  \eTrotter
  \lesssim
  \frac{\Theta}{2}\varDelta t\left[\frac{\osc}{2}\left(b(T)-b(0)\right)+\int_0^T\left(\frac{2\pi}{L}\Gamma_1^{(q)}\sqrt{2\osc\,a\,b^3}+\frac{2\pi^2}{L^2}\Gamma_2^{(q)}\,a\,b\right)dt\right] \ .
\end{equation}
Since $\Theta\leq1$ for every choice of the sampling times, \cref{eq:trotter_error_final} implies \cref{eq:trotter error}, for every sampling scheme, with
\begin{equation}\label{eq:c12_generic}
  c_1=\frac{\pi}{L}\sqrt{2\,\osc}\;\Gamma_1^{(q)} \ ,
  \qquad
  c_2=\frac{\pi^2}{L^2}\,\Gamma_2^{(q)} \ .
\end{equation}
In \cref{eq:trotter_error_final}, all the dependence on $d$ and $q$ is contained in $\Gamma_1^{(q)}$ and $\Gamma_2^{(q)}$, \cref{eq:def_Gamma}.
Their dependence on $q$ is set by the regularity of the cost function at the boundary of the domain.
Through the aliasing formula \cref{eq:aliasing}, the behavior of $\Gamma_i^{(q)}$ for large $q$ is dictated by the decay of the Fourier coefficients $\tilde F(\boldsymbol{\kappa})$ of the periodic extension of $f$ (see \cref{tab:regimes}).

\begin{table}[h]
  \centering
  \caption{Behavior of $\Gamma_1^{(q)}$ and $\Gamma_2^{(q)}$ as functions of $q$, for a fixed cost function $f$, according to the regularity of its periodic extension $F$.
  The constants depend on $f$, and no explicit factor of $d$ appears.
  In the last two rows, factors polynomial in $q$ can be added by the singularities at the edges and corners of the domain.}
  \label{tab:regimes}
  \begin{tabular}{@{}lccc@{}}
    \toprule
    periodic extension $F$ & $\left|\tilde F(\boldsymbol{\kappa})\right|$ decay & $\Gamma_1^{(q)}$ & $\Gamma_2^{(q)}$\\
    \midrule
    smooth & faster than any power & $\mathcal{O}(1)$ & $\mathcal{O}(1)$\\
    continuous, with discontinuous derivative & $\sim\left|\boldsymbol{\kappa}\right|^{-2}$ & $\mathcal{O}(q)$ & $\mathcal{O}\left(2^q\right)$\\
    discontinuous & $\sim\left|\boldsymbol{\kappa}\right|^{-1}$ & $\mathcal{O}\left(2^q\right)$ & $\mathcal{O}\left(4^q\right)$\\
    \bottomrule
  \end{tabular}
\end{table}

For cost functions with a smooth periodic extension, \cref{lem:aliasing_sums} already shows that the coefficients \cref{eq:c12_generic} are bounded independently of $q$, since $\Gamma_i^{(q)}\leq\Gamma_i^{(\infty)}$.
Sharper coefficients are obtained by inserting \cref{eq:normaV,eq:commutator_bound_smooth} into \cref{eq:trotter_error_two_norms}, and proceeding as for \cref{eq:trotter_error_final}:
\begin{align}
  \eTrotter
  &\lesssim
  \frac{\Theta}{4}\varDelta t\int_0^T\left[\osc\dot b+ab\left(2\max\left|\nabla F\right|\sqrt{2\osc \frac{b}{a}}+\max\left|\Delta F\right|+2\delta_q\right)\right]dt \ .\label{eq:trotter_error_smooth}
\end{align}
Since $\delta_q$ vanishes for $q\to\infty$ and $\Theta\leq1$, for large $q$ this bound implies \cref{eq:trotter error}, for every sampling scheme, with $c_1$ and $c_2$ given by \cref{eq:c12_smooth}.
In both cases the bound is $\mathcal{O}\left(R\,\varDelta t^2\right)$ for schedules bounded on $[0,T]$, and it does not depend explicitly on $d$: the dimension enters only through $\osc$, $\max\left|\nabla F\right|$ and $\max\left|\Delta F\right|$, or $\Gamma_i^{(\infty)}$ (which also appears in $\delta_q$), which are properties of the given cost function.

Finally we consider quantum functions and, in general, trigonometric polynomials.
Under the assumptions of \cref{prop:trig}, with $\nu\coloneqq\max_{\ell}\nu_{\ell}$ and $L=2\pi$, we insert \cref{eq:commutator_quantum} into \cref{eq:trotter_error_two_norms}, and proceed as for \cref{eq:trotter_error_final}.
At first order in $\varDelta t$, we obtain
\begin{equation}\label{eq:trotter_error_schedules}
  \eTrotter
  \lesssim
  \frac{\Theta}{4}\varDelta t\,\osc\int_0^T\left[\dot b+\nu(\nu+1)\,a\,b\left(\sqrt{2\,d\,\osc\,\frac{b}{a}}+\frac{2\nu+1}{6}\,d\right)\right]dt \ .
\end{equation}
Since $\Theta\leq1$, \cref{eq:trotter_error_schedules} implies \cref{eq:trotter error}, for every sampling scheme, with the coefficients $c_1$ and $c_2$ of \cref{eq:trotter_error_quantum_main}.
This bound is $\mathcal{O}\left(d\,R\,\varDelta t^2\right)$, with a constant that depends only on the schedules, on $\osc\leq2\norm{\hat{O}}$ and on $\nu$.
Unlike \cref{eq:trotter_error_smooth}, it contains no correction vanishing for large $q$: it holds for every $q>\log_2\nu+1$, and it grows at most linearly in $d$.

We stress that \cref{eq:trotter_error_final,eq:trotter_error_smooth,eq:trotter_error_schedules} are of first order in $\varDelta t$ for every choice of the sampling times, whereas the actual error is of first order only if $t_j$ is not the midpoint of $\mathcal I_j$.
For midpoint sampling, the first term of the Magnus expansion of $\hat U_j$, $-i\int_{\mathcal I_j}\hat H_I(s)\,ds$, differs from $-i\varDelta t\,\hat H_I(t_j)$ only at order $\varDelta t^3$, since $\int_{\mathcal I_j}\left(s-t_j\right)ds=0$, and the second term, which contains the commutators $\left[\hat H_I(s),\hat H_I(s')\right]$ with $s,s'\in\mathcal I_j$, is also of order $\varDelta t^3$: the single-step error is $\mathcal O\left(\varDelta t^3\right)$, and the total error is $\mathcal O\left(R\,\varDelta t^3\right)$.
In our derivation, this cancellation is lost in \cref{eq:duhamel_bound}, where the norm is taken inside the integral over each step, so that $\hat H_I(s)-\hat H_I(t_j)$ is bounded by a quantity proportional to $\left|s-t_j\right|$, whose integral over $\mathcal I_j$ does not vanish.
For midpoint sampling, the bounds of this appendix are therefore worst-case estimates of first order for a second-order method, and they become increasingly pessimistic as $\varDelta t\to0$.

Finally, we remark that the bound $\matrixel{\psi(t)}{\hat T_0}{\psi(t)}\leq\osc\,b/a$ of \cref{lem:kinetic_moments}, which enters all the bounds above through \cref{eq:p_bound}, is a worst-case estimate: for a state localized in the harmonic well around the global minimum one has instead $\matrixel{\psi(t)}{\hat T_0}{\psi(t)}\sim\omega\sqrt{b/a}$, with $\omega=\max_i\left|\omega_i\right|$ as in \cref{eq:risultato_convergenza_QHD}.
In this regime, the term $\sqrt{ab^3}=ab\sqrt{b/a}$ in \cref{eq:trotter_error_schedules} is effectively replaced by $ab\left(b/a\right)^{1/4}$.

\section{Efficient implementation of the kinetic step}
\label{app:kinetic_step}

Here we show that the kinetic step $\hat{T}(t_j)$ defined in \cref{eq:trotter_steps} can be efficiently implemented.
For a given step, we denote by
\[
\alpha\coloneqq A(t_{j+1})-A(t_j)=\int_{t_j}^{t_{j+1}}a(t)\,dt
\]
the integrated kinetic schedule between two consecutive sampling times, i.e.~the kinetic angle $\alpha_j$ of \cref{sec:gate simul}, which is computed classically (see \cref{app:angles}), so that the kinetic step reads
\[
\hat{T}(t_j)=e^{-i\alpha\frac{\hat{\boldsymbol{p}}^2}{2}} = \hat{\boldsymbol{F}}_{\!s}\, e^{-i\alpha\hat{\boldsymbol{L}}} \, \hat{\boldsymbol{F}}_{\!s}^\dagger \ .
\]
First, we observe that, since the terms of $\hat{\boldsymbol{L}}$ in \cref{eq:def_L_operator} commute, one has
\[
e^{-i\alpha\hat{\boldsymbol{L}}} = e^{-i\alpha\sum_{\ell=1}^d \hat{\mathbb{I}}^{\otimes (\ell-1)}\otimes\hat{L}\otimes\hat{\mathbb{I}}^{\otimes (d-\ell)}} = \bigotimes_{\ell=1}^d e^{-i\alpha\hat{L}} \ .
\]
Since $\hat{\boldsymbol{F}}_{\!s}$ is also a tensor product over the $d$ sub-registers $S_\ell$ of $q$ qubits that encode the coordinates $x_\ell$ (see \cref{eq:def_F_operator}), the whole kinetic block can be decomposed into blocks of operators acting on separate registers.
We therefore study how to implement
\[
\hat{F}_{\!s}\, e^{-i\alpha\hat{L}} \, \hat{F}_{\!s}^\dagger \ .
\]

\noindent As for the shifted QFT $\hat{F}_{\!s}$, it can be implemented by exploiting
\begin{equation}
    \hat{F}_{\!s}= \hat{s}\hat{F} \ ,
\end{equation}
where $\hat{F}\coloneqq\sum_{r,s=0}^{N-1}\frac{e^{i\frac{2\pi}{N}rs}}{\sqrt{N}}\ketbra{x_s}{x_r}$ is the quantum Fourier transform (QFT) and
\begin{equation}
    \hat{s}\coloneqq \sum_{r=0}^{N-1}e^{-i\pi r}\ketbra{x_r}{x_r}=\sum_{r=0}^{N-1}(-1)^r\, \ketbra{x_r}{x_r} \ .
\end{equation}
The operator $\hat{s}$ applies $\hat{\mathbb{I}}$ to all computational-basis states with even $r$, namely those for which the least significant qubit $c_0$ is in the state $\ket{0}$, and applies $-\hat{\mathbb{I}}$ to all computational-basis states with odd $r$, namely those for which the least significant qubit $c_0$ is in the state $\ket{1}$.
In other words, on our qubit register, $\hat{s}$ acts equivalently to $\hat{\sigma}_z$ on the qubit $c_0$:
\[
\hat{s}=\hat{\mathbb{I}}_{q-1}\otimes\cdots\otimes\hat{\mathbb{I}}_1\otimes \hat{\sigma}_z \ .
\]
The implementation of the QFT operator is standard and is already available in libraries such as \texttt{Qiskit} and \texttt{PennyLane}.

As for the operator $e^{-i\alpha\hat{L}}$, it can be implemented efficiently because it is diagonal in the computational basis and the phases it applies are structured, rather than generic.
In fact, one needs to implement the operator
\[
e^{-i\alpha\hat{L}}
=\sum_{r=0}^{N-1}e^{-i\alpha\frac{2\pi^2}{L^2}\left(r-\frac{N}{2}\right)^2}\ketbra{x_r}{x_r}\ ,
\]
with $N=2^q$ and $\ket{x_r}=\ket{r_{\text{base}2}}=\ket{c_{q-1}\ldots c_0}$, writing $r=\sum_{j=0}^{q-1}2^j c_j$ with $c_j\in\{0,1\}$.
We first consider the expression of $(r-N/2)^2$ in terms of the bits $\{c_j\}$\footnote{We discard any constant term, i.e.~any term independent of the $c_j$, since it represents a global phase: indeed, being independent of $c_j\ \forall j$, it is the same for every bit string $c_{q-1}\ldots c_0$, namely for every state $\ket{x_r}$.}
\begin{align*}
\left(r-\frac{N}{2}\right)^2
&=\left(\sum_{j=0}^{q-1}2^j c_j-2^{q-1}\right)^2\\
&=\left(\sum_{j=0}^{q-1}2^j c_j\right)^2
-2^q\sum_{j=0}^{q-1}2^j c_j
+2^{2q-2}\\
&\equiv
\sum_{j=0}^{q-1}2^{2j}c_j
+\sum_{0\le j<k\le q-1}2^{j+k+1}c_jc_k
-\sum_{j=0}^{q-1}2^{q+j}c_j\\
&=
\sum_{j=0}^{q-1}\left(2^{2j}-2^{q+j}\right)c_j
+\sum_{0\le j<k\le q-1}2^{j+k+1}c_jc_k\ .
\end{align*}
We then consider the action on a generic computational-basis state $\ket{x_r}=\ket{c_{q-1}\ldots c_0}$:
\begin{align*}
  e^{-i\alpha\hat{L}}\ket{x_r}
&= e^{-i\sum_{j=0}^{q-1}\theta_j\,c_j -i\sum_{0\leq j<k\leq q-1}\theta_{jk}\,c_jc_k}\ket{c_{q-1}\ldots c_0}\\
&=\prod_{j=0}^{q-1}e^{-i\theta_j\,c_j}\ \prod_{0\leq j<k\leq q-1}e^{-i\theta_{jk}\,c_jc_k}\ket{c_{q-1}\ldots c_0} \ ,
\end{align*}
where we have defined
\[
\theta_j \coloneqq \alpha\,\frac{2\pi^2}{L^2}\left(2^{2j}-2^{q+j}\right) \ ,\qquad
  \theta_{jk} \coloneqq \alpha\,\frac{2\pi^2}{L^2}\,2^{j+k+1}\ .
\]
Finally, we obtain
\begin{equation}
  e^{-i\alpha\hat{L}}= \prod_{j=0}^{q-1}\hat{P}_j\left(\theta_j\right)\ \prod_{0\leq j<k\leq q-1}\widehat{CP}_{jk}\left(\theta_{jk}\right) \ ,
\end{equation}
where the operators
\[
\hat{P}(\theta)\coloneqq
\begin{pmatrix}
    1 & 0\\
    0 & e^{-i\theta}
  \end{pmatrix}\ ,
  \qquad\widehat{CP}_{jk}(\theta)\coloneqq
  \begin{pmatrix}
    1 & 0 & 0 & 0\\
    0 & 1 & 0 & 0\\
    0 & 0 & 1 & 0\\
    0 & 0 & 0 & e^{-i\theta}
  \end{pmatrix}
\]
are phase gates and controlled-phase gates that can be implemented efficiently, in a number scaling as $q^2$.

\noindent Overall, the action of the kinetic step factorizes over the $d$ simulation registers, and its implementation is described by \cref{circ:K_step_offline}.
\begin{qcircuit}[circ:K_step_offline]
\begin{quantikz}[row sep=0.6cm, column sep=0.4cm]
\lstick{$S$}& \qwbundle{dq} &\gate{\hat T(t_j)} &\qw
\end{quantikz}
\quad = \quad
\begin{quantikz}[row sep=0.6cm, column sep=0.4cm]
\lstick{$S_1$}
& \qwbundle{q}
& \gate{\hat{s}}
& \gate{\hat{F}^{\dagger}}
& \gate{e^{-i\alpha\hat{L}}}
& \gate{\hat{F}}
& \gate{\hat{s}}
& \qw
\\
\lstick{$\vdots$}
\\
\lstick{$S_d$}
& \qwbundle{q}
& \gate{\hat{s}}
& \gate{\hat{F}^{\dagger}}
& \gate{e^{-i\alpha\hat{L}}}
& \gate{\hat{F}}
& \gate{\hat{s}}
& \qw
\end{quantikz} \ .
\end{qcircuit}
Since the QFT on $q$ qubits also requires $\mathcal O\left(q^2\right)$ gates, the whole kinetic step is implemented with $\mathcal O\left(d\,q^2\right)$ elementary gates.

\section{Choice of schedules and simulation angles}
\label{app:schedules_angles}

In this appendix we derive the optimal gauge (\cref{app:schedules}), discuss the choice of the curve $r(\tau)$ (\cref{app:curve}), and compute the resulting kinetic and potential angles of the Trotterized evolution (\cref{app:angles}).

\subsection{Optimal gauge}
\label{app:schedules}
Let $\varepsilon^*$ be the target error, and let us impose that the bound \cref{eq:trotter error}, neglecting the boundary term $\frac{\osc}{4}\left[b(T)-b(0)\right]$, equals it,
\begin{equation}
  \varepsilon^*= \varDelta t\int_0^T \left[c_1\sqrt{a(t)\,b(t)^3}+c_2\,a(t)\,b(t)\right]dt \ ,
\end{equation}
with $c_1$ and $c_2$ numbers dependent on the cost function's features.
We substitute $\tau=\int_0^t a(s)\,ds$ and $r(\tau)=\frac{b(t(\tau))}{a(t(\tau))}$ in the integral, with $dt=\frac{d\tau}{a(t)}$, $t(0)=0$ and $t(\tau_{\mathrm{f}})=T$:
\begin{align*}
  \varepsilon^*
  &=
  \varDelta t\int_0^{\tau_{\mathrm{f}}} b(t(\tau))\left[c_1\sqrt{r(\tau)}+c_2\right]d\tau \\
  &=
  \varDelta t\int_0^{\tau_{\mathrm{f}}} a(t(\tau))\left[c_1\,r(\tau)^{\frac{3}{2}}+c_2\,r(\tau)\right]d\tau \\
  &=
  \varDelta t\int_0^{\tau_{\mathrm{f}}}a(t(\tau))\,g(r(\tau)) \,d\tau \ .
\end{align*}
Since $T=\int_0^T dt=\int_0^{\tau_{\mathrm{f}}}\frac{d\tau}{a(t(\tau))}$ we have:
\begin{align*}
  R=\frac{T}{\varDelta t}=\frac{\int_0^{\tau_{\mathrm{f}}}\frac{d\tau}{a(t(\tau))}\ \int_0^{\tau_{\mathrm{f}}} a(t(\tau))\,g(r(\tau)) \,d\tau}{\varepsilon^*}\geq \frac{\left[\int_0^{\tau_{\mathrm{f}}}\sqrt{g(r(\tau))}d\tau\right]^2}{\varepsilon^*} \ ,
\end{align*}
by the Cauchy--Schwarz inequality, with equality if and only if $a\sqrt{g(r)}$ is constant along the curve, i.e.~for the optimal gauge $a=\lambda\,g^{-1/2}$, with $\lambda>0$.

In this gauge, we compare the boundary term $\frac{\osc}{4}\left[b(T)-b(0)\right]$ with the integral term $\int_0^{\tau_{\mathrm{f}}}a(t(\tau))\,g(r(\tau))\,d\tau$ in \cref{eq:trotter error}.
Using $b(0)\geq0$, $b=a\,r$ and $g(r)\geq c_2\,r$, we have
\begin{align*}
  \frac{\frac{\osc}{4}\left[b(T)-b(0)\right]}{\int_0^{\tau_{\mathrm{f}}}a(t(\tau))\,g(r(\tau))\,d\tau}
  &\leq
  \frac{\osc\, r(\tau_{\mathrm{f}})\,a(T)}{4\int_0^{\tau_{\mathrm{f}}}a(t(\tau))\,g(r(\tau))\,d\tau}
  \\ &=
  \frac{\osc\,\lambda\, r(\tau_{\mathrm{f}})}{4\sqrt{g(r(\tau_{\mathrm{f}}))}\int_0^{\tau_{\mathrm{f}}}\lambda \sqrt{g(r(\tau))}\,d\tau}
  \\ &\leq
  \frac{\osc\, r(\tau_{\mathrm{f}})}{4\sqrt{c_2\,r(\tau_{\mathrm{f}})}\int_0^{\tau_{\mathrm{f}}}\sqrt{c_2\, r(\tau)}\,d\tau}
  \\ &=
  \frac{\osc \sqrt{r(\tau_{\mathrm{f}})}}{4\,c_2\int_0^{\tau_{\mathrm{f}}}\sqrt{r(\tau)}\,d\tau}\ .
\end{align*}
Along the curve $r(\tau)=r_0\,e^{k\tau}$ of \cref{eq:r_curve}, run up to $r(\tau_{\mathrm{f}})=r_{\mathrm{f}}$, we have $d\sqrt{r}=\frac{k}{2}\sqrt{r}\,d\tau$, so that
\begin{equation*}
  \int_0^{\tau_{\mathrm{f}}}\sqrt{r(\tau)}\,d\tau
  =\frac{2}{k}\left(\sqrt{r_{\mathrm{f}}}-\sqrt{r_0}\right)
  =\frac{2}{k}\left(1-\sqrt{\frac{r_0}{r_{\mathrm{f}}}}\right)\sqrt{r_{\mathrm{f}}} \ .
\end{equation*}
Hence, for a generic cost function,
\begin{equation*}
  \frac{\frac{\osc}{4}\left[b(T)-b(0)\right]}{\int_0^{\tau_{\mathrm{f}}}a(t(\tau))\,g(r(\tau))\,d\tau}
  \leq
  \frac{\osc\,k}{8\,c_2\left(1-\sqrt{r_0/r_{\mathrm{f}}}\right)} \ .
\end{equation*}
If the periodic extension $F$ of $f$ is smooth, $c_2\simeq\frac{1}{4}\max\left|\Delta F\right|$ by \cref{eq:c12_smooth}, and for $r_{\mathrm{f}}\gg r_0$ the ratio is $\lesssim k\,\ell^2/2$, with $\ell^2\coloneqq\osc/\max\left|\Delta F\right|$ the squared length scale over which $f$ varies by its oscillation.
The boundary term is therefore negligible for $k\,\ell^2\ll1$, i.e.~if the kinetic time spent per $e$-fold of $r$ is much longer than $\ell^2$.
For quantum functions, $c_2=\frac{\osc}{24}\,\nu(\nu+1)(2\nu+1)\,d\geq\frac{\osc}{4}\,d$ by \cref{eq:trotter_error_quantum_main}, and the condition becomes universal: the ratio is at most $k/\left[2\,d\left(1-\sqrt{r_0/r_{\mathrm{f}}}\right)\right]$, i.e.~$\simeq0.05/d$ with $k=0.1$, $r_0=10^{-4}$ and $r_{\mathrm{f}}\geq10$, as in our simulations.
In the gauge with constant $b$ that we use for quantum functions (\cref{sec:online_procedure}), the boundary term vanishes identically.
Thus \cref{eq:R_gauge} describes the minimum number of Trotter steps sufficient to guarantee error $\leq\varepsilon^*$.

\subsection{Choice of the curve}
\label{app:curve}

While the gauge only determines how the curve $r(\tau)$ is cut into Trotter steps, the curve determines the ideal dynamics \cref{eq:tau_dynamics}, and hence the result of the algorithm.
It must keep the evolution adiabatic where this matters, and three regimes can be distinguished.
\begin{enumerate}
  \item[(i)] For $r\,\osc\ll\frac{1}{2}\left(2\pi/L\right)^2$, the lowest kinetic excitation energy, the potential is a small perturbation and the uniform state remains close to the instantaneous ground state $\ket{0_r}$ of \cref{eq:tau_dynamics}: nothing happens, and the evolution can start directly from a small $r_0>0$.
  We can choose its value \emph{a priori},
  \begin{equation}\label{eq:r0}
    r_0=\eta\,\frac{2\pi^2}{\osc\,L^2} \ , \qquad \eta\ll1 \ .
  \end{equation}
  At first order in perturbation theory, the uniform state has overlap $\left|\braket{0_{r_0}}{\psi(0)}\right|^2\gtrsim1-\eta^2/4$ with the instantaneous ground state at $r=r_0$,\footnote{To lowest order in $r$, $1-\left|\braket{0_r}{\psi(0)}\right|^2\simeq r^2\sum_{\boldsymbol k\neq\boldsymbol 0}\left|\matrixel{\boldsymbol k}{f(\hat{\boldsymbol x})}{\boldsymbol 0}\right|^2/E_{\boldsymbol k}^2$, where $\ket{\boldsymbol k}$ are the eigenvectors of $\hat{\boldsymbol p}$, with $\ket{\boldsymbol 0}=\ket{\psi(0)}$, and $E_{\boldsymbol k}\geq2\pi^2/L^2$ are the kinetic energies of those with $\boldsymbol k\neq\boldsymbol 0$.
  The sum $\sum_{\boldsymbol k\neq\boldsymbol 0}\left|\matrixel{\boldsymbol k}{f(\hat{\boldsymbol x})}{\boldsymbol 0}\right|^2$ is the variance of $f$ over the grid, which is at most $\osc^2/4$.} so that $\eta\lesssim0.2$ is already sufficient.
  Smaller values are harmless, since the initial part of the curve requires few Trotter steps (see \cref{eq:R_curve} and below).

  \item[(ii)] In the \emph{search window} $r_0\lesssim r\lesssim r_{\mathrm{s}}$, the potential competes with the kinetic energy, the wavefunction redistributes among the valleys of $f$ through tunneling, and the populations of their basins, hence the success probability (\cref{eq:Psuc_def}), are decided.
  The search window ends at the ratio $r_{\mathrm{s}}$ at which the instantaneous ground state $\ket{0_r}$ becomes the ground state of the harmonic approximation of the valley of the global minimum.
  Since $r_{\mathrm{s}}$ depends on the whole landscape, it is not easily estimated \emph{a priori}, and we read it off \emph{a posteriori} from the stabilization of $P_{\mathrm{suc}}$ in simulations of small instances.
  This criterion can be misleading: $P_{\mathrm{suc}}$ may appear stable while the instantaneous ground state has not yet reached this final regime, as in the case discussed below.

  The gaps depend on the landscape, and the optimal curve follows the local adiabatic principle \cite{roland2002quantum},
  \begin{equation}\label{eq:local_adiabatic}
    \frac{dr}{d\tau}\lesssim\delta_{\mathrm{ad}}\,\frac{\Delta(r)^2}{\left|\matrixel{1_r}{f(\hat{\boldsymbol x})}{0_r}\right|} \ ,
  \end{equation}
  where $\Delta(r)$ is the gap between the two lowest eigenstates $\ket{0_r},\ket{1_r}$ of \cref{eq:tau_dynamics} and $\delta_{\mathrm{ad}}\ll1$: kinetic time should be spent where the gap is small.
  Since $\Delta(r)$ is generally unknown, a sub-optimal but robust choice is to distribute the kinetic time uniformly in $\ln r$, as in the curve $r(\tau)=r_0\,e^{k\tau}$ of \cref{eq:r_curve}, where $1/k$ is the kinetic time spent per $e$-fold of $r$, so that the search window takes the kinetic time $\ln\left(r_{\mathrm{s}}/r_0\right)/k$.
  The rate $k$ is the main free parameter of the algorithm, to be decreased until $P_{\mathrm{suc}}$ no longer improves.
  Slowing down the search, however, does not always increase $P_{\mathrm{suc}}$.
  If the global minimum lies in a valley narrower than that of a nearly degenerate local minimum, the larger zero-point energy of the narrow valley makes the instantaneous ground state localize first in the wider one, and move to the global valley only through an avoided crossing with an exponentially small gap.
  If this crossing occurs at a large $r$, where $P_{\mathrm{suc}}$ already appears stable, a slower curve follows the wrong valley more faithfully, so that $P_{\mathrm{suc}}$ decreases as $k$ is reduced until the evolution becomes adiabatic also with respect to this gap, which may require an impractically long time.

  \item[(iii)] For $r\gtrsim r_{\mathrm{s}}$, the wavefunction is localized in the valleys and tunneling between them is exponentially suppressed, so that the success probability is frozen and QHD only refines the position within each valley, with the error \cref{eq:risultato_convergenza_QHD} decreasing as $r^{-1/2}$.
  For classical functions (\cref{sec:Optimization of Classical Functions}), the ratio $r_{\mathrm{f}}$ at the end of the simulation is fixed by the target accuracy $\delta_f$ on the cost function through \cref{eq:risultato_convergenza_QHD} with $\mathcal{N}=1$, which holds when the state is in the ground state of the valley of the global minimum,
  \begin{equation}
    r_{\mathrm{f}}\simeq\left(\frac{d\,\omega}{4\,\delta_f}\right)^2 \ .
  \end{equation}
  Instead, in the case of quantum functions (\cref{sec:Optimization of Quantum Functions}), it is more efficient to stop at a moderate $r_{\mathrm{f}}\gtrsim r_{\mathrm{s}}$ (see \cref{sec:online_procedure}); at that point the state is localized within the basin of the global minimum, so we can measure and refine the solution with a classical optimizer, e.g.~gradient descent with the parameter-shift rule.
\end{enumerate}

We run the curve \cref{eq:r_curve} up to $r(\tau_{\mathrm{f}})=r_{\mathrm{f}}$, i.e.~$\tau_{\mathrm{f}}=\ln\left(r_{\mathrm{f}}/r_0\right)/k$.
Along this curve $d\tau=dr/(k\,r)$, so that, in the optimal gauge of \cref{sec:schedules}, the number of Trotter steps \cref{eq:R_gauge} sufficient to guarantee error $\leq\varepsilon^*$ according to the bound \cref{eq:trotter error} becomes
\begin{equation}\label{eq:R_curve}
  R_{\mathrm{min}}\simeq\frac{1}{\varepsilon^*}\left(\frac{1}{k}\int_{r_0}^{r_{\mathrm{f}}}\frac{\sqrt{g(r)}}{r}\,dr\right)^2 \ .
\end{equation}
The integrand behaves as $\sqrt{c_2/r}$ for small $r$, so that the integral converges for $r_0\to0$: at a fixed rate $k$, the initial part of the curve, where nothing happens, takes a vanishing fraction of the Trotter steps, and $r_0$ can be chosen conservatively.
Since the bound \cref{eq:trotter error} is of first order in $\varDelta t$, while for midpoint sampling the actual error is of second order (see the end of \cref{app:trotter_final}), \cref{eq:R_curve} is a sufficient but pessimistic number of steps.

\subsection{Kinetic and potential angles}
\label{app:angles}

Here we compute the angles $\alpha_j$ and $\varphi_j$ of \cref{eq:trotter_interaction_main}, for midpoint sampling times, along a curve $r(\tau)$ run through in a gauge of the form
\begin{equation}\label{eq:gauge_g}
  a(t(\tau))=\frac{\lambda}{\sqrt{g(r(\tau))}} \ ,
\end{equation}
where $\lambda>0$ and $g$ is a positive function of $r$.
Let
\begin{equation}\label{eq:cost_length}
  s(\tau)\coloneqq\int_0^\tau\sqrt{g(r(\tau'))}\,d\tau'
\end{equation}
be the \emph{cost length} of the curve.
From the definition of the kinetic time $\tau=\int_0^t a(t')\,dt'$ and \cref{eq:gauge_g}, we have $\frac{d\tau}{dt}=a(t)=\frac{\lambda}{\sqrt{g(r(\tau))}}$, hence
\begin{equation}
  dt=\frac{\sqrt{g(r(\tau))}}{\lambda}\,d\tau=\frac{ds}{\lambda} \ ,\label{eq:dt_ds}
\end{equation}
with $\lambda=s(\tau_{\mathrm{f}})/T$: choosing $\lambda$ is equivalent to choosing the total time $T$, and the physical time $t$ is proportional to the cost length $s$.
Since $s(\tau)$ is strictly increasing, it can be inverted, and we denote its inverse by $\tau(s)$; since $s=\lambda t$, the kinetic time is $\int_0^t a(t')\,dt'=\tau(\lambda t)$.
We define
\begin{equation}
  s_j\coloneqq\left(j-\frac{1}{2}\right)\varDelta s \ ,
  \qquad
  \varDelta s\coloneqq\frac{s(\tau_{\mathrm{f}})}{R} \ ,
\end{equation}
for $j=1,\ldots,R$, with $s_0\coloneqq0$ and $s_{R+1}\coloneqq s(\tau_{\mathrm{f}})$, so that $s_j=\lambda t_j$ corresponds to the sampling time $t_j$, and $\tau(s_j)=\int_0^{t_j}a(t)\,dt$.
Then, using $b=a\,r$ and $\lambda\,\varDelta t=\varDelta s$, the angles are
\begin{equation}\label{eq:angles}
  \alpha_j=\tau(s_{j+1})-\tau(s_j) \ ,
  \qquad
  \varphi_j=\varDelta s\,\frac{r\left(\tau(s_j)\right)}{\sqrt{g\left(r\left(\tau(s_j)\right)\right)}} \ .
\end{equation}
They depend only on the curve, on $g$ and on $R$, and not on $\lambda$, in agreement with the symmetry $(a,b,t)\to(\lambda a,\lambda b,t/\lambda)$ discussed in \cref{sec:schedules}.
For a generic $g$, $s(\tau)$ and $\tau(s)$ are obtained by numerical quadrature and inversion, at a negligible classical cost.

In kinetic time, a Trotter step lasts $\alpha_j\simeq\varDelta s/\sqrt{g(r(\tau(s_j)))}$: if $g$ is increasing, the steps are long where $r$ is small, and short where $r$ is large.
Each step covers the same interval $\varDelta s$ of cost length, and therefore contributes equally to any error bound of the form
\begin{equation}
  \varDelta t\int_{\mathcal I_j}a^2\,g(r)\,dt =\varDelta t\int a\,g(r)\,d\tau =\lambda\,\varDelta t\int\sqrt{g(r)}\,d\tau =\varDelta s^2 \ ,
\end{equation}
where the integrals in $\tau$ run over the $j$-th step, so that the bound equals $R\,\varDelta s^2=s(\tau_{\mathrm{f}})^2/R$, as in \cref{eq:R_gauge}.

\section{Online procedure error bound}
\label{app:online_error}

In this appendix we bound the error introduced by the online procedure with respect to the unitary Trotterized evolution \cref{eq:trotter_interaction_main}, in terms of the potential angles $\varphi_j=\varDelta t\,b(t_j)$, with $t_j=\left(j-\frac{1}{2}\right)\varDelta t$ and $j=1,\ldots,R$.
We denote by $\mathcal{T}_j(\rho)\coloneqq\hat{T}(t_{j-1})\,\rho\,\hat{T}^\dagger(t_{j-1})$ the $j$-th kinetic step and by $\mathcal{V}_j(\rho)\coloneqq\hat{V}(t_j)\,\rho\,\hat{V}^\dagger(t_j)$ the exact $j$-th potential step, with $\hat T$ and $\hat V$ defined in \cref{eq:trotter_steps}, so that the unitary Trotterized evolution is $\mathcal{V}_R\circ\mathcal{T}_R\circ\cdots\circ\mathcal{V}_1\circ\mathcal{T}_1$, up to the last kinetic step $\hat{T}(t_R)$, which is unitary and does not change the trace distances below.
The online procedure replaces $\mathcal{V}_j$ with the channel
\begin{equation*}
  \Phi_j(\rho)\coloneqq\mathrm{Tr}_{\!M}\left[\hat{V}_{\!S\!M}(t_j)\left(\rho\otimes\ketbra{0_{\!M}}{0_{\!M}}\right)\hat{V}_{\!S\!M}^\dagger(t_j)\right]=\sum_m\hat{K}_{j,m}\,\rho\,\hat{K}_{j,m}^\dagger \ ,
\end{equation*}
where $\hat{V}_{\!S\!M}$ is defined in \cref{eq:online_potential_operator}, $m$ indexes the basis of the measure register and $\hat{K}_{j,m}\coloneqq\matrixel{m}{\hat{V}_{\!S\!M}(t_j)}{0_{\!M}}$ are Kraus operators.
Since $\hat{W}_{\!S\!M}$ acts on $M$ conditionally on the computational basis of $S$ (\cref{eq:W extended}), the Kraus operators are diagonal in that basis.
Indeed, inserting \cref{eq:online_potential_operator,eq:W extended} with $\varDelta t\,b(t_j)=\varphi_j$, and using the orthogonality of the projectors $\ketbra{\boldsymbol{x}_r}{\boldsymbol{x}_r}$, we get
\begin{align}
  \hat{K}_{j,m}
  &=\matrixel{m}{\hat{W}_{\!S\!M}^\dagger\left(\hat{\mathbb{I}}_{S}\otimes e^{-i\varphi_j\hat{O}_{\!M}}\right)\hat{W}_{\!S\!M}}{0_{\!M}}
  \nonumber\\ &=
  \matrixel{m}{\left(\sum_{r}\ketbra{\boldsymbol{x}_r}{\boldsymbol{x}_r}\otimes\hat{\mathbb{I}}_{M}\right)\hat{W}_{\!S\!M}^\dagger\left(\hat{\mathbb{I}}_{S}\otimes e^{-i\varphi_j\hat{O}_{\!M}}\right)\left(\sum_{s}\ketbra{\boldsymbol{x}_s}{\boldsymbol{x}_s}\otimes\hat{\mathbb{I}}_{M}\right)\hat{W}_{\!S\!M}}{0_{\!M}}
  \nonumber\\ &=
  \sum_r\matrixel{m}{\hat{W}_{\!M}^\dagger(\boldsymbol{x}_r)\,e^{-i\varphi_j\hat{O}_{\!M}}\,\hat{W}_{\!M}(\boldsymbol{x}_r)}{0_{\!M}}\ketbra{\boldsymbol{x}_r}{\boldsymbol{x}_r}
  \nonumber\\ &=
  \sum_r\braket{m}{\chi_r^{(j)}}\ketbra{\boldsymbol{x}_r}{\boldsymbol{x}_r} \ ,
  \label{eq:kraus_online}
\end{align}
where
\begin{equation*}
  \ket{\chi_r^{(j)}}\coloneqq e^{-i\varphi_j\hat{O}_r}\ket{0_{\!M}} \ ,
  \qquad
  \hat{O}_r\coloneqq\hat{W}_{\!M}^\dagger(\boldsymbol{x}_r)\,\hat{O}_{\!M}\,\hat{W}_{\!M}(\boldsymbol{x}_r) \ ,
\end{equation*}
and in the last step we used $\hat{W}_{\!M}^\dagger(\boldsymbol{x}_r)\,e^{-i\varphi_j\hat{O}_{\!M}}\,\hat{W}_{\!M}(\boldsymbol{x}_r)=e^{-i\varphi_j\hat{O}_r}$, which follows from the unitarity of $\hat{W}_{\!M}(\boldsymbol{x}_r)$.
We decompose
\begin{equation*}
  \hat{O}_r\ket{0_{\!M}}=f(\boldsymbol{x}_r)\ket{0_{\!M}}+\ket{\delta_r} \ ,
  \qquad
  \braket{0_{\!M}}{\delta_r}=0 \ ,
\end{equation*}
where $f(\boldsymbol{x}_r)=\matrixel{0_{\!M}}{\hat{O}_r}{0_{\!M}}$ is the value of the cost function at the grid point $\boldsymbol{x}_r$, and
\begin{equation*}
  \norm{\ket{\delta_r}}^2=\matrixel{0_{\!M}}{\hat{O}_r^2}{0_{\!M}}-f(\boldsymbol{x}_r)^2\leq\norm{\hat{O}}^2
\end{equation*}
is the variance of $\hat{O}_{\!M}$ on the state $\hat{W}_{\!M}(\boldsymbol{x}_r)\ket{0_{\!M}}$.
In the following we drop the index $j$ when it is not necessary.

\begin{proposition}[Error of a single online step]\label{prop:online_step}
  For every density matrix $\rho$ on $S$,
  \begin{equation}\label{eq:online_step}
    \frac{1}{2}\norm{\Phi(\rho)-\mathcal{V}(\rho)}_1\leq\varphi^2\max_r\norm{\ket{\delta_r}}^2+\mathcal{O}\left(\varphi^3\norm{\hat{O}}^3\right) \ .
  \end{equation}
\end{proposition}

\begin{proof}
To second order in $\varphi$, $\ket{\chi_r}=\ket{0_{\!M}}-i\varphi\,\hat{O}_r\ket{0_{\!M}}-\frac{\varphi^2}{2}\hat{O}_r^2\ket{0_{\!M}}+\mathcal{O}\left(\varphi^3\norm{\hat{O}}^3\right)$.
Since $\braket{m}{\delta_r}=\matrixel{m}{\hat{O}_r}{0_{\!M}}$ for $m\neq0$, the Kraus operators \cref{eq:kraus_online} are
\begin{equation*}
  \hat{K}_0=\hat{\mathbb{I}}-i\varphi\,f(\hat{\boldsymbol{x}})-\frac{\varphi^2}{2}\hat{B}+\mathcal{O}\left(\varphi^3\right) \ ,
  \qquad
  \hat{K}_{m\neq0}=-i\varphi\,\hat{A}_m+\mathcal{O}\left(\varphi^2\right) \ ,
\end{equation*}
with
\begin{equation*}
  \hat{B}\coloneqq\sum_r\matrixel{0_{\!M}}{\hat{O}_r^2}{0_{\!M}}\ketbra{\boldsymbol{x}_r}{\boldsymbol{x}_r} \ ,
  \qquad
  \hat{A}_m\coloneqq\sum_r\braket{m}{\delta_r}\ketbra{\boldsymbol{x}_r}{\boldsymbol{x}_r} \ .
\end{equation*}
Therefore
\begin{align*}
  \Phi(\rho)&=\hat{K}_0\,\rho\,\hat{K}_0^\dagger+\sum_{m\neq0}\hat{K}_m\,\rho\,\hat{K}_m^\dagger
  \\ &=
  \rho-i\varphi\left[f(\hat{\boldsymbol{x}}),\rho\right]+\varphi^2f(\hat{\boldsymbol{x}})\,\rho\,f(\hat{\boldsymbol{x}})-\frac{\varphi^2}{2}\left\{\hat{B},\rho\right\}+\varphi^2\sum_{m\neq0}\hat{A}_m\,\rho\,\hat{A}_m^\dagger+\mathcal{O}\left(\varphi^3\right) \ ,
  \\
  \mathcal{V}(\rho)&=e^{-i\varphi f(\hat{\boldsymbol{x}})}\,\rho\,e^{i\varphi f(\hat{\boldsymbol{x}})}
  \\ &=
  \rho-i\varphi\left[f(\hat{\boldsymbol{x}}),\rho\right]+\varphi^2f(\hat{\boldsymbol{x}})\,\rho\,f(\hat{\boldsymbol{x}})-\frac{\varphi^2}{2}\left\{f(\hat{\boldsymbol{x}})^2,\rho\right\}+\mathcal{O}\left(\varphi^3\right) \ .
\end{align*}
Since $\braket{0_{\!M}}{\delta_r}=0$, we have
\begin{equation*}
  \hat{B}-f(\hat{\boldsymbol{x}})^2=\sum_{m\neq0}\hat{A}_m^\dagger\hat{A}_m=\sum_r\norm{\ket{\delta_r}}^2\ketbra{\boldsymbol{x}_r}{\boldsymbol{x}_r}\eqqcolon\hat{G} \ ,
\end{equation*}
so that $\Phi(\rho)-\mathcal{V}(\rho)=\varphi^2\mathcal{D}(\rho)+\mathcal{O}\left(\varphi^3\right)$, with the dissipator
\begin{equation*}
  \mathcal{D}(\rho)\coloneqq\sum_{m\neq0}\hat{A}_m\,\rho\,\hat{A}_m^\dagger-\frac{1}{2}\left\{\hat{G},\rho\right\} \ .
\end{equation*}
The first term of $\mathcal{D}(\rho)$ is a positive operator, whose trace norm equals its trace, $\mathrm{Tr}\left(\hat{G}\rho\right)\leq\norm{\hat{G}}_\infty$.
By H\"older's inequality, $\norm{\hat{G}\rho}_1\leq\norm{\hat{G}}_\infty\norm{\rho}_1=\norm{\hat{G}}_\infty$, and similarly for $\rho\hat{G}$.
By the triangle inequality, $\frac{1}{2}\norm{\mathcal{D}(\rho)}_1\leq\norm{\hat{G}}_\infty=\max_r\norm{\ket{\delta_r}}^2$.
\end{proof}

\begin{proposition}[Online procedure error]\label{prop:online_global}
  Let $\rho^{\mathrm{on}}(T)$ be the final state of the simulation register with the online procedure, and $\ket{\tilde{\psi}(T)}=\hat{U}_{\mathrm{Trotter}}(T,0)\ket{\psi(0)}$ the final state of the unitary Trotterized evolution.
  Then
  \begin{equation}\label{eq:online_global}
    \don\coloneqq\frac{1}{2}\norm{\rho^{\mathrm{on}}(T)-\ketbra{\tilde{\psi}(T)}{\tilde{\psi}(T)}}_1
    \leq\norm{\hat{O}}^2\sum_{j=1}^R\varphi_j^2+\mathcal{O}\left(R\max_j\varphi_j^3\norm{\hat{O}}^3\right) \ .
  \end{equation}
\end{proposition}

\begin{proof}
Let $\ket{\psi_j}$ be the state of the unitary Trotterized evolution just before the $j$-th potential step, i.e.~$\ketbra{\psi_j}{\psi_j}=\mathcal{T}_j\circ\mathcal{V}_{j-1}\circ\cdots\circ\mathcal{V}_1\circ\mathcal{T}_1\left(\ketbra{\psi(0)}{\psi(0)}\right)$.
The telescoping identity
\begin{multline*}
  \Phi_R\circ\mathcal{T}_R\circ\cdots\circ\Phi_1\circ\mathcal{T}_1-\mathcal{V}_R\circ\mathcal{T}_R\circ\cdots\circ\mathcal{V}_1\circ\mathcal{T}_1
  \\
  =\sum_{j=1}^R\Phi_R\circ\mathcal{T}_R\circ\cdots\circ\Phi_{j+1}\circ\mathcal{T}_{j+1}\circ\left(\Phi_j-\mathcal{V}_j\right)\circ\mathcal{T}_j\circ\mathcal{V}_{j-1}\circ\cdots\circ\mathcal{V}_1\circ\mathcal{T}_1 \ ,
\end{multline*}
applied to $\ketbra{\psi(0)}{\psi(0)}$, the triangle inequality and the contractivity of the trace norm under quantum channels give
\begin{equation*}
  \don\leq\sum_{j=1}^R\frac{1}{2}\norm{\left(\Phi_j-\mathcal{V}_j\right)\left(\ketbra{\psi_j}{\psi_j}\right)}_1 \ .
\end{equation*}
Each term is bounded by \cref{eq:online_step}, with $\max_r\norm{\ket{\delta_r}}^2\leq\norm{\hat{O}}^2$.
\end{proof}
For a Pauli string $\norm{\hat{O}}=1$, and \cref{eq:online_global} gives \cref{eq:online_bound_main}.
Since $\varphi_j=b(t_j)\,\varDelta t$, the sum is a Riemann sum, and at first order in $\varDelta t$
\begin{equation}\label{eq:online_error_schedules}
  \don\lesssim\sum_{j=1}^R\left[b(t_j)\,\varDelta t\right]^2\simeq\varDelta t\int_0^T b^2\,dt \ .
\end{equation}
This bound is of second order in the potential angles, so it requires $\varphi_j\ll1$.

Finally, for pure states, with $c\coloneqq\left|\braket{\psi}{\phi}\right|$,
\begin{align*}
  \frac{1}{2}\norm{\ketbra{\psi}{\psi}-\ketbra{\phi}{\phi}}_1
  &=\sqrt{1-c^2}
  =\sqrt{\left(1-c\right)\left(1+c\right)}
  \\ &\leq\sqrt{2\left(1-c\right)}
  \\ &=\min_\theta\norm{e^{i\theta}\ket{\psi}-\ket{\phi}} \ .
\end{align*}
Hence, defining the QHD error via the online procedure, with respect to the exact final state $\ket{\psi(T)}$, as
\begin{equation*}
  \eonline\coloneqq\frac{1}{2}\norm{\rho^{\mathrm{on}}(T)-\ketbra{\psi(T)}{\psi(T)}}_1 \ ,
\end{equation*}
the triangle inequality gives $\eonline\leq\eTrotter+\don$, and \cref{eq:error QHD} follows from \cref{eq:trotter_error_schedules} with $\Theta=1$ and \cref{eq:online_error_schedules}.

\section{Quantum function instances}
\label{app:quantum_instances}

The quantum functions of \cref{sec:results} are defined as
\begin{equation}
  f(\boldsymbol x)=\matrixel{0}{\hat W^\dagger(\boldsymbol x)\,\hat{O}\,\hat W(\boldsymbol x)}{0} \ ,
  \qquad \boldsymbol x\in[0,2\pi)^d \ ,
\end{equation}
where the preparation circuit $\hat W(\boldsymbol x)$ acts on $m$ qubits initialized in $\ket{0}^{\otimes m}$ and the observable $\hat{O}$ is a single Pauli string.
We denote by $\hat X$, $\hat Y$, $\hat Z$ the Pauli matrices, by $\hat R_\sigma(x)\coloneqq e^{-ix\hat\sigma/2}$, with $\sigma\in\{X,Y,Z\}$, the rotations, and by $\hat R_{ZZ}(x)\coloneqq e^{-ix\hat Z\otimes\hat Z/2}$ the two-qubit $ZZ$ rotation; qubit $0$ is the top line of the circuits and the leftmost factor of the Pauli strings.

Examples 1 and 2 share the one-parameter ring circuit of \cref{circ:ring}, in which the parameter $x$ is re-uploaded on all the qubits, between entangling gates and fixed rotations; they differ only in the observable, $\hat{O}=\hat{\mathbb I}\otimes\hat Y\otimes\hat Z$ for Example 1 and $\hat{O}=\hat X\otimes\hat X\otimes\hat X$ for Example 2.
\begin{qcircuit}[circ:ring]
  \begin{quantikz}[row sep=0.3cm, column sep=0.25cm]
    \lstick{$\ket{0}$} & \gate{\hat R_Y(x)} & \ctrl{1} & \qw & \qw & \qw & \qw & \qw & \targ{} & \qw \\
    \lstick{$\ket{0}$} & \qw & \targ{} & \gate{\hat R_X\left(\frac{3\pi}{4}\right)} & \gate{\hat R_Y(x)} & \ctrl{1} & \qw & \qw & \qw & \qw \\
    \lstick{$\ket{0}$} & \qw & \qw & \qw & \qw & \targ{} & \gate{\hat R_X\left(\frac{\pi}{6}\right)} & \gate{\hat R_Y(x)} & \ctrl{-2} & \qw
  \end{quantikz}
\end{qcircuit}%
Example 3 is a hardware-efficient ansatz with two qubits and two layers, in which each parameter is re-uploaded on the same qubit, with observable $\hat{O}=\hat X\otimes\hat{\mathbb I}$:
\begin{qcircuit}[circ:hea2]
  \begin{quantikz}[row sep=0.3cm, column sep=0.3cm]
    \lstick{$\ket{0}$} & \gate{\hat R_Y(x_1)} & \ctrl{1} & \gate{\hat R_Y(x_1)} & \ctrl{1} & \qw \\
    \lstick{$\ket{0}$} & \gate{\hat R_Y(x_2)} & \targ{} & \gate{\hat R_Y(x_2)} & \targ{} & \qw
  \end{quantikz}
\end{qcircuit}%
Example 4 is a hardware-efficient ansatz with three qubits and two layers, with the parameter $x_1$ on all the qubits of the first layer and $x_2$ on those of the second one, and observable $\hat{O}=\hat Z\otimes\hat X\otimes\hat{\mathbb I}$:
\begin{qcircuit}[circ:hea3]
  \begin{quantikz}[row sep=0.3cm, column sep=0.3cm]
    \lstick{$\ket{0}$} & \gate{\hat R_Y(x_1)} & \ctrl{1} & \qw & \gate{\hat R_Y(x_2)} & \ctrl{1} & \qw & \qw \\
    \lstick{$\ket{0}$} & \gate{\hat R_Y(x_1)} & \targ{} & \ctrl{1} & \gate{\hat R_Y(x_2)} & \targ{} & \ctrl{1} & \qw \\
    \lstick{$\ket{0}$} & \gate{\hat R_Y(x_1)} & \qw & \targ{} & \gate{\hat R_Y(x_2)} & \qw & \targ{} & \qw
  \end{quantikz}
\end{qcircuit}%
Example 5 is a QAOA-like circuit, with a layer of $ZZ$ and $Z$ rotations that depend on $x_1$ and a layer of $X$ rotations that depend on $x_2$, and observable $\hat{O}=\hat X\otimes\hat Z\otimes\hat Z$:
\begin{qcircuit}[circ:qaoa]
  \begin{quantikz}[row sep=0.3cm, column sep=0.3cm]
    \lstick{$\ket{0}$} & \gate{\hat H} & \gate[2]{\hat R_{ZZ}(x_1)} & \gate{\hat R_Z(x_1)} & \gate{\hat R_X(x_2)} & \qw \\
    \lstick{$\ket{0}$} & \gate{\hat H} & & \gate[2]{\hat R_{ZZ}(x_1)} & \gate{\hat R_X(x_2)} & \qw \\
    \lstick{$\ket{0}$} & \gate{\hat H} & \qw & & \gate{\hat R_X(x_2)} & \qw
  \end{quantikz}
\end{qcircuit}%

The corresponding functions, obtained from the exact Fourier coefficients of the circuits and checked against their simulation, are
\begin{align}
  f_1(x)&=\frac{\sqrt2}{8}\sin2x\left(2+\cos x\right) \ ,\\
  f_2(x)&=-\frac{\sqrt2}{8}\sin2x\left(1+2\cos x\right) \ ,\\
  f_3(\boldsymbol x)&=\sin x_1\cos x_2\left(\cos x_1+\sin x_2+\cos x_1\sin x_2\right) \ ,\\
  f_4(\boldsymbol x)&=\cos x_1\sin x_1\cos^3x_2+\cos^2x_1\cos x_2\sin^2x_2 \nonumber\\
  &\quad-\cos x_1\sin^2x_1\sin^3x_2-\sin x_1\left(1-\cos x_1\right)\cos^2x_2\sin x_2 \ ,\\
  f_5(\boldsymbol x)&=\frac{1}{4}\sin2x_1\left(1+\cos x_1\right)\sin2x_2 \ .
\end{align}
They are trigonometric polynomials of degree at most $3$ in each variable, as expected from the number of rotations that depend on each parameter (\cref{app:trotter_quantum}): $\nu=2$ for Example 3 and $\nu=3$ for the other examples, so that the condition $q>\log_2\nu+1$ of \cref{eq:trotter_error_quantum_main} is satisfied by the values of $q$ in \cref{tab:quantum_results}.
The numbers of their local minima in $[0,2\pi)^d$, reported in \cref{tab:quantum_results}, are obtained from these expressions.

\bibliographystyle{quantum}
\bibliography{references}

@preamble{ "% DoNotMakeWarningsErrors" }

@misc{leng2023quantum,
  title         = {{Quantum Hamiltonian Descent}},
  author        = {Leng, Jiaqi and Hickman, Ethan and Li, Joseph and Wu, Xiaodi},
  year          = {2023},
  eprint        = {2303.01471},
  archivePrefix = {arXiv},
  primaryClass  = {quant-ph},
  doi           = {10.48550/arXiv.2303.01471},
  url           = {https://arxiv.org/abs/2303.01471}
}

@article{wibisono2016variational,
  title     = {A variational perspective on accelerated methods in optimization},
  author    = {Wibisono, Andre and Wilson, Ashia C. and Jordan, Michael I.},
  journal   = {Proceedings of the National Academy of Sciences},
  volume    = {113},
  number    = {47},
  pages     = {E7351--E7358},
  year      = {2016},
  publisher = {National Academy of Sciences},
  doi       = {10.1073/pnas.1614734113}
}

@article{childs2022quantum,
  title     = {Quantum simulation of real-space dynamics},
  author    = {Childs, Andrew M. and Leng, Jiaqi and Li, Tongyang and Liu, Jin-Peng and Zhang, Chenyi},
  journal   = {Quantum},
  volume    = {6},
  pages     = {860},
  year      = {2022},
  publisher = {Verein zur F{\"o}rderung des Open Access Publizierens in den Quantenwissenschaften},
  doi       = {10.22331/q-2022-11-17-860}
}

@article{jamil2013literature,
  title     = {A literature survey of benchmark functions for global optimisation problems},
  author    = {Jamil, Momin and Yang, Xin-She},
  journal   = {International Journal of Mathematical Modelling and Numerical Optimisation},
  volume    = {4},
  number    = {2},
  pages     = {150--194},
  year      = {2013},
  publisher = {Inderscience Publishers},
  doi       = {10.1504/IJMMNO.2013.055204}
}

@misc{jamil_yang_testfunctions,
  author = {Gavana, Andrea},
  title  = {Test Functions Index --- {AMPGO} 0.1.0 documentation},
  year   = {2013},
  howpublished = {\url{https://infinity77.net/global_optimization/test_functions.html}},
  url    = {https://infinity77.net/global_optimization/test_functions.html},
  note   = {Consulted on 26th January 2026}
}

@misc{surjanovic2013vlse,
  author = {Surjanovic, Sonja and Bingham, Derek},
  title  = {Virtual Library of Simulation Experiments: Test Functions and Datasets},
  year   = {2013},
  howpublished = {\url{https://www.sfu.ca/~ssurjano/}},
  url    = {https://www.sfu.ca/~ssurjano/},
  note   = {Consulted on 21st November 2025}
}

@article{cerezo2021variational,
  title     = {Variational quantum algorithms},
  author    = {Cerezo, Marco and Arrasmith, Andrew and Babbush, Ryan and Benjamin, Simon C. and Endo, Suguru and Fujii, Keisuke and McClean, Jarrod R. and Mitarai, Kosuke and Yuan, Xiao and Cincio, Lukasz and others},
  journal   = {Nature Reviews Physics},
  volume    = {3},
  number    = {9},
  pages     = {625--644},
  year      = {2021},
  publisher = {Nature Publishing Group},
  doi       = {10.1038/s42254-021-00348-9}
}

@article{albash2018adiabatic,
  title     = {Adiabatic quantum computation},
  author    = {Albash, Tameem and Lidar, Daniel A.},
  journal   = {Reviews of Modern Physics},
  volume    = {90},
  number    = {1},
  pages     = {015002},
  year      = {2018},
  publisher = {American Physical Society},
  doi       = {10.1103/RevModPhys.90.015002}
}

@article{mcclean2018barren,
  title     = {Barren plateaus in quantum neural network training landscapes},
  author    = {McClean, Jarrod R. and Boixo, Sergio and Smelyanskiy, Vadim N. and Babbush, Ryan and Neven, Hartmut},
  journal   = {Nature Communications},
  volume    = {9},
  number    = {1},
  pages     = {4812},
  year      = {2018},
  publisher = {Nature Publishing Group},
  doi       = {10.1038/s41467-018-07090-4}
}

@article{cerezo2022challenges,
  title     = {Challenges and opportunities in quantum machine learning},
  author    = {Cerezo, Marco and Verdon, Guillaume and Huang, Hsin-Yuan and Cincio, Lukasz and Coles, Patrick J.},
  journal   = {Nature Computational Science},
  volume    = {2},
  number    = {9},
  pages     = {567--576},
  year      = {2022},
  publisher = {Nature Publishing Group},
  doi       = {10.1038/s43588-022-00311-3}
}

@misc{Bennett.jl,
  title  = {Bennett.jl: Compiling {J}ulia to Reversible Circuits},
  author = {Tobias Osborne},
  howpublished = {\url{https://github.com/tobiasosborne/Bennett.jl}},
  url    = {https://github.com/tobiasosborne/Bennett.jl},
  year   = {2026}
}

@article{Low2019hamiltonian,
  doi = {10.22331/q-2019-07-12-163},
  url = {https://doi.org/10.22331/q-2019-07-12-163},
  title = {Hamiltonian {S}imulation by {Q}ubitization},
  author = {Low, Guang Hao and Chuang, Isaac L.},
  journal = {{Quantum}},
  issn = {2521-327X},
  publisher = {{Verein zur F{\"{o}}rderung des Open Access Publizierens in den Quantenwissenschaften}},
  volume = {3},
  pages = {163},
  month = jul,
  year = {2019}
}

@article{li2026learning,
  title={Learning parameterized quantum circuits with quantum gradient},
  author={Li, Keren and Wang, Yuanfeng and Gao, Pan and Zheng, Shenggen},
  journal={npj Quantum Information},
  volume={12},
  number={1},
  pages={59},
  year={2026},
  doi={10.1038/s41534-025-01179-7},
  publisher={Nature Publishing Group UK London}
}

@article{berry2020time,
  title={Time-dependent {H}amiltonian simulation with {$L^1$}-norm scaling},
  author={Berry, Dominic W and Childs, Andrew M and Su, Yuan and Wang, Xin and Wiebe, Nathan},
  journal={Quantum},
  volume={4},
  pages={254},
  year={2020},
  doi={10.22331/q-2020-04-20-254},
  publisher={Verein zur F{\"o}rderung des Open Access Publizierens in den Quantenwissenschaften}
}

@article{chen2021quantum,
  title={Quantum algorithm for time-dependent {H}amiltonian simulation by permutation expansion},
  author={Chen, Yi-Hsiang and Kalev, Amir and Hen, Itay},
  journal={PRX Quantum},
  volume={2},
  number={3},
  pages={030342},
  year={2021},
  doi={10.1103/PRXQuantum.2.030342},
  publisher={APS}
}

@article{an2022time,
  title={Time-dependent {H}amiltonian simulation of highly oscillatory dynamics and superconvergence for {S}chr{\"o}dinger equation},
  author={An, Dong and Fang, Di and Lin, Lin},
  journal={Quantum},
  volume={6},
  pages={690},
  year={2022},
  doi={10.22331/q-2022-04-15-690},
  publisher={Verein zur F{\"o}rderung des Open Access Publizierens in den Quantenwissenschaften}
}

@article{low2018hamiltonian,
  title={Hamiltonian simulation in the interaction picture},
  author={Low, Guang Hao and Wiebe, Nathan},
  journal={arXiv preprint arXiv:1805.00675},
  year={2018},
  eprint={1805.00675},
  archivePrefix={arXiv}
}

@article{kieferova2019simulating,
  title={Simulating the dynamics of time-dependent {H}amiltonians with a truncated {D}yson series},
  author={Kieferov{\'a}, M{\'a}ria and Scherer, Artur and Berry, Dominic W},
  journal={Physical Review A},
  volume={99},
  number={4},
  pages={042314},
  year={2019},
  doi={10.1103/PhysRevA.99.042314},
  publisher={APS}
}

@article{poulin2011quantum,
  title={Quantum Simulation of Time-Dependent {H}amiltonians and the Convenient Illusion of {H}ilbert Space},
  author={Poulin, David and Qarry, Angie and Somma, Rolando and Verstraete, Frank},
  journal={Physical Review Letters},
  volume={106},
  number={17},
  pages={170501},
  year={2011},
  doi={10.1103/PhysRevLett.106.170501},
  publisher={APS}
}

@article{mitarai2018quantum,
  title={Quantum circuit learning},
  author={Mitarai, Kosuke and Negoro, Makoto and Kitagawa, Masahiro and Fujii, Keisuke},
  journal={Physical Review A},
  volume={98},
  number={3},
  pages={032309},
  year={2018},
  publisher={APS},
  doi={10.1103/PhysRevA.98.032309}
}

@article{schuld2020effect,
  title={Effect of data encoding on the expressive power of variational quantum-machine-learning models},
  author={Schuld, Maria and Sweke, Ryan and Meyer, Johannes Jakob},
  journal={Physical Review A},
  volume={103},
  number={3},
  pages={032430},
  year={2021},
  doi={10.1103/PhysRevA.103.032430}
}

@article{schuld2019evaluating,
  title={Evaluating analytic gradients on quantum hardware},
  author={Schuld, Maria and Bergholm, Ville and Gogolin, Christian and Izaac, Josh and Killoran, Nathan},
  journal={Physical Review A},
  volume={99},
  number={3},
  pages={032331},
  year={2019},
  publisher={APS},
  doi={10.1103/PhysRevA.99.032331}
}

@article{rebentrost2019quantum,
  title={Quantum gradient descent and {N}ewton’s method for constrained polynomial optimization},
  author={Rebentrost, Patrick and Schuld, Maria and Wossnig, Leonard and Petruccione, Francesco and Lloyd, Seth},
  journal={New Journal of Physics},
  volume={21},
  number={7},
  pages={073023},
  year={2019},
  publisher={IOP Publishing},
  doi = {10.1088/1367-2630/ab2a9e}
}

@inproceedings{gilyen2019quantum,
  title={Quantum singular value transformation and beyond: exponential improvements for quantum matrix arithmetics},
  author={Gily{\'e}n, Andr{\'a}s and Su, Yuan and Low, Guang Hao and Wiebe, Nathan},
  booktitle={Proceedings of the 51st Annual ACM SIGACT Symposium on Theory of Computing},
  pages={193--204},
  year={2019},
  doi={10.1145/3313276.3316366}
}

@article{bharti2022noisy,
  title={Noisy intermediate-scale quantum algorithms},
  author={Bharti, Kishor and Cervera-Lierta, Alba and Kyaw, Thi Ha and Haug, Tobias and Alperin-Lea, Sumner and Anand, Abhinav and Degroote, Matthias and Heimonen, Hermanni and Kottmann, Jakob S and Menke, Tim and others},
  journal={Reviews of Modern Physics},
  volume={94},
  number={1},
  pages={015004},
  year={2022},
  doi={10.1103/RevModPhys.94.015004},
  publisher={APS}
}

@article{roland2002quantum,
  title     = {Quantum search by local adiabatic evolution},
  author    = {Roland, J{\'e}r{\'e}mie and Cerf, Nicolas J.},
  journal   = {Physical Review A},
  volume    = {65},
  number    = {4},
  pages     = {042308},
  year      = {2002},
  doi       = {10.1103/PhysRevA.65.042308}
}

@article{blanes2009magnus,
  title         = {The {Magnus} expansion and some of its applications},
  author        = {Blanes, Sergio and Casas, Fernando and Oteo, Jos{\'e} A. and Ros, Jos{\'e}},
  journal       = {Physics Reports},
  volume        = {470},
  number        = {5--6},
  pages         = {151--238},
  year          = {2009},
  doi           = {10.1016/j.physrep.2008.11.001},
  eprint        = {0810.5488},
  archivePrefix = {arXiv}
}

@article{abbas2023quantum,
  title={On quantum backpropagation, information reuse, and cheating measurement collapse},
  author={Abbas, Amira and King, Robbie and Huang, Hsin-Yuan and Huggins, William J and Movassagh, Ramis and Gilboa, Dar and McClean, Jarrod},
  journal={Advances in Neural Information Processing Systems},
  volume={36},
  pages={44792--44819},
  year={2023},
  doi={10.52202/075280-1940}
}

@misc{leng2025nonsmooth,
  title         = {{Quantum Hamiltonian Descent} for Non-smooth Optimization},
  author        = {Leng, Jiaqi and Zheng, Yufan and Jia, Zhiyuan and Fan, Lei and Zhao, Chaoyue and Peng, Yuxiang and Wu, Xiaodi},
  year          = {2025},
  eprint        = {2503.15878},
  archivePrefix = {arXiv},
  primaryClass  = {math.OC},
  doi           = {10.48550/arXiv.2503.15878},
  url           = {https://arxiv.org/abs/2503.15878}
}

@mastersthesis{burato2026thesis,
  author = {Burato, Marco},
  title  = {Ottimizzazione di funzioni classiche e quantistiche tramite {Quantum Hamiltonian Descent}},
  school = {University of Florence},
  year   = {2026},
  note   = {Defended on 22 April 2026},
  url = {https://www.fis-astro-lm.unifi.it/vp-251-archivio-delle-lauree-discusse-2024-2025.html}
}

@misc{acedo2026coherent,
  author        = {Acedo, Ignacio B. and Gonzalez-Conde, Javier and Rodriguez-Grasa, Pablo and Sanders, Barry C. and Pira, Lirand{\"e}},
  title         = {Quantum {H}amiltonian Evolution for Coherent Quantum Learning},
  year          = {2026},
  eprint        = {2609.03640},
  archivePrefix = {arXiv},
  primaryClass  = {quant-ph}
}

@article{banchi2021measuring,
    doi = {10.22331/q-2021-01-25-386},
    url = {https://doi.org/10.22331/q-2021-01-25-386},
    title = {Measuring {A}nalytic {G}radients of {G}eneral {Q}uantum {E}volution with the {S}tochastic {P}arameter {S}hift {R}ule},
    author = {Banchi, Leonardo and Crooks, Gavin E.},
    journal = {{Quantum}},
    issn = {2521-327X},
    publisher = {{Verein zur F{\"{o}}rderung des Open Access Publizierens in den Quantenwissenschaften}},
    volume = {5},
    pages = {386},
    month = jan,
    year = {2021}
}

@article{banchi2026overshifted,
    doi = {10.1088/2058-9565/ae73ad},
    url = {https://doi.org/10.1088/2058-9565/ae73ad},
    year = {2026},
    month = {jun},
    publisher = {IOP Publishing},
    volume = {11},
    number = {3},
    pages = {035031},
    author = {Banchi, Leonardo and Branford, Dominic and Waghela, Chetan},
    title = {Overshifted parameter-shift rules: optimizing complex quantum systems with few measurements},
    journal = {Quantum Science and Technology}
}

@article{anschuetz2022quantum,
  title={Quantum variational algorithms are swamped with traps},
  author={Anschuetz, Eric R and Kiani, Bobak T},
  journal={Nature Communications},
  volume={13},
  number={1},
  pages={7760},
  year={2022},
  publisher={Nature Publishing Group UK, London},
  doi ={10.1038/s41467-022-35364-5}
}

@misc{tognini2025solving,
  title         = {Solving {MNIST} with a globally trained mixture of quantum experts},
  author        = {Tognini, Paolo Alessandro Xavier and Banchi, Leonardo and De Palma, Giacomo},
  year          = {2025},
  eprint        = {2505.14789},
  archivePrefix = {arXiv},
  primaryClass  = {quant-ph},
  doi           = {10.48550/arXiv.2505.14789}
}

@article{Ostaszewski2021structure,
  doi = {10.22331/q-2021-01-28-391},
  url = {https://doi.org/10.22331/q-2021-01-28-391},
  title = {Structure optimization for parameterized quantum circuits},
  author = {Ostaszewski, Mateusz and Grant, Edward and Benedetti, Marcello},
  journal = {{Quantum}},
  issn = {2521-327X},
  publisher = {{Verein zur F{\"{o}}rderung des Open Access Publizierens in den Quantenwissenschaften}},
  volume = {5},
  pages = {391},
  month = jan,
  year = {2021}
}

@article{arrasmith2022equivalence,
  title         = {Equivalence of quantum barren plateaus to cost concentration and narrow gorges},
  author        = {Arrasmith, Andrew and Holmes, Zo{\"e} and Cerezo, M. and Coles, Patrick J.},
  journal       = {Quantum Science and Technology},
  volume        = {7},
  number        = {4},
  pages         = {045015},
  year          = {2022},
  doi           = {10.1088/2058-9565/ac7d06},
  eprint        = {2104.05868},
  archivePrefix = {arXiv}
}

@article{banchi2020convex,
  title={Convex optimization of programmable quantum computers},
  author={Banchi, Leonardo and Pereira, Jason and Lloyd, Seth and Pirandola, Stefano},
  journal={npj Quantum Information},
  volume={6},
  number={1},
  pages={42},
  year={2020},
  doi={10.1038/s41534-020-0268-2},
  publisher={Nature Publishing Group UK, London}
}

@article{pirandola2017fundamental,
  title={Fundamental limits of repeaterless quantum communications},
  author={Pirandola, Stefano and Laurenza, Riccardo and Ottaviani, Carlo and Banchi, Leonardo},
  journal={Nature communications},
  volume={8},
  number={1},
  pages={15043},
  year={2017},
  doi={10.1038/ncomms15043},
  publisher={Nature Publishing Group UK, London}
}

@article{campbell2019random,
  title   = {Random Compiler for Fast {H}amiltonian Simulation},
  author  = {Campbell, Earl},
  journal = {Physical Review Letters},
  volume  = {123},
  number  = {7},
  pages   = {070503},
  year    = {2019},
  doi     = {10.1103/PhysRevLett.123.070503}
}

\end{document}